\documentclass[11p,a4paper]{article}
\pdfoutput=1
\usepackage{graphicx}
\usepackage{jheppub}
\usepackage[utf8]{inputenc}
\usepackage{latexsym,amsfonts,amsmath,amscd,amssymb,epsf}
\usepackage{hyperref}
\usepackage{soul,xcolor}
\usepackage{natbib}
\usepackage{enumitem}
\usepackage{graphicx}
\usepackage{subfig}

\newcommand{\beq}{\begin{equation}}
\newcommand{\eeq}{\end{equation}}

\DeclareMathAlphabet{\pazocal}{OMS}{zplm}{m}{n}

\renewcommand{\v}[1]{\mathbf{#1}}
\usepackage{bm}
\let\v=\bm

\title{A Coherent View of the Quark-Gluon Plasma from Energy Correlators}

\author[a]{Carlota Andres,}
\author[b]{Fabio Dominguez,}
\author[a]{Jack Holguin,}
\author[a]{Cyrille Marquet,}
\author[c]{Ian Moult}
\emailAdd{carlota.andres-casas@polytechnique.edu}
\emailAdd{fabio.dominguez@usc.es}
\emailAdd{jack.holguin@polytechnique.edu}
\emailAdd{cyrille.marquet@polytechnique.edu} 
\emailAdd{ian.moult@yale.edu} 

\affiliation[a]{CPHT, CNRS, \'Ecole polytechnique,
Institut Polytechnique de Paris, 91120 Palaiseau, France}
\affiliation[b]{Instituto Galego de F\'isica de Altas Enerx\'ias (IGFAE), Universidade de Santiago de Compostela, Santiago de Compostela 15782, Spain}
\affiliation[c]{Department of Physics, Yale University, New Haven, CT 06511}
\newcommand{\comment}[1]{}

\newcommand{\td}{\mathrm{d}}
\newcommand{\SO}[1]{\mathrm{SO}(#1)}

\newcommand{\TT}[1]{\mathrm{#1}}

\renewcommand{\>}{\right\rangle}
\newcommand{\<}{\left\langle}
\newcommand{\lkl}{\left|}
\newcommand{\rkl}{\right|}

\newcommand{\As}{\alpha_{\mathrm{s}}}

\newcommand{\cE}{\mathcal{E}}

\begin{document}

\abstract{The ability to measure detailed aspects of the substructure of high-energy jets traversing the quark-gluon plasma (QGP) has provided a new window into its internal dynamics. However, drawing robust conclusions from traditional jet substructure observables has been difficult.
In this manuscript we expand on a new approach to jet substructure in heavy-ion collisions based on the study of correlation functions of energy flow operators (energy correlators).
We compute the two-point energy correlator of an in-medium massless quark jet and perform a detailed numerical analysis of the produced spectra. Our calculation incorporates vacuum radiation resummed at next-to-leading log accuracy together with the leading order contribution in medium-induced splittings evaluated through the BDMPS-Z multiple scattering and GLV single scattering formalisms for a static brick of QGP.
Our analysis demonstrates how particular features of the modifications of in-medium splittings are imprinted in the correlator spectra, particularly showing  how energy correlators may be used to extract  the onset of colour coherence. We further present a comprehensive discussion on the accuracy and limitations of our study emphasizing how it can be systematically improved. This work sets the foundations for a rich program studying energy correlators in heavy-ion collisions.
}

\maketitle

\section{Introduction}
\label{sec:intro}

The successful results of the heavy-ion runs at the Relativistic Heavy Ion Collider (RHIC) and at the Large Hadron Collider (LHC) \cite{PHOBOS:2004zne,Muller:2006ee,Muller:2012zq} have opened up a rich program in advancing the understanding of the strongly interacting matter created under extreme conditions. For recent reviews see \cite{Connors:2017ptx,Busza:2018rrf,Dexheimer:2020zzs,Cunqueiro:2021wls,Apolinario:2022vzg}. This strongly interacting matter, referred to as the quark-gluon plasma (QGP), provides a unique opportunity to study free quarks and gluons, as well as the phase structure of Quantum Chromodynamics (QCD), and may have applications ranging from understanding neutron stars, to the dynamics of the early universe \cite{Dexheimer:2020zzs}.

One of the most recent exciting advances in collider physics, both theoretically and experimentally, has been the development of the field of jet substructure, which uses the detailed internal structure of highly energetic jets to determine the properties of the underlying microscopic collisions. While jet substructure had its origins in searches for physics beyond the Standard Model in p-p collisions (see \cite{Dasgupta:2013ihk,Larkoski:2013eya,Larkoski:2017jix,Asquith:2018igt,Marzani:2019hun} for reviews), it also provides an excellent tool to study the QGP created in heavy-ion collisions. Indeed, in heavy-ion collisions, high-energy quarks and gluons are produced in the initial hard scattering and propagate through the evolving strongly interacting system. An imprint of the system's evolution is then left in the internal structure of the reconstructed jets. For these reasons, jet substructure has attracted significant interest from the heavy-ion community, rapidly becoming one of the most promising approaches to studying the QGP \cite{Salgado:2003rv,Andrews:2018jcm,Cunqueiro:2021wls,CMS:2013lhm,CMS:2018zze,Chien:2015hda,Chien:2016led,Connors:2017ptx,Apolinario:2017qay,Caucal:2018dla,Ringer:2019rfk,Vaidya:2020lih,Caucal:2021cfb,Mehtar-Tani:2021fud,Milhano:2022kzx,Andres:2022ovj}.

Jet substructure studies in heavy-ion collisions aim at disentangle the properties of the QGP by looking at the modifications of the jets' inner structure in A-A with respect to p-p collisions (vacuum). This places high demands on the theoretical understanding of QCD both in vacuum and in medium. Due to the extraordinary complexity of the system created in these collisions, several phenomena emerge, such as  colour coherence \cite{Mehtar-Tani:2010ebp,Mehtar-Tani:2011hma,Casalderrey-Solana:2011ule,Mehtar-Tani:2012mfa} and medium response \cite{Cao:2020wlm,CMS:2021otx}, which complicate the extraction of robust conclusions about the inner dynamics of the QGP from traditional jet substructure observables. Ideally, one would like to formulate a program of jet substructure measurements in heavy-ion collisions based on a set of observables which simultaneously have simple theoretical properties and are sensitive in clean ways to specific features of the QGP. Unfortunately, this has been proven to be far from straightforward for standard jet shape observables.

In this paper we formulate the study of jet substructure in heavy-ion collisions in terms of a simple class of \emph{inclusive} observables known as energy correlators, which are correlation functions of energy flow operators $\langle \cE(\vec n_1)\cdots\cE(\vec n_k)\rangle$ \cite{Basham:1977iq,Basham:1978bw,Basham:1978zq,Basham:1979gh,Hofman:2008ar}, where $\cE(\vec n_1)$ measures the asymptotic energy flux in the direction $\vec n_1$ \cite{Hofman:2008ar,Ore:1979ry,Korchemsky:1999kt,Belitsky:2013xxa}. In vacuum, the observables  present a clear separation between the non-perturbative and perturbative regimes, with the latter being known to very high precision (typically NNLL \cite{Dixon:2019uzg}). Within their perturbative regime, they exhibit a smooth power law behaviour \cite{Hofman:2008ar,Lee:2022ige}, with sharp transitions arising at the presence of any additional scale in the problem \cite{Holguin:2022epo,Craft:2022kdo}.
Additionally, due to their inclusivity, their theoretical description keeps some simplicity compared to more exclusive observables, even when applied to very complicated environments with potentially large backgrounds. Indeed, it is for this reason that close relatives to energy correlators, temperature correlators, are used to study the cosmic-microwave background (CMB) \cite{Planck:2018vyg}, the big bang, and inflationary cosmology \cite{Arkani-Hamed:2018kmz}. In all, this makes energy correlators an ideal candidate for jet substructure in heavy-ion collisions, opening the door for robustly identifying the dynamics associated with specific scales of the QGP, whilst precisely controlling other independent factors such as quark/gluon fractions. 

In order to illustrate how the features of a given jet quenching formalism are imprinted into the energy correlators spectra, we study in this manuscript the two-point correlator (EEC) of a massless quark-initiated jet,  which we compute perturbatively as a spray of partons. The calculation allows us to access colour coherence properties of the interaction between the jet and the medium. We compute the EEC spectra within different jet quenching formalisms for a static brick of QGP. First of all, we make use of a semi-hard implementation \cite{Dominguez:2019ges, Isaksen:2020npj} of the multiple scattering BDMPS-Z formalism \cite{Baier:1996kr,Baier:1996sk,Zakharov:1996fv,Zakharov:1997uu} for medium-induced radiation using two different parton-medium interaction models: a gaussian (also referred to as the Harmonic Oscillator) and a Yukawa (Gyulassy-Wang) model. Independently, we obtain the two-point correlator within the complete single scattering GLV framework \cite{Gyulassy:2000fs,Gyulassy:2000er,Wiedemann:2000za,Ovanesyan:2011kn} using a Yukawa parton-medium interaction model. We find that while the specific details of the medium-induced formalism employed are imprinted into the detailed behaviour of the two-point correlator, the ability of this observable to identify the onset of colour coherence is independent of the jet quenching formalism employed. This illustrates how the energy correlators can isolate the dynamics of the QGP at a particular scale. A brief presentation of our results for the Harmonic Oscillator approach was previously given in \cite{Andres:2022ovj}, here we significantly extend the discussion by considering all the jet quenching frameworks mentioned above. 

The analysis presented in this work is meant to be an initial theoretical exploration of energy correlators in a heavy-ion environment, aiming at showing how the features of the underlying formalism for jet-medium interactions are reflected into an EEC-type observable. For this purpose, we adopted a simple model for the QGP, which needs to be upgraded to a much more realistic implementation before any meaningful comparisons to experimental data could be considered. Given the current limitations and scope of the study, we refrain from delving into the specifics of potential measurements and how the results might differ for different event selections. These aspects can be more thoroughly addressed in future publications.  Nevertheless, it is useful to keep in mind that the processes of greatest interest to this observable are the associated production of $\gamma$+jet or $Z$+jet, as previously noted in \cite{Andres:2022ovj}, since the hard scale necessary to measure the EEC  can be taken as either the energy or $p_{\rm T}$ of the tagged $\gamma$/$Z$.  We note, however, that we will not apply in this study any jet algorithm, rather the jet is the collimated spray of hadrons antipodal to the axis of the tagged $\gamma$/$Z$ (referred in the literature as hemisphere-jets \cite{Dasgupta:2001sh,Chien:2010kc}). Therefore any considerations about jet radii and $p_T$ cuts are postponed for future works.

An outline of this paper is as follows. In section~\ref{sec:EEC_promo} we provide a general overview of the use of energy correlators in jet substructure, highlighting their  advantages for elucidating the inner dynamics of the QGP. In section~\ref{sec:two-point} we present in detail the calculation of the two-point energy correlator of an in-medium massless quark jet, emphasising the approximations used, and describing the different theoretical approaches taken for the treatment of the medium-induced radiation. In section~\ref{sec:results} we present and discuss the numerical evaluation of this two-point correlator, showing that a consistent picture of colour coherence emerges from the different jet quenching formalisms considered. Section~\ref{sec:discuss} is dedicated to a technical discussion on the limitations of our current theoretical framework, particularly highlighting the parametric size of subleading corrections not accounted for. Finally, we summarise and conclude in section~\ref{sec:conclusions}. We further provide two appendices: in appendix~\ref{app:app_1} we give a detailed review of the EEC distribution in vacuum, and in appendix~\ref{app:figs} we expand on the numerical results presented in section~\ref{sec:results}.

\section{Energy Correlators as a Probe of the Quark-Gluon Plasma}
\label{sec:EEC_promo}

\begin{figure}
\begin{center}
\includegraphics[scale=0.3]{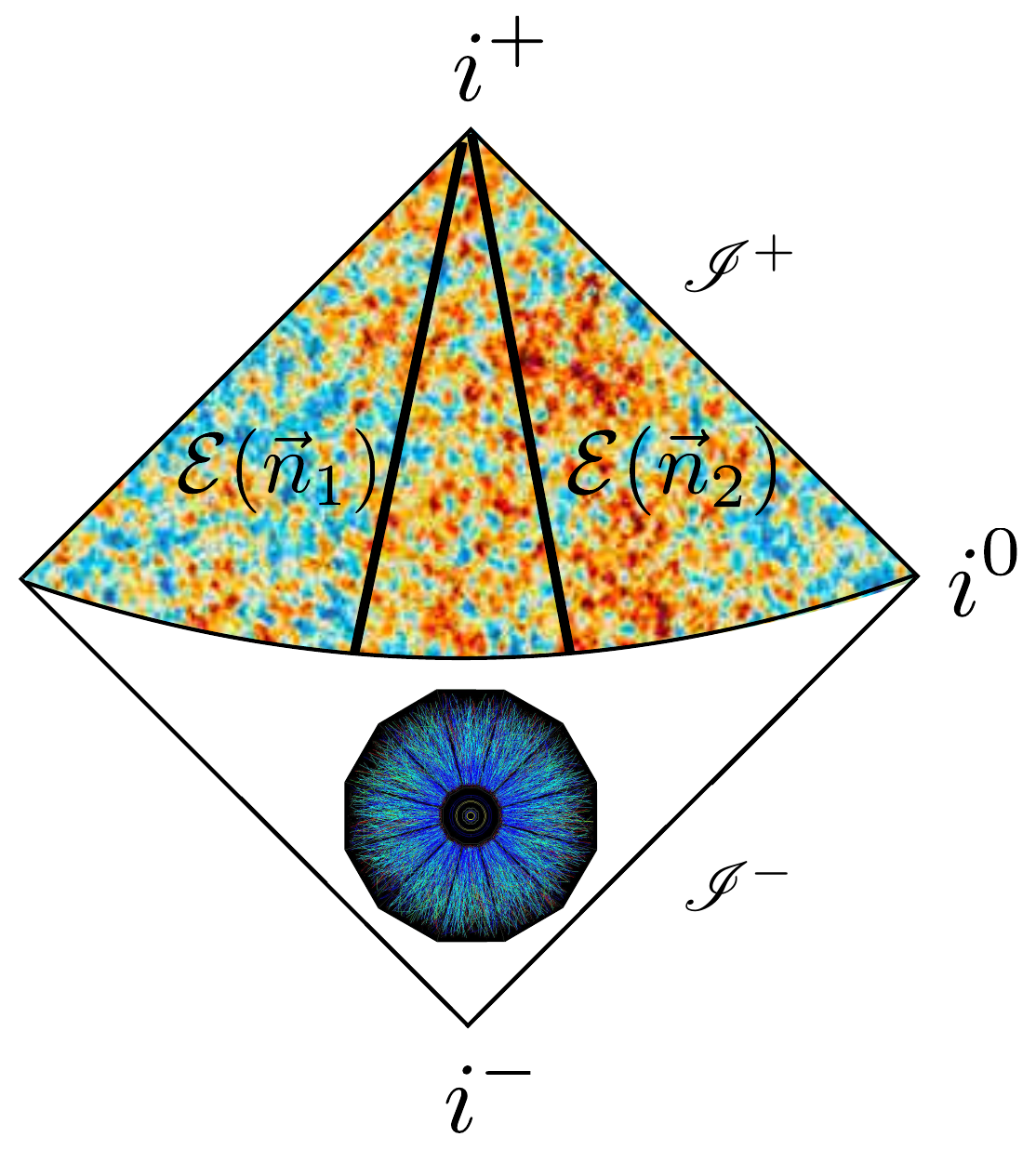}
\end{center}
\caption{Just as the CMB radiation from the Big Bang propagates to null infinity, where correlations in the temperature are measured \cite{Planck:2018vyg}, so does the radiation produced in the ``Little Bang" in relativistic heavy-ion collisions. In this case, the radiation is measured by energy flux operators $\mathcal{E}(\vec n)$ at null infinity. \emph{Time} scales in the microscopic collision are imprinted in \emph{angular} scales of the energy distribution on the celestial sphere.}
\label{fig:intro_EEC}
\end{figure}

The standard approach to studying  jet substructure in collider experiments is through the use of \emph{jet shapes}, which are observables that are sensitive in some manner to the shape of radiation in a jet. Examples of jet shape observables include thrust, jet mass, angularities, etc. Jet shapes received significant attention in early work on jet substructure in p-p collisions, and thus studying their modifications in the presence of a QGP became a natural approach to studying jets in heavy-ion collisions. However, addressing how a given \emph{shape} is modified by the jet interactions with the medium is extremely difficult, since a jet's shape contains competing dynamics at many different scales, each of which may be modified in a different way. We would instead like to have jet substructure observables that are sensitive to the QGP's dynamics at a given scale, which requires a completely novel approach to jet substructure.

Instead of considering the shapes or splitting histories of a jet, we can think of the pattern of energy flux deposited on the detectors, which can be viewed as the celestial sphere. This is illustrated in figure~\ref{fig:intro_EEC} in the form of a Penrose diagram. Averaged over many events, this produces a density field of energy flux, much like the CMB, or like a two-dimensional condensed matter system. Statistical properties of the energy flux can be measured using a light-ray operator, $\mathcal{E}(\vec{n}_{1})$, which measures the asymptotic energy flux  in the direction $\vec{n}_{1}$ \cite{Ore:1979ry,Korchemsky:1999kt,Hofman:2008ar,Belitsky:2013xxa}, and is defined as
\beq
    \cE(\vec n_1) = 
    \lim_{r\rightarrow \infty} \int \mathrm{d}t \,r^2 n_1^i \,
    T_{0i}(t,r\vec{n}_1)\,,
    \label{eq:energy_flux}
\eeq
where $T_{\mu \nu}$ is the stress-energy tensor of the field theory. In the Penrose diagram in figure~\ref{fig:intro_EEC} these operators are shown as lines, illustrating the integral over time. Multi-point correlation functions of the energy flow operators, $\langle \mathcal{E}(\vec{n}_{1}) \mathcal{E}(\vec{n}_{2})\cdots \mathcal{E}(\vec{n}_{k}) \rangle$ \cite{Basham:1979gh,Basham:1978zq,Basham:1978bw,Basham:1977iq,Hofman:2008ar}, then describe the structure of the energy flux. Due to recent advances in field theory, these correlation functions can now be computed over a large hierarchy of angles, and have been measured in Open Data inside high-energy jets in p-p collisions \cite{Komiske:2022enw}.

\begin{figure}
\begin{center}
\subfloat[]{
\includegraphics[scale=0.25]{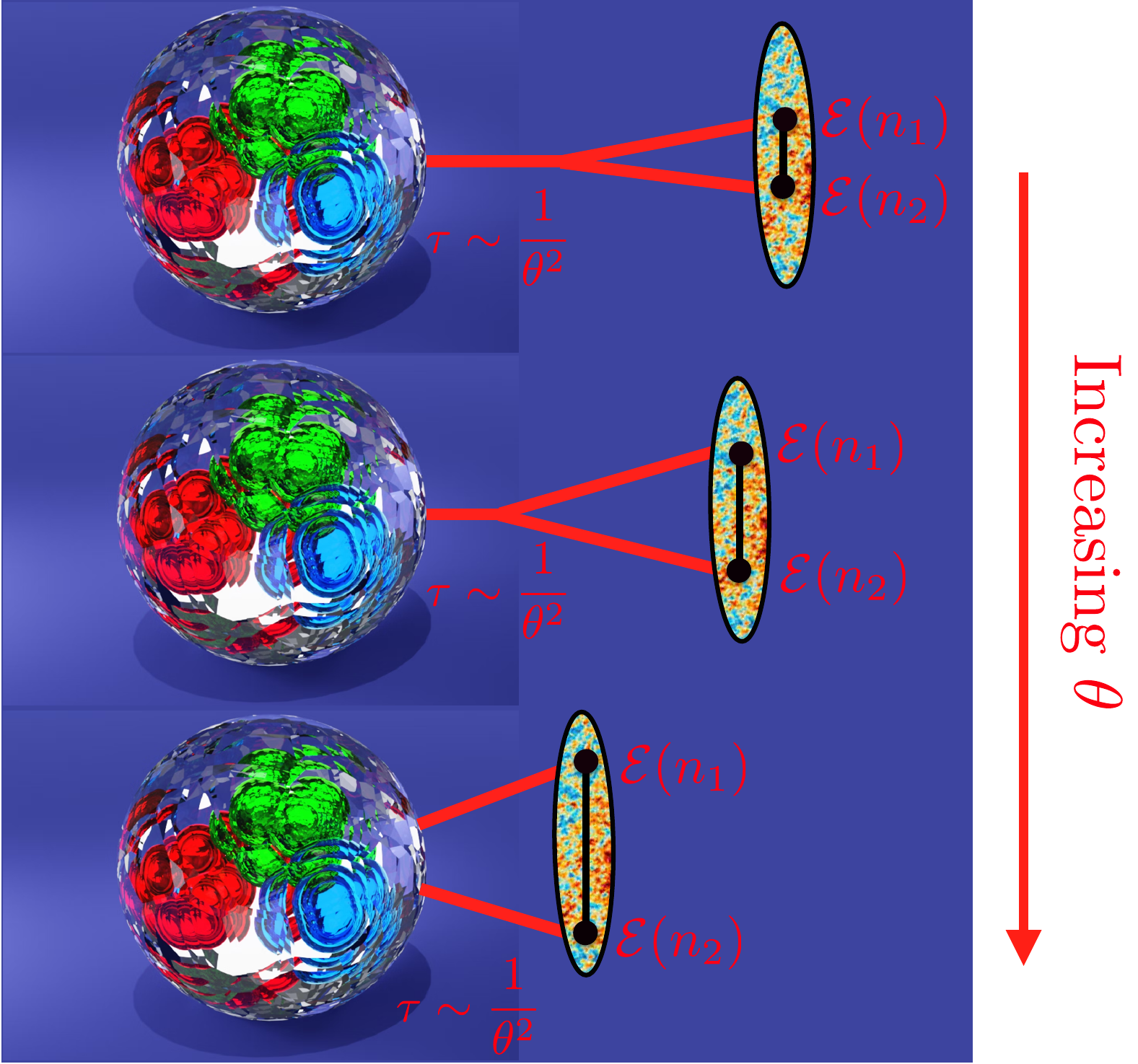}\label{fig:resolve_a}}\qquad
\subfloat[]{
\includegraphics[scale=0.3]{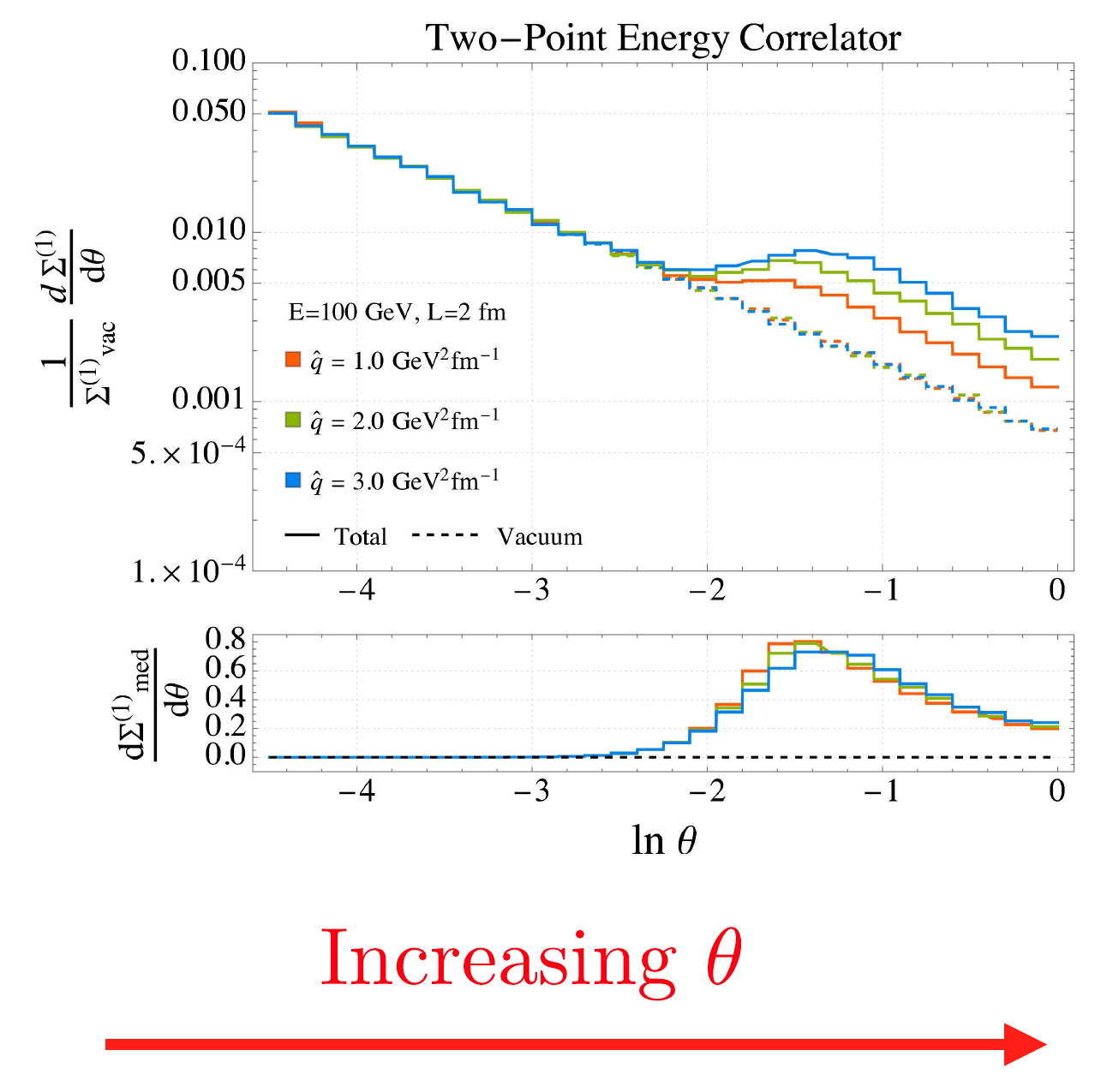}\label{fig:resolve_b}}\quad
\end{center}
\caption{(a) Angular scales in the two-point energy correlator map the time evolution of the jet. (b) The dynamics associated with the interactions with the QGP are clearly imprinted at a particular angular scale. Panel (b) first presented in \cite{Andres:2022ovj}.}
\label{fig:resolve}
\end{figure}

The key observation is that time scales in the evolution of the jet are imprinted into \emph{angular} scales in the asymptotic energy flux. Therefore, by studying the structure of the correlators at a given angular region, one can access the dynamics at a specific time scale in the evolution of the jet.\footnote{For other recent approaches to accessing different timescales of the system's evolution
with jet quenching observables see e.g.~\cite{Apolinario:2017sob,Andres:2019eus,Apolinario:2020uvt,Andres:2022bql,Apolinario:2022guz}.} For a variety of applications of the energy correlators for identifying scales in QCD systems, see \cite{Komiske:2022enw,Holguin:2022epo,Craft:2022kdo,Andres:2022ovj,Liu:2022wop,Liu:2023aqb,Cao:2023rga}. This is of course not an unfamiliar idea, since due to precisely the same reason correlation functions in the CMB are used to access particular stages in the time evolution of the universe. Modifications to the energy flux within a jet, associated with different scales in the QGP, will therefore be imprinted at different angular regions in the correlators, as  schematically illustrated for the two-point correlator of an in-medium jet in figure~\ref{fig:resolve}. 

Energy correlators present several other characteristics which may make them excellent candidates for a jet substructure program in heavy-ion physics.  First, since the modifications of jet substructure observables due to the presence of the QGP are analysed with respect to a vacuum baseline, it is highly desirable for this baseline to be featureless and independent of non-perturbative effects. In vacuum, normalised energy correlators exhibit a clear angular separation between the non-perturbative and perturbative regimes \cite{Komiske:2022enw}, presenting in the latter a featureless power-law behaviour indicative of massless QCD being asymptotically conformal \cite{Hofman:2008ar,Lee:2022ige}. This power law is known perturbatively for both quark and gluon jets at NLO+NNLL accuracy, see \cite{Dixon:2018qgp,Dixon:2019uzg} and appendix~\ref{app:app_1} and obeys rigorously understood factorisation theorems \cite{Chen:2020vvp}. Thus, in heavy-ions the presence of additional scales due to the QGP formation is expected to result in clear changes in this power-law behaviour that cannot be attributed to other modifications, such as quark/gluon fractions. 

Second, energy correlators isolate single logarithmic collinear physics, which is expected to be robust to the underlying soft backgrounds. Furthermore, they can be weighted by additional powers $n$ of the energy, $\mathcal{E}^n(\vec{n}_1)$, to further suppress soft radiation while remaining theoretically tractable. For the same reason, they can be computed efficiently on tracks \cite{Chen:2020vvp,Li:2021zcf,Jaarsma:2022kdd,Chang:2013rca,Chang:2013iba}.  They achieve this insensitivity to soft radiation without the need for grooming algorithms, due to the fact that they are energy weighted, and inclusive over extra radiation, so they are not Sudakov observables.  This is particularly interesting in the heavy-ion context, where grooming techniques have not been tailored to the presence of a large underlying event, which can lead to a substantial number of splittings to be misidentified \cite{Mulligan:2020tim}.

It is also worth comparing the energy correlator with other observables based on declustering algorithms \cite{Larkoski:2015lea,Larkoski:2017bvj}, such as the groomed energy fraction $z_{\rm g}$ and groomed angle $\theta_{\rm g}$. The groomed energy fraction $z_{\rm g}$ gives access to the collinear dynamics of the jet while being insensitive to soft radiation. However, the energy fraction is not associated with a particular time scale in the jet, making it not ideal for studying the dynamics of the QGP, which is more naturally mapped into an angular variable. Contrary to $z_{\rm g}$, which is extraordinarily insensitive to non-perturbative and higher order contributions \cite{Larkoski:2015lea},  $\theta_{\rm g}$ may be associated with particular time scales, but it is highly sensitive to non-perturbative and higher order contributions, and is not well described by parton showers and analytic calculations in p-p collisions \cite{Tripathee:2017ybi}.  Additionally, there have also been attempts to be  further differential by studying in-medium modifications to the primary Lund plane of the jet radiation \cite{Andrews:2018jcm,Cunqueiro:2022svx}. This approach can be roughly understood as differential in both $\theta_{\rm g}$ and $z_{\rm g}$, and thus, whilst potentially accessing novel physics, it also comes with the downside that the drawbacks of both observables are simultaneously present. In contrast to $\theta_{\rm g}$ and the Lund plane, the power-law behaviour displayed by energy correlators in vacuum is insensitive to non-perturbative effects and higher order corrections for angles larger than $\sim \Lambda_{\rm QCD}/Q$, where $Q$ is a hard scale usually taken as the initial jet energy or $p_T$ \cite{Chen:2022jhb,Chen:2022swd}, thus providing a robust angular variable \cite{Andres:2022ovj}, which is otherwise very hard to find. 


\section{Calculating the Two-Point Correlator of an In-Medium Jet}
\label{sec:two-point}

Having described our general approach to probing the dynamics of the QGP using energy correlators, in this section we focus on the specific case of the simplest two-point energy correlator of an in-medium jet. We describe in detail how this observable can be computed incorporating interactions with the medium within several jet quenching formalisms.

\subsection{The Collinear Limit of the Energy Correlators}
\label{sec:errors}

The $n$-th  weighted normalised two-point correlator can be written in terms of the inclusive cross-section $\sigma_{ij}$  to produce two hadrons ($i,j$) as
\beq
    \frac{\< \cE^{n}(\vec n_1) \cE^{n}(\vec n_2) \>}{Q^{2n}}  
    =\frac{1}{\sigma} \sum_{ij} \int \frac{\td \sigma_{ij}}{\td \vec n_i \td \vec n_j } \frac{E^{n}_{i} E^{n}_{j}}{Q^{2n}} \delta^{(2)}(\vec n_i - \vec n_{1})\,\delta^{(2)}(\vec n_j - \vec n_{2})\,,
\eeq
where $E_{i}$ is the lab-frame energy of a final state hadron $i$, and $Q$ is an appropriate hard scale. $\sigma(Q)$ is the integrated cross-section for the studied process. In an isotropic environment, 3 of the degrees of freedom in $\{\vec n_1,\vec n_2\}$ correspond $\SO{3}$ symmetries. Hence, we will study the distribution
\beq
    \frac{\td \Sigma^{(n)}}{\td \theta} = \int \td \vec n_{1,2}\, \frac{\< \cE^{n}(\vec n_1) \cE^{n}(\vec n_2) \>}{Q^{2n}} \,\delta(\vec n_2 \cdot \vec n_{1} - \cos \theta)\,. 
\eeq
Here we have introduced the short-hand notation $\td x_{1,2} = \td x_{1} \td x_{2}$.

Let us a consider a situation where a highly energetic massless quark-initiated jet propagates through the QGP. Here the relevant hard scale, $Q$, is the jet energy. In an experimental realisation of an EEC measurement, that scale can only be determined by inference from the detected final state particles. Typically this is done by using a jet algorithm, however doing so transforms the measurement from an inclusive observable to an exclusive one. Therefore, the processes of greatest interest to this observable will be the associated production of $\gamma+$jet or $Z+$jet, in which case $Q$ would be either the energy or $p_{T}$ of the tagged $\gamma/Z$.\footnote{If a jet algorithm must be used, the goal will be to let the radius be as large as possible (ideally twice the angular scale studied) so that edge effects from the cone boundary are small \cite{Komiske:2022enw}, thus letting the energy weighting in the correlator tame the backgrounds rather than the cone size. The hard scale should always be defined by the vector boson to remove the effects of a bias on the jet energy scale.} We are interested in the relatively wide angle region of medium modification. As the average momentum exchange between the two correlator points goes as $\sim \theta Q$, the wide angle region where $\theta Q \gg \Lambda_{\mathrm{QCD}}$ is largely determined by perturbative physics. We therefore write the observable as a sum over inclusive partonic cross-sections:
\begin{align}
    \frac{\td \Sigma^{(n)}}{\td \theta} =& \frac{1}{\sigma}\int  \td E_{q,g}  \frac{\td \hat{\sigma}_{qg}}{\td \theta \td E_{q} \td E_{g} } \frac{E_{g}^n E_{q}^n}{Q^{2n}} + \frac{1}{\sigma}\int  \td E_{g_{1},g_{2}}  \frac{\td \hat{\sigma}_{g_{1}g_{2}}}{\td \theta \td E_{g_{1}} \td E_{g_{2}} } \frac{E_{g_{1}}^n E_{g_{2}}^n}{Q^{2n}} \nonumber \\
    &+ \frac{1}{\sigma}\int  \td E_{q_{1},q_{2}}  \frac{\td \hat{\sigma}_{q_{1}q_{2}}}{\td \theta \td E_{q_{1}} \td E_{q_{2}} } \frac{E_{q_{1}}^n E_{q_{2}}^n}{Q^{2n}} + (\mathrm{perm.}~ q \leftrightarrow \bar{q} ) + \mathcal{O}\left(\frac{\Lambda_{\mathrm{QCD}}}{\theta ~ Q}\right)\,,
\end{align}
where $\hat{\sigma}_{ij}$ is the inclusive cross-section to produce final state partons $i,j$.\footnote{Note that the given result is for single flavour QCD. Additional sums over quark flavours should be included for multi-flavour QCD.} $\hat{\sigma}_{ij}$ still contains initial state non-perturbative physics, such as (n)PDFs. This factorisation of final state perturbative and non-perturbative physics is of Collins-Soper-Sterman \cite{Collins:1985ue,Collins:1988ig} form and has been proven for energy correlator observables with hadronic initial states \cite{Chen:2020vvp}. In principle one can convolute the given partonic result with fragmentation functions or track functions as appropriate for a collider measurement.

We define a partition between the vacuum and medium physics in the partonic cross-sections as
\beq
    \frac{\td \hat{\sigma}_{ij}}{\td \theta \td E_{i} \td E_{j} } = \left(1 + F^{(ij)}_{\rm med}( E_{i}, E_{j},\theta) \right)\,\frac{\td \hat{\sigma}^{\mathrm{vac}}_{ij}}{\td \theta \td E_{i} \td E_{j}}\,,
\eeq
where $F^{(ij)}_{\rm med}$ is defined to be the medium modification to a given $i,j$ vacuum partonic cross-section. Using this definition, we can resum the vacuum inclusive-jet contributions to the observable (labelled $J^{(n)}_{\mathrm{EEC}}$ in the literature \cite{Chen:2019bpb}) into a closed form expression using the celestial operator product expansion (OPE):
\begin{align}
    \frac{1}{\sigma}\sum_{ij \in \{g,q,\bar{q}\}} \int  \td E_{i,j}  \frac{\td \hat{\sigma}^{\mathrm{vac}}_{ij}}{\td \theta \td E_{i} \td E_{j} } \frac{E_{i}^n E_{j}^n}{Q^{2n}} \bigg|_{n=1}\equiv J^{(1)}_{\mathrm{EEC}}  = C \frac{1}{\theta^{1-\gamma(3)}} + \mathcal{O}(\theta^{0}) \,, \label{eq:vacPart}
\end{align}
where $\gamma(3)$ is the twist-2 spin-3 QCD anomalous dimension at fixed coupling.\footnote{The running coupling slightly breaks the simple exponentiation of the anomalous dimension~\cite{Chen:2021gdk}, see \cite{Chen:2022jhb} and appendix \ref{app:app_1}.} The expression becomes more complicated with $n>1$, however crucially one still gets the $\theta^{-1}$ scaling behaviour at leading order in $\As$. We provide in appendix \ref{app:app_1} a review of the celestial OPE and the relevant results. $C$ is a constant (up to the running coupling) that will not be relevant for our analysis. For completeness:
$$ C = \frac{\As }{2 \pi} \frac{9 C_{\mathrm{F}}}{2}+ \mathcal{O}(\As^2)\,.$$

Now we must turn our attention to the terms dependent on $F^{(ij)}_{\rm med}$. Since we are here focusing on the propagation of a quark jet through the medium, we expect $\td \hat{\sigma}_{qg}$ to provide the leading contribution to the medium enhancement. Specifically, relative to $\td \hat{\sigma}_{qg}$, the other cross-sections will be suppressed by at least a factor of $\As(\theta Q)$. We are therefore motivated to introduce a momentum fraction $z=E_{g}/Q$ and a scale $\mu_{\mathrm{s}}$ such that
\begin{align}
    \int  \td E_{q,g} ~ F^{(qg)}_{\rm med} \frac{\td \hat{\sigma}^{\mathrm{vac}}_{qg}}{\td \theta \td E_{q} \td E_{g} } \frac{E_{q}^n E_{g}^n}{Q^{2n}} = \int  \td z\, \td \mu_{\mathrm{s}} ~ F^{(qg)}_{\rm med} \frac{\td \hat{\sigma}^{\mathrm{vac}}_{qg}}{\td \theta \td z \td \mu_{\mathrm{s}} }z^n (1 - z - \mu_{\mathrm{s}}/Q)^n.
\end{align}
Written in this form, $\mu_{\mathrm{s}}$ can be interpreted as the energy scale of the radiation over which $F^{(qg)}_{\rm med}$ is inclusive for a given $(z,\theta)$. By momentum conservation, one necessarily has that 
\begin{align}
    F^{(g_{1}g_{2})}_{\rm med} \frac{\td \hat{\sigma}^{\mathrm{vac}}_{g_{1}g_{2}}}{\td \theta \td E_{g_{1}} \td E_{g_{2}} } \frac{E_{g_{1}}^n E_{g_{2}}^n}{Q^{2n}} \leq  F^{(g_{1}g_{2})}_{\rm med} \frac{\td \hat{\sigma}^{\mathrm{vac}}_{g_{1}g_{2}}}{\td \theta \td E_{g_{1}} \td E_{g_{2}} }z^n (\mu_{\mathrm{s}}/Q)^n, \\
    F^{(q_{1}q_{2})}_{\rm med} \frac{\td \hat{\sigma}^{\mathrm{vac}}_{q_{1}q_{2}}}{\td \theta \td E_{q_{1}} \td E_{q_{2}} } \frac{E_{q_{1}}^n E_{q_{2}}^n}{Q^{2n}} \leq  F^{(q_{1}q_{2})}_{\rm med} \frac{\td \hat{\sigma}^{\mathrm{vac}}_{q_{1}q_{2}}}{\td \theta \td E_{q_{1}} \td E_{q_{2}} }(1-z)^n (\mu_{\mathrm{s}}/Q)^n.
\end{align}
Following conventional arguments of strong ordering, the $\td \hat{\sigma}_{gg}$ and $\td \hat{\sigma}_{qq}$ type cross-sections are largest when $\mu_{\mathrm{s}}/Q \ll z$ and $\ln \theta^{-1} \gg 1$. Hence, we can identify each term which depends on $F^{(ij)}_{\rm med}$ for $(i,j) \neq (q,g)$ as sub-leading, maximally scaling as $\mathcal{O}(\As(\theta Q) \ln \theta ~ \mu^{n}_{\mathrm{s}}/Q^{n})$ relative to $\td \hat{\sigma}_{qg}$. And so, we write the medium contribution to the observable as
\begin{align}
    &\sum_{ij \in \{g,q,\bar{q}\}} \int  \td E_{i,j}   ~ F^{(ij)}_{\rm med} ~ \frac{\td \hat{\sigma}^{\mathrm{vac}}_{ij}}{\td \theta \td E_{i} \td E_{j} } \frac{E_{i}^n E_{j}^n}{Q^{2n}} \nonumber \\ &~~~~~~= \int  \td z ~ F^{(qg)}_{\rm med} \frac{\td \hat{\sigma}^{\mathrm{vac}}_{qg}}{\td \theta \td z}z^n (1 - z)^n \left(1+ \mathcal{O}\left(\frac{\bar\mu_{\mathrm{s}}}{Q}\right)+ \mathcal{O}\left(\As(\theta Q) \ln \theta \frac{\bar\mu^{n}_{\mathrm{s}}}{Q^{n}}\right)\right)\,. \label{eq:medPart}
\end{align}
Here the first quoted error comes from neglecting complete momentum conservation in the energy weighting of the $q\rightarrow q g$ term, note that $\td \hat{\sigma}_{ij}$ still keeps complete energy conservation. $\bar\mu_{\mathrm{s}}$ is the cross-section weighted average inclusivity scale ($\bar\mu_{\mathrm{s}} = \frac{1 }{\hat\sigma}\int \frac{\td \hat\sigma}{\td \mu_{\mathrm{s}}\td z} \mu_{\mathrm{s}} \td \mu_{\mathrm{s}} $). Typically, in the high energy limit,  $\td (F_{\rm med} \hat{\sigma}^{\mathrm{vac}}_{ij})/\td \omega \sim (\omega)^{-3/2}\omega_{c}^{1/2} \Theta(Q >\omega \gtrsim \mu)$  where $\omega$ is the energy of a quanta of radiation over which our computation is inclusive, $\omega_{c}$ is the characteristic scale for soft in-medium radiation,\footnote{In the harmonic oscillator approximation of the soft spectrum one finds that $\omega_{c} = \hat{q}L^{2}/2$ where $\hat{q}$ is the transport coefficient and $L$ is the medium length.} and $\mu$ is the screening mass of the medium \cite{Andres:2020vxs}. Therefore, we are led to assume that $\bar \mu_{\mathrm{s}}/Q \sim \sqrt{\mu/Q}$. We treat the $\mathcal{O}(\bar \mu_{\mathrm{s}}/Q)$ corrections as small in this paper. At this point we stress that, due to the inclusivity of the observable, neglecting the additional energy radiated by the jet through interactions with the medium ($\bar\mu_{\mathrm{s}}$) is not equivalent to neglecting the energy loss in typical jet substructure observables employed in heavy-ions, such as $z_{\rm g}$ or $\theta_{\rm g}$ which are exclusive. We will further discuss this point and outline how to systematically improve the approximations made here in section~\ref{sec:discuss}.\footnote{See also \cite{Barata:2023vnl} for the first estimates of the impact of energy loss effects on the in-medium EEC.} From here on we will drop the second error term as it will always be parametrically smaller than the first in the perturbative region of the observable (where $\As \ln \theta^{-1} <1 $). We will also drop the $(qg)$ superscript on $F^{(qg)}_{\rm med}$.

Upon combining the vacuum resummation in eq.~\eqref{eq:vacPart} with eq.~\eqref{eq:medPart}, we reach the master equation for our analysis,
\begin{align}
    &\frac{\td \Sigma^{(n)}}{\td \theta} = \frac{1}{\sigma}\int \td z  \left(g^{(n)}(\theta,\As) + F_{\rm med}(z,\theta) \right) \frac{\td \hat\sigma^{\rm vac}_{qg}}{\td \theta \td z }  z^n (1-z)^n \left(1  + \mathcal{O}\left(\frac{\bar \mu_{\rm s} }{Q} \right)\right) + \mathcal{O}\left(\frac{\Lambda_{\mathrm{QCD}}}{\theta ~ Q}\right) \,,  \label{eq:master}
\end{align}
in which we can use $F_{\rm med}(z,\theta)$ obtained within different jet quenching formalisms. Here $g^{(n)}$ contains the vacuum resummation, i.e. $g^{(1)} = \theta^{\gamma(3)} + \mathcal{O}(\theta)$ at fixed coupling given $\td\hat\sigma^{\rm vac}_{qg}$ at $\mathcal{O}(\alpha_s)$:
\beq
    \frac{1}{\sigma}\frac{\td \hat\sigma^{\rm vac }_{qg}}{\td \theta \td z } = \frac{\As(\theta Q)}{\pi}  ~ C_{\rm F} \frac{1+(1-z)^2}{z ~ \theta} + \mathcal{O}(\As^{2},\theta^{0})\,.
\eeq
See appendix~\ref{app:gn} for complete expressions for $g^{(n)}$ including the running coupling. Note that by construction the two-point correlator in p-p collisions, ${\rm d}\Sigma^{(n)}_{\rm vac}/{\rm d}\theta$, is achieved by setting $F_{\rm med} =0$ in \eqref{eq:master}.

Before moving on, a word on model dependent errors. In this paper, to simplify the computation of $F_{\rm med}$, we will be using two further assumptions. Firstly, that $F_{\rm med}$ is dominated by the initial hardest $q\rightarrow qg$ splitting, ignoring the resummation of multiple medium-induced emissions. This will introduce a multiplicative error to the integrand of eq.~\eqref{eq:master} of the form $\mathcal{O}\left(\As \ln \theta_{\TT{onset}}^{-1}\right)$ where $\theta_{\TT{onset}}$ is the smallest angle at which $|F_{\rm med}|$ is parametrically $\gtrsim 1$. Secondly, when computing  $F_{\rm med}$ within jet quenching formalisms which account for multiple in-medium scatterings, we will make use of a \emph{semi-hard approximation} for the parton-medium scatterings in which all partons propagate eikonally undergoing medium-induced colour rotations. This is valid provided $1-\mu_{\mathrm{s}}/Q>z>\mu_{\mathrm{s}}/Q$, and so we introduce a further multiplicative error in the integrand of eq.~\eqref{eq:master} of the form $\mathcal{O}\left(\bar \mu_{\mathrm{s}}/zQ\right)$. For approaches which only include a single medium scattering, instead of resumming multiple eikonal interactions, an $\mathcal{O}\left(n_0 L \right)$ error arises with $n_0$ being the linear density of scattering centers and $L$ the medium length.

\subsection{Multiple Scattering Framework for In-Medium Interactions}
\label{sec:theory_mult}

The usual approach to calculate medium-induced radiation accounting for multiple scatterings is through the BDMPS-Z perturbative QCD formalism \cite{Baier:1996kr,Baier:1996sk,Zakharov:1996fv,Zakharov:1997uu}. In this setup, one considers the high-energy limit in which the momentum transfer between the probe and the medium is only transverse,\footnote{For recent generalisations accounting for full momentum transfer between the medium and the probe see \cite{Sadofyev:2021ohn,Barata:2022krd,Andres:2022ndd,Barata:2022utc,Barata:2023qds}.} and thus multiple interactions can be resummed in terms of an in-medium propagator taking the following form\footnote{In this section we use light-cone coordinates ($p^+$, $p^-$, $\v p$), where $p^{\pm}\equiv(p^0\pm p^3)/\sqrt{2}$ and $\v p$ is the transverse momentum. Analogously, in coordinate space we use ($x^+$, $x^-$, $\v x$), where $x^{\pm}\equiv(x^0\pm x^3)/\sqrt{2}$ and $\v x$ is the transverse position.}
\beq
\mathcal{G}_R(\v{x}_1,t_1;\v{x}_0,t_0;p^+) =\int_{\v r(t_0)=\v x_0}^{\v r (t_1)=\v x_1} \mathcal{D}\v{r} \,
\exp \left[ i\frac{p^+}{2} \int_{t_0}^{t_1} {\rm d}s \, \v \dot{\v r}^2(s) \right]
 W_R(t_1, t_0; [\v r])\,,
 \label{eq:propagator}
\eeq
where $W_R$ is a Wilson line in the representation $R$ given by
\beq
W_R(t_1, t_0; [\v r]) = \mathcal{P} 
\exp \left [ig \int^{t_1}_{t_0} {\rm d}t\, T^a_R\, A^{-,a}(t, \v r(t)) \right]\,,
\eeq
with $T^a_R$ a colour matrix in the representation $R$ and $A^{-,a}$ a classical background describing the medium.

These propagators are used to calculate cross sections for fixed configurations of the background field, then one must take a weighted average over all possible such configurations to produce an observable. A gaussian average is assumed, where the only non-trivial correlation is the two-point function, which explicitly depends on the medium parameters,
\beq
\left\langle A^{-,a}(t,\v{r})A^{\dagger -,b}(t',\v{r}')\right\rangle = \delta^{ab}\,\delta(t-t')\,n(t)\,\gamma(\v{r}-\v{r}')\,,
\label{eq:gamma}
\eeq
with $n$ the linear density of scattering centers and $\gamma(\v r)$ the Fourier transform of the collision rate. When taking averages of combinations of in-medium propagators one must expand to all orders, consider all possible field pairings, and resum the final result. For the inclusive cross section to produce two partons the resummation is far from straightforward, as has been shown in \cite{Apolinario:2014csa,Blaizot:2012fh,Sievert:2018imd} where partial results have been achieved. Given that we need to keep track of both the splitting energy fraction $z$ and the splitting angle $\theta$, we {\it cannot make use} of previous approaches used for energy loss where only the soft limit was considered ($z\to 0$) \cite{Wiedemann:2000za,Mehtar-Tani:2006vpj,Andres:2020vxs}, or for branching rates where the angular dependence is lost \cite{Zakharov:1996fv,Zakharov:1997uu,Jeon:2003gi,Caron-Huot:2010qjx,Schlichting:2021idr}. Instead, we use a semi-hard approximation \cite{Dominguez:2019ges, Isaksen:2020npj} which is expected to give accurate results as long as the daughter partons are not too soft. In that setup, all partons are sufficiently energetic to propagate eikonally, thus following straight-line trajectories in coordinate space, which considerably simplifies the calculation of the medium averages. The in-medium propagator in \eqref{eq:propagator} then takes the form
\beq
\mathcal{G}(\v{x}_1,t_1;\v{x}_0,t_0;p^+) \approx 
\mathcal{G}_0(\v{x}_1,t_1;\v{x}_0,t_0;p^+) 
W_{R} \left(t_1,t_0; [\v x_{\rm cl}(t)] \right)\,,
\label{eq:prop_tilted}
\eeq
where $\v x_{\rm cl} = \frac{t_1-t}{t_1-t_0}\v x_0 + \frac{t-t_0}{t_1-t_0}\v x_1$ is the classical trajectory and
\beq
\mathcal{G}_0(\v{x}_1,t_1;\v{x}_0,t_0;p^+) = \frac{p^+}{2\pi i (t_1-t_0)}e^{\frac{ip^+(\v{x}_1-\v{x}_0)^2}{2(t_1-t_0)}}\,,
\eeq
is the vacuum version of the propagator. Following \cite{Dominguez:2019ges}, the high-energy limit of the transverse Fourier transform is taken as
\beq
\mathcal{G}(\v{p}, t_1;\v{p}_0, t_0;E) \approx
 (2\pi)^2 \delta^2(\v p -\v p_0)\,
e^{-i \frac{\v p_1^2}{2E}(t_1-t_0)}\,
W_{ R}(t_1,t_0, [\v n t])\,,
\label{eq:prop_tilted2}
\eeq
with $\v n=\v p/E$.

The approximations taken to arrive at the simplified expression for the propagator in \eqref{eq:prop_tilted2} might limit the range of applicability of this approach. Nevertheless, the fact that the energy correlators give more weight to splittings where none of the daughter partons is too soft plays to our advantage since that is precisely the region where the semi-hard approximation works best (see section \ref{sec:discuss} for a more detailed discussion). The simplicity of the propagator in \eqref{eq:prop_tilted2} allows us to obtain analytical results for the two-particle inclusive cross section for simplified models of the medium averages, thus allowing us to identify the relevant scales entering the calculation and explore their relation with those appearing as changes in the shape of the energy correlators.

Let us consider a $1\to 2$ splitting with the initial parton having energy $E$ and the daughter partons having energies $zE,(1-z)E$ and transverse momentum $\v{p}_1,\v{p}_2$ respectively. Using the in-medium propagators one can easily construct the corresponding amplitude for this process, square it and then take the average over medium configurations \cite{Dominguez:2019ges,Isaksen:2020npj}. At leading colour, the cross section can be expressed in terms of averages of only two (dipole) or four (quadrupole) Wilson lines in the fundamental representation, accounting for the multiple scatterings of all partons along their fixed trajectories, both on the amplitude and the conjugate amplitude. We take these trajectories as
\begin{align}
&\v r_0(s) = 0\,, \notag\\
&\v r_1(s) = \v n_1(s-t)\,, \notag\\
&\v r_{\bar 1}(s) = \v n_1(s-\bar t)\,, \notag \\
&\v r_2(s) = \v n_2(s-t)\,, \notag\\
&\v r_{\bar 2}(s) = \v n_2(s - \bar t)\,,
\end{align}
with $\v n_1= \v p_1/((1-z)E)$, $\v n_2= \v p_2/(zE)$, and $t$ and $\bar t$ the splitting times in the amplitude and conjugate amplitude respectively.

The dipoles appearing in this calculation can be written as
\beq
S_{IJ} (t_b, t_a)  
\equiv \frac{1}{N_c}\langle \text{tr} [W_F(t_b,t_a;[\v{r}_I]) W_F^{\dagger}(t_b,t_a;[\v{r}_J])] \rangle
= \exp \left[
-\frac{1}{2} \int_{t_a}^{t_b}
\mathrm{d}s \, n(s) \,\sigma_{IJ}(s)
\right]\,,
\label{eq:S_IJ}
\eeq
with $I,J \in \{0,1,2,\bar 1,\bar 2\}$, $\sigma_{IJ}(s)=\sigma(\v{r}_I(s)-\v{r}_J(s))$, and $\sigma$ is the so-called dipole cross section given by
\beq
\sigma(\v{r}) = g^2\left[\gamma(\v{0})-\gamma(\v{r})\right]\,.
\label{eq:sigma}
\eeq
We note that in order to simplify the notation in \eqref{eq:S_IJ}, we have dropped the explicit dependence on $t$ or/and $\bar t$ that comes through $\sigma_{IJ}$.

On the other hand, only one quadrupole appears, which reads
\beq
Q(t_b, t_a) 
= S_{1\bar 1} (t_b, t_a)\, S_{2\bar 2} (t_b, t_a) +
\int_{t_a}^{t_b} \mathrm{d}s\,
S_{1\bar 1} (t_b,s)
S_{2\bar 2} (t_b,s)\,
T(s)\,
S_{12} (s,t_a)
S_{\bar 1\bar 2} (s,t_a)\,,
\label{eq:Q}
\eeq
with the transition matrix $T$ being 
\beq
T(s) =- \frac{n(s)}{2}
\Big( 
\sigma_{12}(s) +\sigma_{\bar 1 \bar 2}(s) - \sigma_{1\bar 2}(s) -\sigma_{\bar 1 2}(s)
\Big)\,.
\label{eq:T}
\eeq

As shown in \cite{Isaksen:2020npj}, $F_{\rm med}$ in the semi-hard approximation takes the form
\beq
F_{\rm med}(z,\theta) =
2\int_0^L \frac{\mathrm{d}t}{t_{\rm f}}
\left [ \int_t^L \frac{\mathrm{d}\bar t}{t_{\rm f}}\,
\cos{\left(\frac{\bar t-t}{t_{\rm f}}\right)}
C^{(4)}(L, \bar t,t) \,C^{(3)}(\bar t, t) - \sin{\left(\frac{L-t}{t_{\rm f}}\right)}
\,C^{(3)}(L,t) \right]\,,
\label{eq:Fmed}
\eeq
where the formation time $t_{\rm f}$ is given by
\beq
t_{\rm f} =\frac{2}{z(1-z)E\theta^2}\,,
\eeq
and $C^{(n)}(t_a,t_b)$ are $n$-particle correlators which can be written in terms of the dipoles and quadrupole above, but whose explicit form varies depending on the identity of the partons in the splitting.

For the rest of this manuscript we will focus on the $q\to qg$ splitting. In this case and for leading colour, the three- and four-point functions read (see \cite{Dominguez:2019ges, Isaksen:2020npj})
\beq
C^{(3)}(\bar t,t) = S_{02}(\bar t,t)\,S_{12}(\bar t,t) \,,
\label{eq:C3_qg}
\eeq
\beq
C^{(4)}(L, \bar t,t) = Q(L, \bar t)\, S_{2\bar 2} (L, \bar t)\,,
\label{eq:C4_qg}
\eeq
where we have re-established the $t$ dependence in $C^{(4)}$ coming from the trajectories $\v{r}_1,\v{r}_2$.

In order to continue with the evaluation of $F_{\rm med}$, which we will perform at fixed coupling, we must specify the details of the parton-medium interaction encoded in $\sigma(\v{r})$. We will restrict ourselves to the case of a  static QGP of length $L$ with constant linear density of scatterings $n(s)=n_0\Theta(L-s)$, the so-called ``brick'' case. This simplification allows us to recognise the relevant scales entering the calculation which will play an important role when analysing the changing shape of the energy correlators after numerical evaluation. Extensions of this formalism to more realistic medium profiles will be explored in future publications.

In this static medium case, it is easy to see that $C^{(4)}(L,\bar t,t)$ actually depends only on the differences $L-\bar t$ and $\bar t - t$, while $C^{(3)}(\bar t,t)$ depends only on $\bar t-t$. This will be explicitly showed in the following subsections, where we will specify the functional form of $\sigma$ and derive the corresponding $n$-point functions. Defining the following re-scaled $n$-point functions
\beq
C^{(3)}(\bar t,t) =\tilde C^{(3)}\left(\frac{\bar t-t}{L}\right),\qquad C^{(4)}(L,\bar t,t) = \tilde C^{(4)}\left(\frac{L-\bar t}{L},\frac{\bar t-t}{L}\right)\,,
\eeq
the expression for $F_{\rm med}$ in \eqref{eq:Fmed} can then be written as
\beq
F_{\rm med}(z,\theta) =
 \frac{2L}{t_{\rm f}}\int_0^1 \mathrm{d}\tau_L
\left[ -\sin\left(\frac{L}{t_{\rm f}}\tau_L\right)
\,\tilde C^{(3)}(\tau_L) + \frac{L}{t_{\rm f}} \int_0^{1-\tau_L} \mathrm{d}\tau\,
\cos\left(\frac{L}{t_{\rm f}}\tau\right)\,
\tilde C^{(4)}(\tau_L, \tau) \,\tilde C^{(3)}(\tau)  \right]\,,
\label{eq:Fmed2}
\eeq
which can be directly plugged into our master equation \eqref{eq:master} to obtain the two-point energy correlator of an in-medium massless quark jet.

The expression above reflects the fact that emissions with a formation time larger than the length of the medium are not affected by it, and thus $F_{\rm med}$ is suppressed by an  overall $(L/t_{\rm f})^2$ factor for $t_{\rm f}\gg L$. The separation between kinematical regions with formation times either larger or smaller than the length of the medium defines an angular scale
\beq
\theta_{\rm L} = \frac{1}{\sqrt{EL}}\,,
\label{eq:theta_L}
\eeq
which parametrically indicates the minimum angle for emissions to be sensitive to medium modifications.


\subsubsection{Harmonic Oscillator Approximation}
\label{subsec:HO}

We first consider the Harmonic Oscillator (HO) approximation, which uses a simplified form of the dipole cross section $\sigma$ and allows us to compute analytically the $n$-point functions entering the expression for $F_{\rm med}$ \cite{Dominguez:2019ges,Isaksen:2020npj}. This approach has been shown to be a reasonable approximation to more realistic interaction models in the regions of phase space where the multiple scatterings with the medium play an important role \cite{Andres:2020vxs,Andres:2020kfg}.

In this setup, the dipole cross section in coordinate space is quadratic and reads 
\beq
n_0\sigma_{\mathrm{HO}}(\v{r})
= \frac{1}{2} \hat{q}\,\v{r}^2 \,,
\label{eq:sigma_HO}
\eeq
where $\hat q$  is the so-called jet quenching parameter that characterises the average transverse momentum squared transferred from the medium per unit path length.

Plugging this expression into \eqref{eq:S_IJ} one can show that the three- and four-point correlators at leading colour within the HO approach are given by (see \cite{Isaksen:2020npj})
\beq
\tilde C^{(3)}_{\rm HO}(\tau) = 
\exp \left[-\frac{1}{12}\left(\frac{\theta}{\theta_{\rm c}}\right)^2\tau^3 (1+(1-z)^2)\right]\,,
\label{eq:C3_HO}
\eeq
and 
\begin{align}
\tilde C^{(4)}_{\rm HO}(\tau_L,\tau)=&
\,e^{ -\frac{1}{4}\left(\frac{\theta}{\theta_{\rm c}}\right)^2\tau_L\tau^2(z^2 + 2(1-z)^2)} \nonumber \\
&\times \left (1 -\frac{1}{2}\left(\frac{\theta}{\theta_{\rm c}}\right)^2z(1-z)\,\tau^2  \int_0^{\tau_L} {\rm d}\tau_s\,
e^{-\frac{1}{12} \left(\frac{\theta}{\theta_{\rm c}}\right)^2
\tau_s^2\,(2\tau_s +3\tau)}
e^{-\frac{1}{2} \left(\frac{\theta}{\theta_{\rm c}}\right)^2z(1-z)\,\tau_s\tau^2}
\right)\,,
\label{eq:C4_HO}
\end{align}
with
\beq
\theta_{\rm c}=\frac{1}{\sqrt{\hat q L^3}}\,.
\label{eq:theta_c}
\eeq
We can plug these expressions directly into \eqref{eq:Fmed2} and perform the integrations numerically to get $F_{\rm med}$, as done in \cite{Isaksen:2020npj}.

As noted in the analysis presented in \cite{Dominguez:2019ges}, one can easily see that for $\theta\ll\theta_{\rm c}$ both $\tilde C^{(3)}$ and $\tilde C^{(4)}$ are very close to 1 and thus there is a complete cancellation between the two terms in \eqref{eq:Fmed2}. This suppression of $F_{\rm med}$ yields to the interpretation of $\theta_{\rm c}$ as the critical angle determining whether the medium can resolve the two daughter partons or if it sees them as a single colour charge. Of course, this distinction is only relevant when $t_{\rm f}$ is short enough to actually have a formed pair inside the medium. Therefore, if $\theta_{\rm L}\gg\theta_{\rm c}$ all emissions occurring inside the medium are automatically resolved, and $\theta_{\rm c}$ becomes irrelevant. We will refer to this case as the \emph{decoherent} (DC) limit. In the opposite case, $\theta_{\rm L}\ll\theta_{\rm c}$, there is an intermediate region in which emissions can occur inside the medium while not being resolved by it, thus yielding a smaller $F_{\rm med}$ than those emissions which are fully resolved. This case will be referred as the \emph{partially coherent} (PC) limit. This clear distinction between the possible orderings of the angular scales is a direct consequence of colour coherence effects in the splitting process and will be used in section~\ref{sec:results} when analysing the behaviour of the energy correlators for different sets of values of the medium and jet parameters.

\subsubsection{Yukawa Collision Rate}
\label{subsec:Yuk}

Even though the HO approximation is very useful, as it allows us to clearly identify the relevant angular scales, it has well known shortcomings, mainly the absence of the large transverse momentum tails. In the weakly coupled picture of the QGP being considered here, the medium has point-like constituents with Coulomb-like interactions at short distances which manifest as a power-like distribution for large transverse momentum transfers instead of the exponential fall of the HO approximation.

In order to incorporate the effect of such tails in our analysis, we now consider a more realistic model for parton-medium interactions. We use a Yukawa collision rate as implemented in the Gyulassy-Wang model \cite{Gyulassy:1993hr} through 
\beq
V_{\mathrm{Y}}(\v q) = \frac{8 \pi \mu^2}{\left(\v q^2 + \mu^2\right)^2}\,,
\label{eq:V_yuk}
\eeq
where $\mu^2$ is the screening mass. This collision rate is directly related to the average of two fields $\gamma$ in \eqref{eq:gamma} through a Fourier transform, $V(\v{q}) = g^2\int {\rm d}^2\v{r}\,e^{-i\v{r}\cdot\v{q}}\,\gamma(\v{r})$. The corresponding dipole cross-section in coordinate space can then be easily obtained, yielding
\beq
\sigma_{\mathrm{Y}}(\v r) = 2 \left[ 1 - \mu |\v r |K_1(\mu |\v r|)\right]\,,
\label{eq:sigma_yuk}
\eeq
where $K_1$ is the modified Bessel function of the second kind. This can then be employed to calculate the dipoles $S_{IJ}$ in \eqref{eq:S_IJ} and the quadrupole $Q$ in \eqref{eq:Q} to be used in the $n$-point functions. These take the following form:
\beq
\tilde C_{\rm Y}^{(3)}(\tau) = 
\exp \left\{
-  \,n_0 L \left(
2\tau 
- \int_0^{\tau} \mathrm{d}\tau_s\, \mu \,\theta L\,\tau_s 
\Big[
 (1-z)  \,K_1\left(\mu\, (1-z) \,\theta\, \tau_s \,L\right) +
K_1\left(\mu \,\theta \,\tau_s \,L \right)
\Big]
\right)
\right\}
\,,
\label{eq:C3_qg_yuk_tau}
\eeq
and
\begin{align}
\tilde C_{\rm Y}^{(4)}(\tau_L,\tau) &= \exp \left\{
-  n_0 L\,\tau_L\, \left[3 - 2\mu\,(1-z)\,\theta\, \tau\, L \, K_1\left(\mu\,(1-z)\, \theta\, \tau \,L\right) - \mu\, z\,\theta\, \tau\, L \, K_1\left(\mu \,z\theta\, \tau \,L\right)\right]
\right\} \notag\\
& \quad\times\left(1 + \int_0^{\tau_L}\mathrm{d}\tau_s \, L \, T(\tau_s, \tau)\,\exp\Big\{-n_0\,\mu\,\theta\,\tau_s\,\tau \,L^2\right.
\notag\\ 
& \qquad\times
\big [(1-z)\,K_1\left(\mu\, (1-z) \,\theta \,\tau L\right)+
 z \,K_1\left(\mu\, z \,\theta \,\tau \, L \right)
 \big] \Big\}\notag\\
&\left.\qquad\times\exp\Big\{ n_0L\,\mu\,\theta\,
\int_0^{\tau_s} \mathrm{d}\tau_s'\,  \big[
(\tau_s' + \tau) \, L \,K_1\left(\mu \,\theta \,( \tau_s' + \tau) \,L \right) + 
\tau_s'\, L \,K_1\left(\mu\, \theta \,\tau_s'\, L\right)\big]\Big\}\right)\,,
\end{align}
where the transition matrix $T(\tau_s,\tau)$ reads
\beq
\begin{aligned}
T(\tau_s,\tau) = &- n_0 L\,\mu\,\theta \,
\Big[ 
 -(\tau_s+\tau)\, K_1\left(\mu\, \theta \,(\tau_s+\tau)\,L \right) -\tau_s\,K_1\left(\mu \, \theta\, \tau_s \,L \right)
\\
&+ \, \,[(1-z)\, (\tau_s+\tau) +  z \,\tau_s]\,K_1\left(\mu\,\theta \,[(1-z)\, (\tau_s+ \tau) +  z \,\tau_s] \,L\right) \\ 
&+ \,[(1-z) \,\tau_s +  z \,(\tau_s+\tau)]\,K_1(\mu\,\theta \,[(1-z) \,\tau_s + z\, (\tau_s+\tau)]\,L )
\Big]\,.
\end{aligned}
\label{eq:T_yuk}
\eeq

All integrations in the formulas above must be performed numerically, which means that we cannot extract directly from the formulas the relevant angular scales as in the HO case. Nevertheless, the discussion about the two competing angular scales, $\theta_{\rm L}$ and $\theta_{\rm c}$, is still valid and we anticipate seeing the distinction between the two regimes, DC and PC, described in the previous subsection.

The dynamics in this model is expected to be dominated by the behaviour of the dipole cross-section at small distances,
\beq
\sigma_{\rm Y}(\v{r}) \overset{\v{r}\to 0}{\approx} \mu^2\v{r}^2\ln \frac{1}{|\v{r}|}\,.
\eeq
Neglecting the logarithmic dependence and comparing to \eqref{eq:sigma_HO} we see that parametrically the jet quenching parameter $\hat q$ appearing in the HO approximation is related to the parameters entering the Yukawa interaction model by $\hat q \sim n_0\mu^2$. This correspondence will be used to define the angular scale $\theta_{\rm c}$ for the Yukawa collision rate since its definition in \eqref{eq:theta_c} is given in terms of $\hat q$.


\subsection{Single Scattering Framework for In-Medium Interactions (GLV)}
\label{subsec:GLV}

Alternatively, for dilute media, one can consider the evaluation of the two-particle cross-section in an opacity expansion, where a series is defined in terms of the number of scatterings between the probe and the medium. This approach has been widely used in the soft case ($z\to 0$), where analytical results to first order in opacity are available \cite{Gyulassy:2000fs,Wiedemann:2000za} and some numerical analyses for higher orders have been explored \cite{Gyulassy:2000er, Wicks:2008ta}. Going beyond the soft limit, the splitting functions for arbitrary values of $z$ at first order in opacity were calculated in \cite{Ovanesyan:2011kn} and a method to recursively obtain higher orders was derived in \cite{Sievert:2018imd,Sievert:2019cwq}, where it is clearly seen that the complexity of the formulas grows very fast when the number of scatterings is increased.

The downside of this approach is that it gives incorrect results in regions of phase space where one expects multiple scatterings to be important. It does not fully incorporate the Landau-Pomeranchuk-Migdal (LPM) effect where several scatterings act in a coherent way during a single emission, thus yielding to unitarity issues which can manifest as either too large or negative cross sections. On the other hand, in this approach there is no need to use the semi-hard approximation employed in the multiple scattering case of section~\ref{sec:theory_mult}, and thus one can keep the transverse momentum broadening of the daughter partons.

In this single scattering framework, we can calculate $F_{\rm med}$ for the $q \to qg$ splitting using the formulas in \cite{Ovanesyan:2011kn} adapted to our notation. As in the multiple scattering case, we use a uniform medium with constant linear density $n(s)=n_0\Theta(L-s)$, yielding
\begin{align}
F_{\rm med}(z,\theta) = &\,\frac{2\pi\, n_0}{z(1-z)E} \int_{\v k} \delta\left(|\v k|-z(1-z)E\theta\right)
\int_0^L {\rm d}t \int_{\v q} V_{\rm Y}(\v q) \nonumber\\
&\times\v k^2\left[\frac{\v k + z \v q}{(\v k + z \v q)^2}\cdot \left(\frac{2C_{\rm F}}{N_c}\frac{\v k + z \v q}{(\v k + z \v q)^2} - \frac{\v k -  (1-z) \v q}{(\v k -  (1-z) \v q)^2} + \frac{1}{N_c^2}\frac{\v k}{\v k^2}\right) \left[1-\cos{(\Omega_{1} t)}\right] 
\right.  \nonumber\\
&\quad +  \frac{\v k -  (1-z) \v q}{(\v k -  (1-z) \v q)^2} \cdot \left(\frac{2(\v k -  (1-z) \v q)}{(\v k -  (1-z) \v q)^2}-\frac{\v k + z \v q}{(\v k + z \v q)^2} - \frac{\v k}{\v k^2}\right) \left[1-\cos{(\Omega_{2} t)} \right] \nonumber \\
&\quad +\frac{(\v k + z \v q)\cdot(\v k -  (1-z) \v q)}{(\v k + z \v q)^2 (\v k -  (1-z) \v q)^2}\left[1-\cos{(\Omega_{3} t)} \right] \nonumber \\
&\left.\quad + \frac{\v k}{\v k^2}\cdot \left(\frac{\v k - \v q}{(\v k - \v q)^2}- \frac{\v k}{\v k^2}\right)\left[1-\cos{(\Omega_{4} t)} \right] - \frac{\v k\cdot(\v k - \v q)}{\v k^2(\v k - \v q)^2} \left[1-\cos{(\Omega_{5} t)} \right]
\right]\,, 
\label{eq:GLVspectrum}
\end{align}
where we have used the shorthand $\int_{\v q}= \int {\rm d}^2\v q/(2\pi)^2$ and $V_{\rm Y}$ is the Yukawa collision rate given in \eqref{eq:V_yuk}. We have defined $\v k$ as the relative transverse momentum of the  daughter partons and
\begin{align}
& \Omega_{1} = \frac{(\v k + z \v q)^2}{2z(1-z)E}\,,\quad
\Omega_{2} = \frac{(\v k -  (1-z) \v q)^2}{2z(1-z)E} \,, \quad
\Omega_{3} = \frac{(\v k -  (1-z) \v q)^2 -(\v k + z \v q)^2}{2z(1-z)E} \,,\nonumber\\
& \Omega_{4} = \frac{\v k^2}{2z(1-z)E} \,,\quad
 \Omega_{5} = \frac{\v k^2- (\v k - \v q)^2}{2z(1-z)E} \,.
 \label{eq:GLVphases}
\end{align}

The formula above for $F_{\rm med}$ shows clear differences with the results for the multiple scattering case. In particular, we can see that the angular scales we extracted from the previous analyses do not appear as clear cut here. The lack of medium enhancement for splittings with long formation times in the multiple scattering approaches was related to the arguments of the sine and cosine in \eqref{eq:Fmed2} being the same, whereas here the arguments of all the cosines are different. Similarly, the distinction between coherent and incoherent emissions was a consequence of the exponential suppression factors in the $n$-point correlators, which are absent here as well. The physical principles behind these features of the medium enhancement are still valid, and thus we still expect to see differences between the different regimes where the splitting occurs outside of the medium or where the medium cannot resolve the individual daughter partons. Nevertheless, the boundaries between those regions are expected to be loosely defined and the transition regions to be larger.

\section{Numerical Results}
\label{sec:results}

\begin{figure}
\centering
\includegraphics[width=0.50\textwidth]
{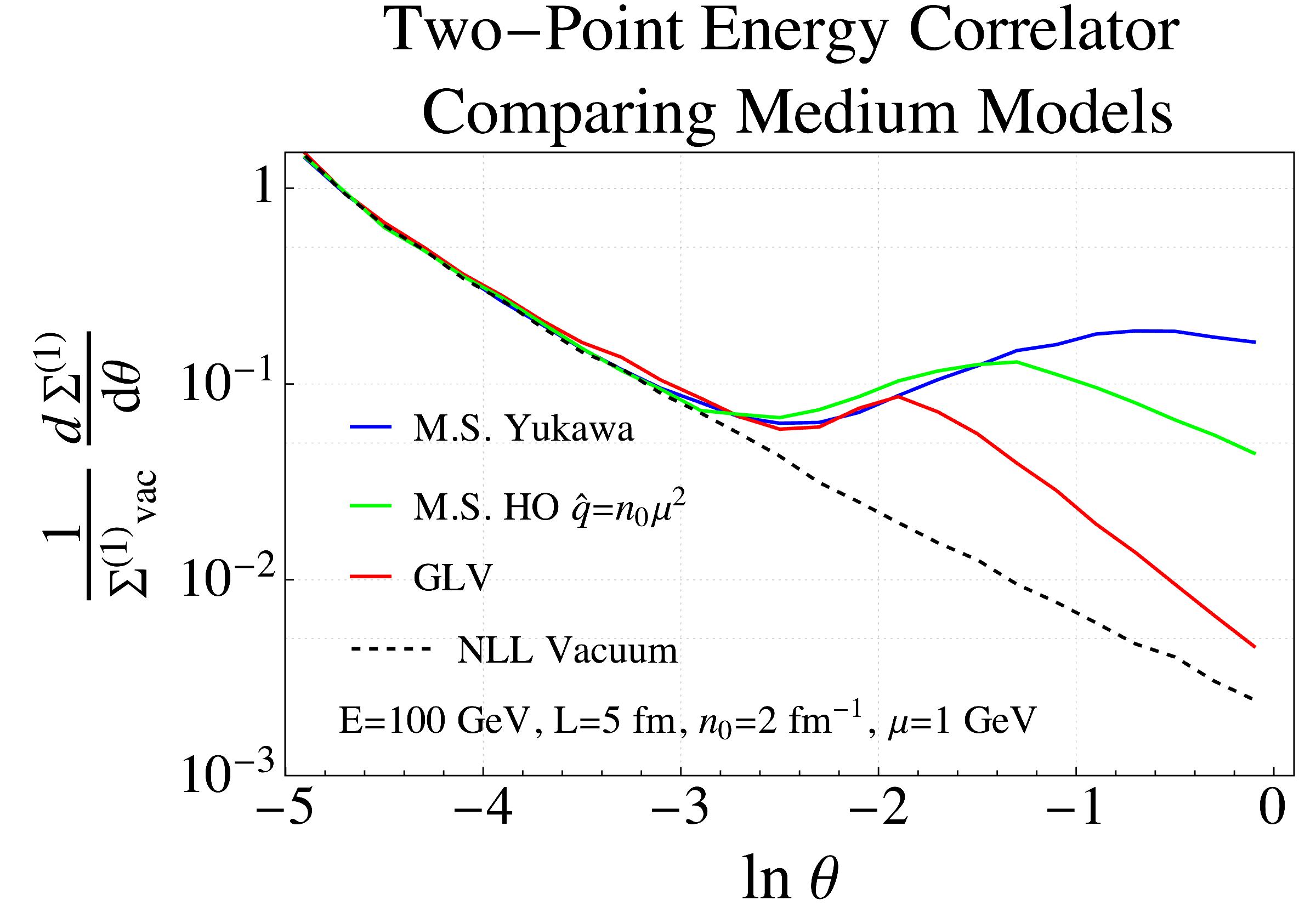}\includegraphics[width=0.50\textwidth]{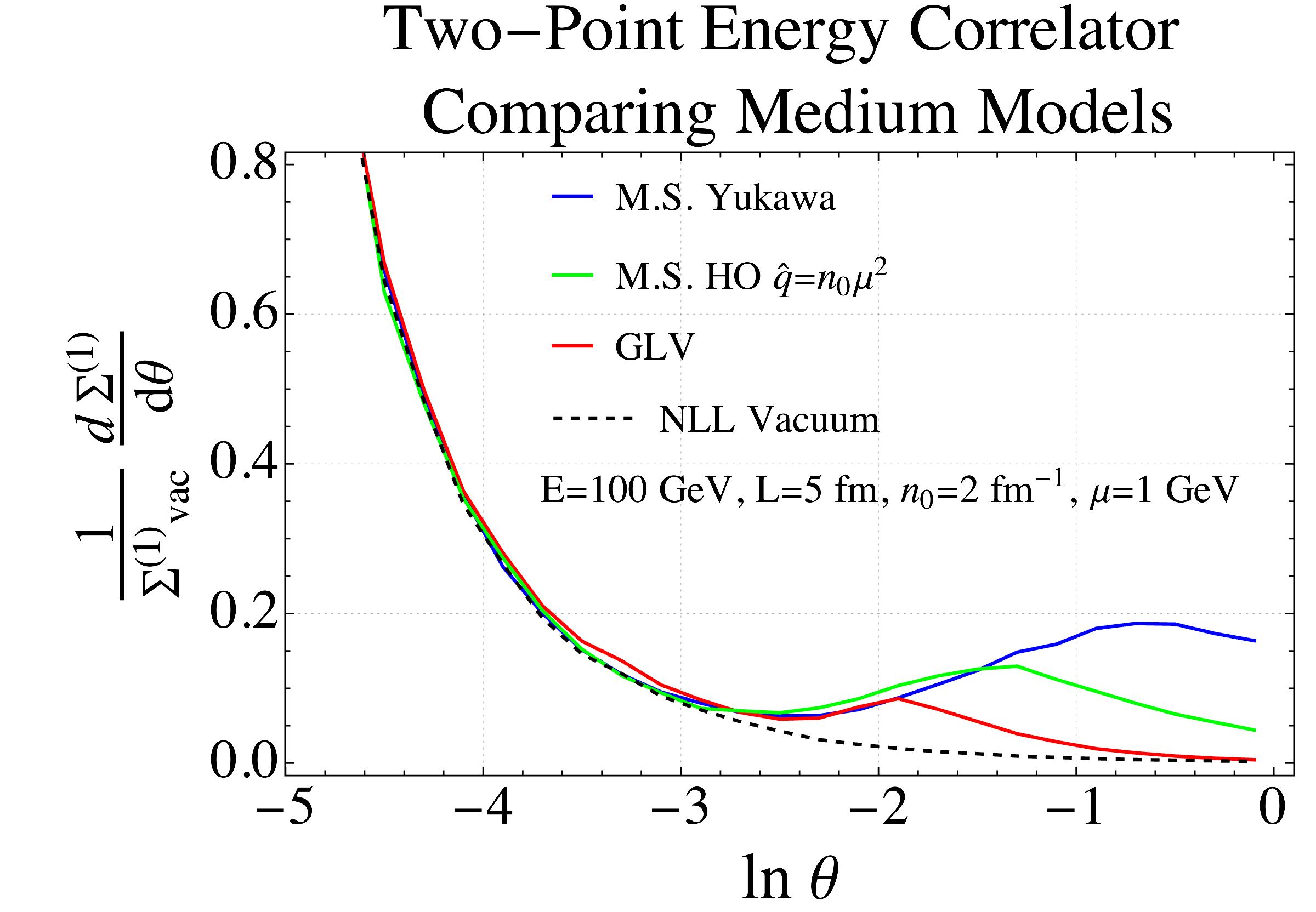}\newline \includegraphics[width=0.50\textwidth]{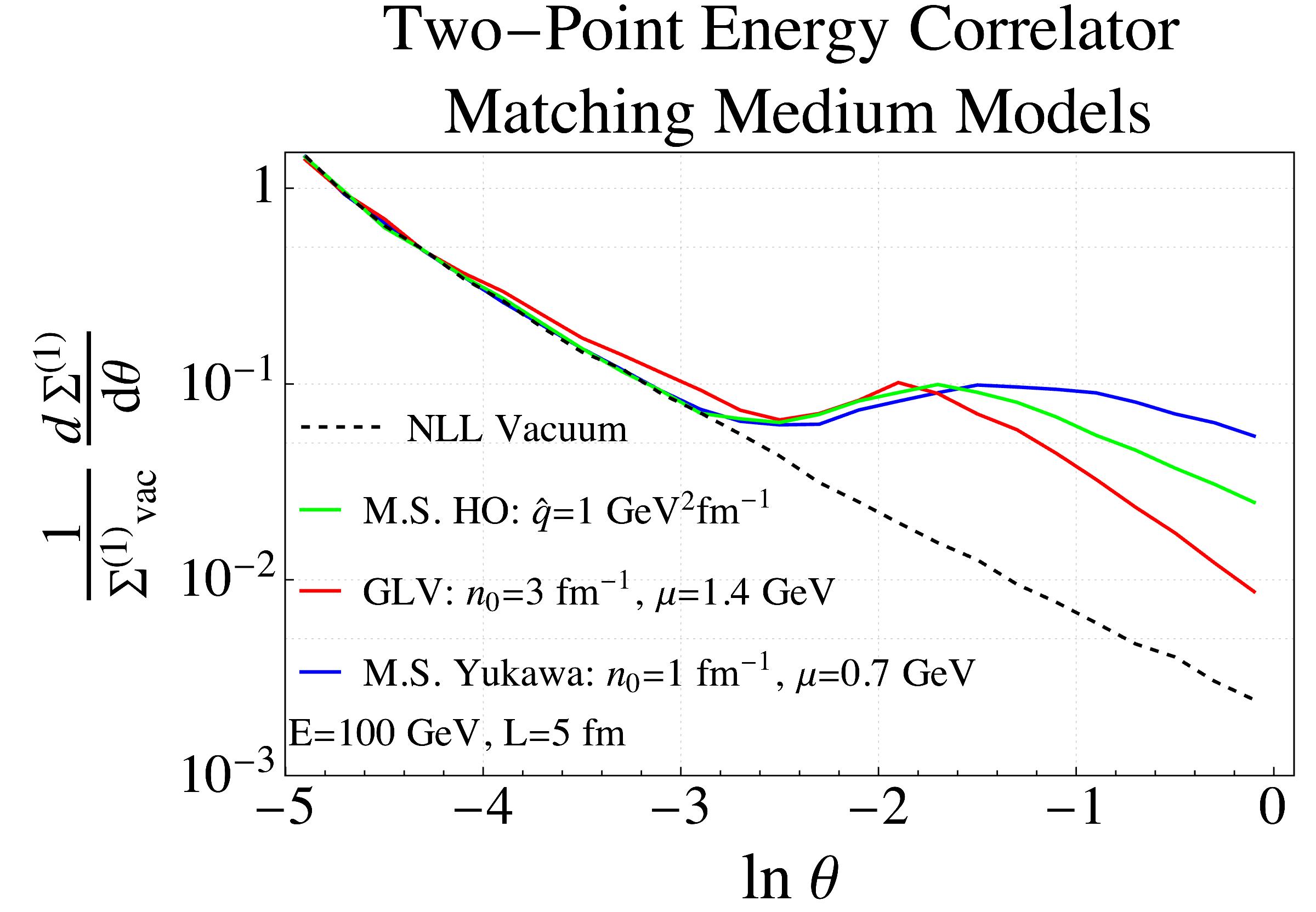}\includegraphics[width=0.50\textwidth]{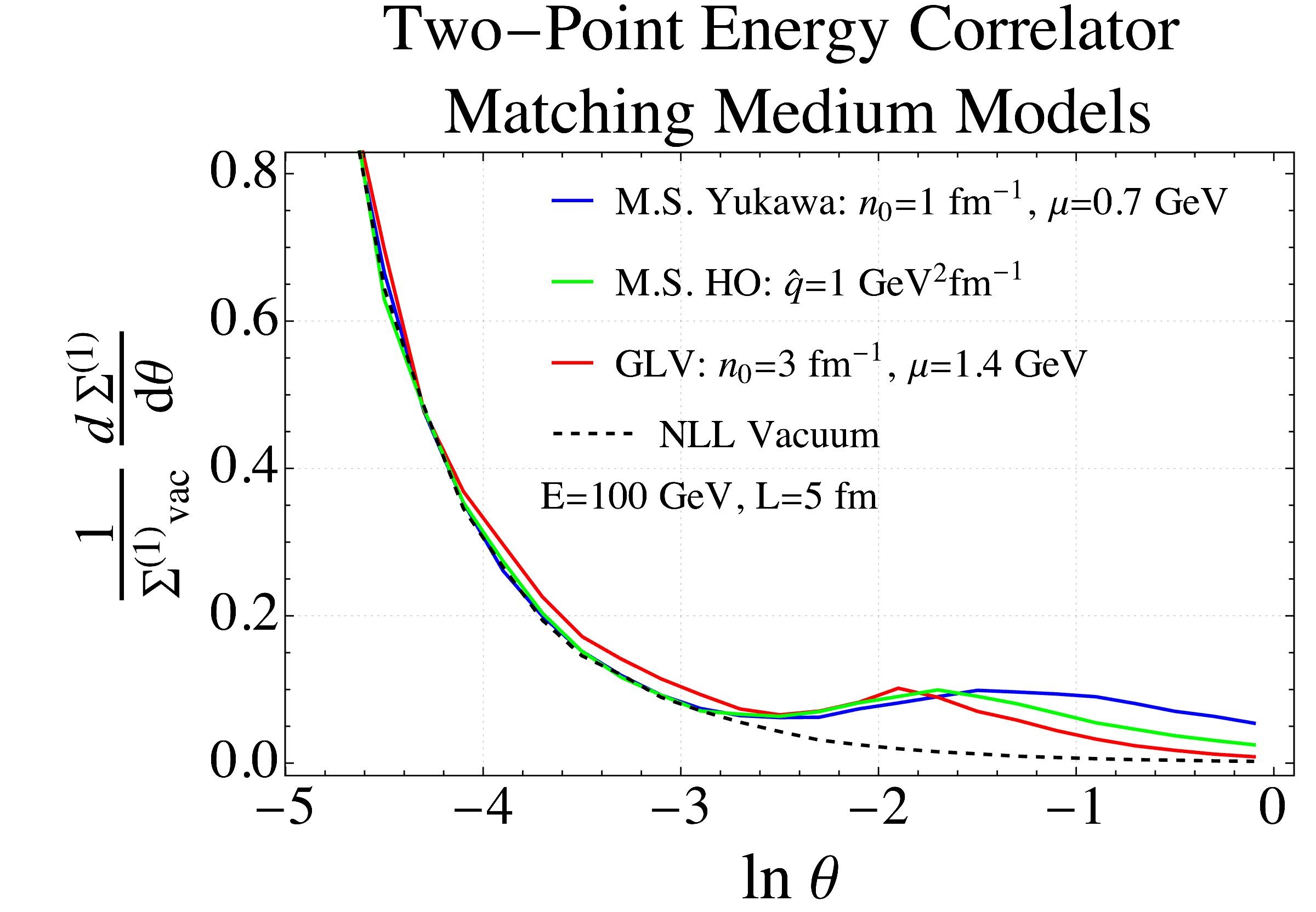}
\caption{Top: The $n=1$ EEC in \eqref{eq:master} of a $E=100$\,GeV quark jet  evaluated through the multiple scattering HO (green), multiple scattering Yukawa (blue), and single scattering GLV (red) approaches compared to the vacuum NLL result (black dashed).  Both plots show the same curves, either with (left) or without (right) log scaling on the y-axis. Bottom: Same as the top panel for different sets of medium parameters chosen so the amplitude and onset of the medium enhancement qualitatively match among the different models. All curves are normalised by  $\Sigma^{(1)}_{\rm vac} \equiv \int^{0}_{-5} {\rm d} \Sigma^{(1)}_{\rm vac} /{\rm d }\theta \, { \rm d}(\ln \theta).$}
\label{fig:combined_bump} 
\end{figure}

To further proceed in the computation of the two-point energy correlator given in eq.~\eqref{eq:master}, one must numerically evaluate $F_{\mathrm{med}}$ in the single and multiple in-medium scatterings approaches described in the previous section. Then, $F_{\mathrm{med}}$ is combined with the analytic NLL expressions for $g^{(n)}$ given in appendix~\ref{app:gn} and the remaining $z$ integral in~\eqref{eq:master} is performed via Monte Carlo integration.

Figure~\ref{fig:combined_bump} shows a sample of the results for the $n=1$ energy correlator, ${\rm d}\Sigma^{(1)}/{\rm d}\theta$, for each jet quenching formalism and for reasonable values of the medium parameters. All the curves are almost identical to the vacuum result for small angles. This follows our expectations as the small angle structure is dominated by splittings with a formation time much larger than the length of the medium and so do not have a significant medium modification. This argument is model independent, which agrees with what is seen in figure~\ref{fig:combined_bump}, where the point at which the in-medium EEC deviates from the vacuum baseline does not depend on the particular model used for the interactions with the medium. At larger angles, the in-medium EEC presents an excess with respect to the vacuum curve due to medium-induced radiation. The particular features of this enhancement depend on the formalism used to obtain $F_{\mathrm{med}}$, with larger differences between models at larger angles (see the top panels of  figure~\ref{fig:combined_bump}). This too follows intuition, since wider angle structures in the EEC are formed from events with a large overall transverse momentum transfer from the medium, which is the region where the different medium-induced radiation approaches differ the most. In contrast, smaller angle structures in the medium enhancement are dominated by the more universal harder collinear physics in the limit where the in-medium scatterings are very soft relative to the jet partons, causing only small amounts of broadening. It is, however, worth noticing that slightly varying the medium parameters allows us to bring the results yielded by the different jet quenching approaches into close qualitative agreement across a much broader angular range, as shown in the bottom panel of figure~\ref{fig:combined_bump}, and thus one must be careful of not drawing conclusions before exploring the full dependence of the shape of the EEC on all the parameters. For completeness, we show the results on the EEC with an energy weight $n = 2$ in figure~\ref{fig:combined_bump_n2} of appendix~\ref{app:figs}, which are in qualitatively agreement with the  $n=1$ results presented here.

\begin{figure}
\centering
\includegraphics[scale=0.45]{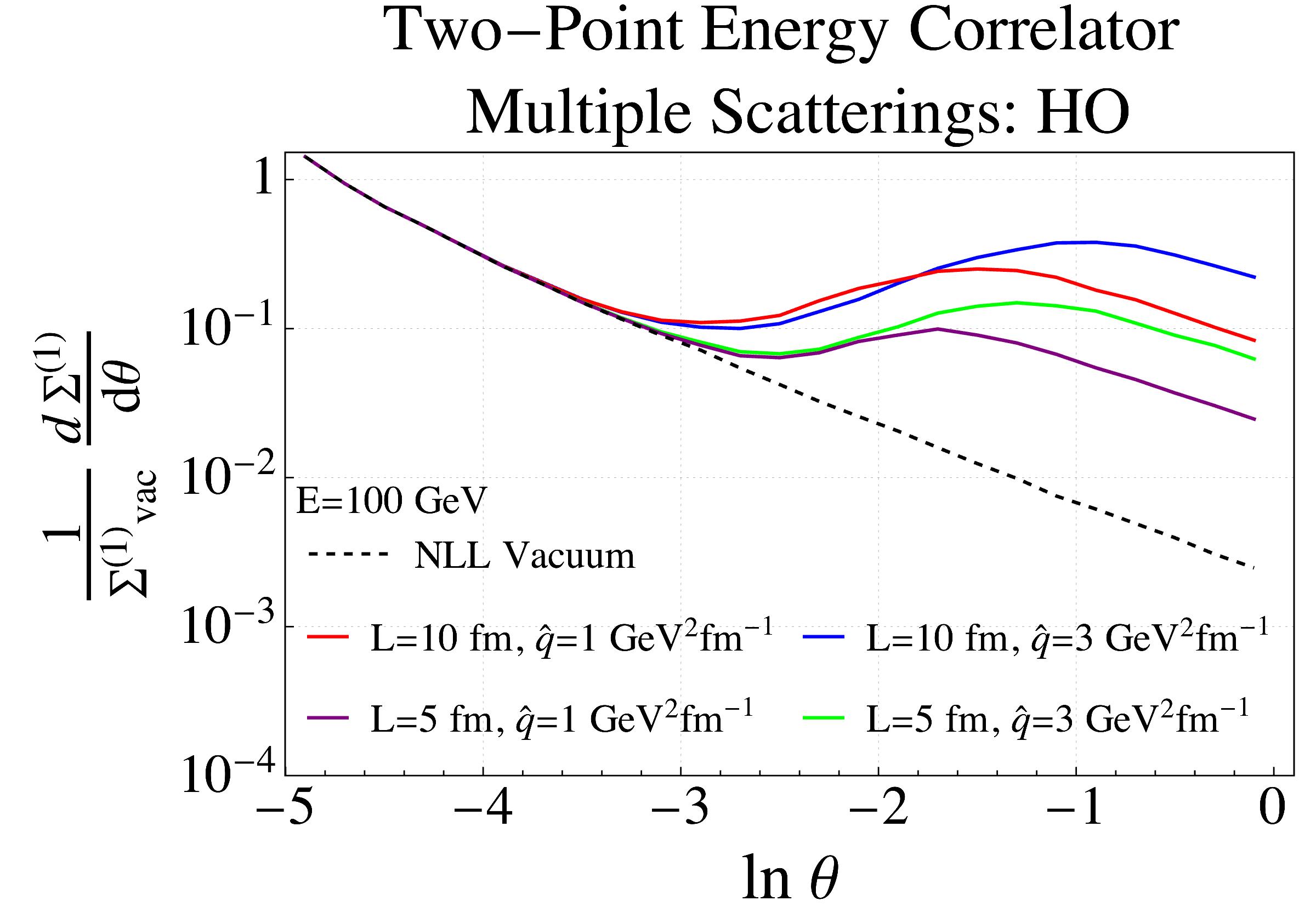}
\caption{The $n=1$ EEC in \eqref{eq:master} of a $100$ GeV  quark  jet evaluated within the multiple scattering HO approach for several values of the medium parameters compared to the vacuum NLL result, as indicated in the legend. All curves are normalised by the integrated vacuum result $\Sigma^{(1)}_{\rm vac}$.}
\label{fig:HOvaryingParameters}
\end{figure}

To illustrate the dependence of the shape of the EEC on the medium parameters, we show in figure~\ref{fig:HOvaryingParameters} the $n=1$ EEC spectrum computed within the multiple scattering HO approach for different sets of medium parameters.\footnote{See appendix~\ref{app:figs} for additional figures showing the two-point correlator within all the jet quenching models considered in this paper and different sets of parameters.} One can observe that for the simplified brick of length $L$ used in this manuscript the angle at which the in-medium EEC starts to deviate from the vacuum result (the onset angle) is mostly dependent on $L$, whilst the area of the medium enhancement (amplitude) depends on both $\hat{q}$ and $L$. We would like to quantitatively determine how features of the EEC spectra scale with the medium and jet parameters, with the ultimate goal of studying the correlator spectra's sensitivity to the dynamics of colour coherence. To this end, we generalise the fitting procedure used for the EEC computed within the HO approach in \cite{Andres:2022ovj}, so it allows us to study the signatures of coherence within all the medium-induced radiation approaches considered in this manuscript.\footnote{We do not study in this manuscript the onset angle, previously explored in \cite{Andres:2022ovj}, as it is not necessary to determine the emergence of colour coherence and its extraction is challenging in the GLV approach due to the more gradual vacuum to medium enhancement transition.} This improved fitting method, which we outline in the following section, increases both the robustness and model independence of our analysis.

\subsection{Analysis Procedure}
\label{sec:procedure}

\begin{figure}
\centering
\includegraphics[scale=0.45]{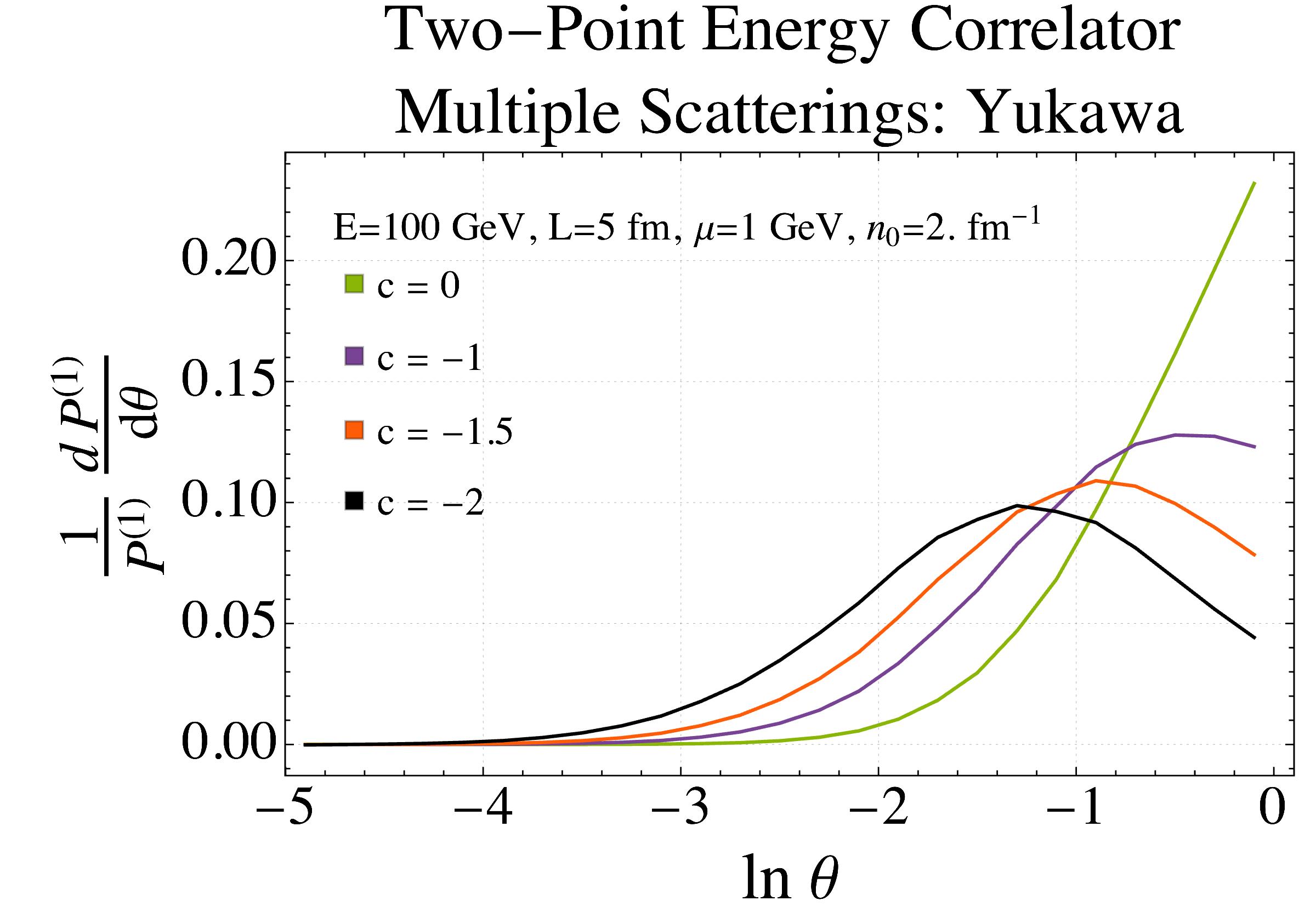}
\caption{The $\td P^{(1)}/\td \theta$ in eq.~\eqref{eq:distribution}  distribution normalised by its area $P^{(1)}$ evaluated through the multiple scattering Yukawa approach (see section~\ref{subsec:Yuk}) for a quark jet of energy $E=100$ GeV  in a medium of $L=5$\,fm, $\mu=1$\,GeV and $n_0=2\,{\rm fm}^{-1}$. Each curve corresponds to a given value of the parameter $c$ in \eqref{eq:distribution}, as indicated in the legend.}
\label{fig:Yukawa_dP/dtheta}
\end{figure}

We first introduce the basic distribution for our analysis, which allows us to determine the positions of the peak $\theta_{\rm peak}$ of the medium enhancements observed in figure~\ref{fig:combined_bump} as well as to access the gradient of the enhancement to the left of the peak. This distribution is given by
\begin{align}
    \frac{\td P^{(n)}}{\td \theta} = \theta^{c}\left[\frac{\td \Sigma^{(n)}}{\td \theta} \left(\frac{\td \Sigma^{(n)}_{\rm vac}}{\td \theta} \right)^{-1} - 1 \right]\,,
    \label{eq:distribution}
\end{align}
where the constant $c$ is used as a tool to explore different angular regions in the correlator spectra, since using $c<0$ allows us to shift the peak position to smaller angles by an amount inversely proportional to the gradient of the enhancement near the peak. In the following we will set  $c= -2,-1.5,-1,0$, in order to achieve shifts of different sizes. We note that for $c=-1$ the peak position in $\td P^{(n)}/\td \theta$ is equal to the peak position in $\td \Sigma^{(n)}/\td \theta$ up to very small logarithmic corrections, and thus by fixing $c=-1$ we recover the analysis method employed in \cite{Andres:2022ovj}. We illustrate our current procedure in figure~\ref{fig:Yukawa_dP/dtheta}, where we present the $\td P^{(1)}/\td \theta$ distribution evaluated through the multiple scattering approach with a Yukawa parton-medium interaction model described in section~\ref{subsec:Yuk} for several values of $c$. We clearly observe that the smaller the value of $c$, the smaller the angle at which the peak appears. Indeed, for this particular set of parameters, we can see that for the largest value of $c$ ($c=0$) the peak is not visible, since the distribution peaks at $\theta >1$.

In order to determine how $\theta_{\rm peak}$ scales in terms of the medium parameters and jet energy, in a given jet quenching approach, we proceed as follows. For a comprehensive span of the jet and medium parameters, we compute the $\td P^{(n)}/\td \theta$ distributions with $n=1,2$ and $c= -2,-1.5,-1,0$ and obtain the peak position $\theta_{\mathrm{peak}}$ by fitting the peaked region of these distributions using a quintic polynomial. The positions of the peaks are then fitted to the parameters using both a \emph{no-coherence} and a \emph{coherence} power-law ansatz. The no-coherence approach consists of performing the power-law fit over the complete set of parameters for which the EEC curves were generated, and thus it assumes that the dependence of the peak position on the parameters is the same for all the sets of parameters considered. In contrast, the coherence ansatz assumes that 
the position of the peak scales differently with the medium parameters in the DC and PC limits  defined at the end of section \ref{subsec:HO}. We separate the regions through the critical energy $E_{\rm c}$ at which $\theta_{\rm c}=\theta_{\rm L}$, with the DC limit corresponding to $E\ll E_{\rm c}$ and the PC limit to $E\gg E_{\rm c}$. We perform the fits in the DC region using only the subsets for which  $E< 0.8 E_{\rm c}$ (as a representation of  $E \ll E_{\rm c}$). For the fits in the PC regime, we employ only the $E> 1.2 E_{\rm c}$ parameter subsets.

Whilst we find the wide angle tails of the $\td P^{(n)}/\td \theta$ distributions to strongly depend on the energy weight $n$, we do not observe any significant dependence on $n$ in the peak position within the accuracy of our numerical convergence. In the present numerical analysis we use the $n=2$ distributions, since they yield the best extraction of $\theta_{\rm peak}$ due to the slightly sharper shape of their peaks. We have checked that consistent fits are found from the $n=1$ distributions.

All the power-law fits were performed using the Wolfram Mathematica \texttt{FindFit} and \newline\texttt{NonlinearModelFit} functions using the ``Principle Axis'' method. We did not enforce the fits to be dimensionally correct, so we can use the mass dimension of the resulting power-law as a cross-check for the validity of the fit. The obtained scaling laws for each of the jet quenching formalisms, shown in section~\ref{subsec:scalings}, are the core result of our numerical analysis. Finally, we present in section~\ref{subsec:visualising} a \emph{jet energy sweep} which enables to clearly observe these scalings in a graphical form. 


\subsection{Distributions and Scaling Properties for Different Jet Quenching Formalisms}
\label{subsec:scalings}

\subsubsection{Approaches with Multiple in-medium Scatterings}
\label{subsec:scalings_HOYuk}

\begin{table}[]
\centering
\begin{tabular}{|c|c|c|}
    \hline \multicolumn{3}{|c|}{\textbf{Multiple scatterings: HO }} \\
    \hline \textbf{Scaling of $\boldsymbol{\theta_{\mathrm{peak}}}$} & $c = -2$ & $c = -1.5$ \\ \hline
   no-coherence  & $\sim E^{-0.65}L^{-0.10}\hat{q}^{0.22}$ $[0.11]$ & $\sim E^{-0.70}L^{0.00}\hat{q}^{0.27}$ $[0.12]$ \\
   $E \ll E_{\rm c}$ & $\sim E^{-0.79}L^{0.06}\hat{q}^{0.30}$ $[0.04]$ & $\sim E^{-0.85}L^{0.18}\hat{q}^{0.35}$ $[0.02]$ \\
   $E \gg E_{\rm c}$ & $\sim E^{-0.55}L^{-0.37}\hat{q}^{0.07}$ $[0.03]$ & $\sim E^{-0.58}L^{-0.32}\hat{q}^{0.10}$ $[0.03]$ \\ \hline
   \textbf{Scaling of $\boldsymbol{\theta_{\mathrm{peak}}}$} & $c = -1$ & $c = 0$ \\ \hline
   no-coherence  & $\sim E^{-0.45}L^{0.86}\hat{q}^{-0.06}$ $[-1.48]$& $\sim E^{-0.77}L^{-0.16}\hat{q}^{0.36}$ $[0.14]$ \\
   $E \ll E_{\rm c}$ & $\sim E^{-0.89}L^{0.25}\hat{q}^{0.40}$ $[0.05]$ & $\sim E^{-0.91}L^{0.34}\hat{q}^{0.44}$ $[0.07]$\\
   $E \gg E_{\rm c}$ & $\sim E^{-0.59}L^{-0.32}\hat{q}^{0.09}$ $[0.00]$ & $\sim E^{-0.62}L^{-0.26}\hat{q}^{0.12}$ $[0.02]$\\\hline
\end{tabular}
\caption{Scaling of the peak position of the $\td P^{(2)}/\td \theta$ distribution in \eqref{eq:distribution}  for $c=-2,-1.5,-1,0$ computed within the multiple scattering HO approach (see section~\ref{subsec:HO}). The no-coherence rows show the power-law obtained when all sets of parameters are fitted, while the other two rows correspond to the coherent anstaz separating the $E \ll E_{\rm c}$  and $E \gg E_{\rm c}$ sets, with $E_{\rm c} = \hat{q}L^{2}$. In square brackets we show the mass dimension of each fit, as its proximity to zero provides a measure of the fit quality. For each fit in which the peak angle power-law presents a deviation of the mass dimension from zero smaller than $1$, the errors on the power of the parameters are smaller or equal to $\pm 0.01$.}\label{tbl:HO}
\end{table}

\emph{Harmonic Oscillator Approximation---} In order to analyse the presence of colour coherence in the multiple scattering HO approach described in section~\ref{subsec:HO}, we followed the above mentioned procedure generating the $\td P^{(n)}/\td \theta$ distributions for 332 sets of parameters within the ranges: $E\in[20,500]$ GeV, $L\in[1,10]$ fm, and $\hat{q}\in[1,3]$\,GeV$^{2}$fm$^{-1}$. The results of power-law fits to the peak angle are presented in table~\ref{tbl:HO}, where the number in square brackets in each row represents the mass dimension of the fit. The fits performed assuming the coherence ansatz, with critical energy given by $E_{\rm c}=\hat q L^2$, are found to provide a good description of the peak position for all values of $c$. This is in striking contrast with the fits performed assuming the no-coherence ansatz, which fail for $c=-1$, as clearly indicated by the quoted mass dimension. This is an unequivocal indication of the ability of the EEC to be sensitive to the presence of colour coherence dynamics within the HO approach. It is worth noticing that, although for $c=-2,-1.5,0$ the peak position can be approximately described by the no-coherence fits, the coherent scaling laws are still substantially preferred. While the scaling with respect to each parameter is individually sensitive to the dynamics of colour coherence, we can observe that the greatest sensitivity is manifested by the medium length. This agrees with our expectations from the dependence of $\theta_{\rm c}$ with the medium parameters shown in eq.~\eqref{eq:theta_c}.

Having extracted these scaling laws, we can now appreciate that the different behaviour of the gradient to the left of the peak in the DC (blue and red) curves  w.r.t the PC (green and purple) ones in figure~\ref{fig:HOvaryingParameters}, points out to the emergence of colour coherence dynamics. This transition to coherence can be further observed in figure~\ref{fig:HO_EEC_app} of appendix~\ref{app:figs}.\\

\begin{table}[]
\centering
\begin{tabular}{|c|c|c|}
    \hline \multicolumn{3}{|c|}{\textbf{Multiple scatterings: Yukawa}} \\
    \hline \textbf{Scaling of $\boldsymbol{\theta_{\mathrm{peak}}}$} & $c = -2$ & $c = -1.5$ \\ \hline
   no-coherence  & $\sim E^{-0.50}L^{4.4}\mu^{0.45}n_0^{~0.36}$ $[-4.1]$ & failed to converge \\
   $E \ll E_{\rm c}$ & $\sim E^{-0.88}L^{0.34}\mu^{0.77}n_0^{~0.47}$ $[0.03]$ & $\sim E^{-0.92}L^{0.44}\mu^{0.87}n_0^{~0.50}$ $[0.00]$ \\
   $E \gg E_{\rm c}$ & $\sim E^{-0.64}L^{-0.2}\mu^{0.27}n_0^{~0.15}$ $[-0.01]$ & $\sim E^{-0.69}L^{-0.10}\mu^{0.38}n_0^{~0.21}$ $[0.00]$ \\ \hline
   \textbf{Scaling of $\boldsymbol{\theta_{\mathrm{peak}}}$} & $c = -1$ & $c = 0$ \\ \hline
   no-coherence  & $\sim E^{-0.30}L^{3.7}\mu^{0.27}n_0^{~0.31}$ $[-3.39]$& peaks absent \\
   $E \ll E_{\rm c}$ & $\sim E^{-0.97}L^{0.47}\mu^{0.93}n_0^{~0.52}$ $[0.01]$ & peaks absent \\
   $E \gg E_{\rm c}$ & $\sim E^{-0.73}L^{-0.01}\mu^{0.46}n_0^{~0.24}$ $[-0.01]$ & peaks absent \\\hline
\end{tabular}
\caption{Scaling of the peak position of the $\td P^{(2)}/\td \theta$ distribution in \eqref{eq:distribution} with $c=-2,-1.5,-1,0$ computed through the multiple scattering approach with a Yukawa interaction model (see section~\ref{subsec:Yuk}). The critical energy is given by $E_{\rm c} = n_0\mu^2 L^{2}$. In square brackets we show the mass dimension of each fit. For $c=0$, $\td P^{(n)}/\td \theta$ with $n=1,2$ does not peak for $\theta \leq 1$, see figure~\ref{fig:Yukawa_dP/dtheta}. For each fit in which the peak angle power law has a deviation of the mass dimension from zero smaller than $1$, the errors on the power of the parameters are smaller or equal to $\pm 0.01$.}
\label{tbl:Yukawa}
\end{table}

\noindent\emph{Yukawa collision rate ---} We now present in table~\ref{tbl:Yukawa} the corresponding results obtained using the multiple scattering formalism with a Yukawa parton-medium interaction model described in section~\ref{subsec:Yuk} for $266$ sets of parameters within the ranges: $E\in[50,1000]$ GeV, $L\in[2,10]$ fm, $\mu\in[0.7,1.4]$ GeV, and $n_{0}\in[1,4]$ fm$^{-1}$. The slightly reduced sample size relative to the HO approach is due to the increased computational time needed to evaluate $F_{\mathrm{med}}$ using a Yukawa parton-interaction model. As for the HO approach, we performed both the coherence and no-coherence power-law fits, with the critical energy defined as $E_{\rm c} = n_{0}\mu^{2}L^{2}$ due to the $\hat q \sim n_0 \mu^2$ matching already discussed at the end of section~\ref{subsec:Yuk}. We can see that for $c=-2,-1.5,-1$ the coherent power-law fits are strongly favoured over the no-coherence fits, for which either the fit does not converge or the resulting mass dimensions substantially deviate from $0$, indicating the bad quality of the fit. Peaks could not be identified for $c=0$, as it was already seen for $n=1$ in figure~\ref{fig:Yukawa_dP/dtheta}, and so no power-law fits were performed. The greatest sensitivity to the dynamics of coherence is found for the $c=-1.5$ spectra. It is interesting to note that when matching parameters with the HO curves  the $c=-1.5$ Yukawa peak angle overlaps with the $c=-1$ HO peak angle, for which the most sensitivity to coherence was observed in table~\ref{tbl:HO}. This seems to indicate that the angular location of the region of the EEC spectrum sensitive to colour coherence is nearly model independent. Also, as for the HO approach, the scaling parameter with the greatest sensitivity to the dynamics of coherence is the medium length.

\subsubsection{The Single Scattering Approach (GLV)}
\label{subsec:scalingsGLV}

\begin{table}[]
\centering
\begin{tabular}{|c|c|c|}
    \hline \multicolumn{3}{|c|}{\textbf{Single scattering (GLV) }} \\
    \hline \textbf{Scaling of $\boldsymbol{\theta_{\mathrm{peak}}}$} & $c = -2$ & $c = -1.5$ \\ \hline
   no-coherence  & failed to converge & $\sim E^{-0.54}L^{-0.50}\mu^{0.07}n_0^{~0.05}$ $[0.10]$ \\
   $E \ll E_{\rm c}$ & failed to converge & $\sim E^{-0.56}L^{-0.50}\mu^{0.07}n_0^{~0.08}$ $[0.10]$ \\
   $E \gg E_{\rm c}$ & failed to converge & $\sim E^{-0.51}L^{-0.52}\mu^{0.06}n_0^{~0.01}$ $[0.06]$ \\ \hline
   \textbf{Scaling of $\boldsymbol{\theta_{\mathrm{peak}}}$} & $c = -1$ & $c = 0$ \\ \hline
   no-coherence  & $\sim E^{-0.57}L^{-0.37}\mu^{0.20}n_0^{~0.03}$ $[0.00]$& $\sim E^{-0.55}L^{-0.40}\mu^{0.17}n_0^{~-0.01}$ $[0.01]$ \\
   $E \ll E_{\rm c}$ & $\sim E^{-0.62}L^{-0.37}\mu^{0.24}n_0^{~0.00}$ $[0.01]$ & $\sim E^{-0.60}L^{-0.34}\mu^{0.25}n_0^{~-0.01}$ $[0.02]$ \\
   $E \gg E_{\rm c}$ & $\sim E^{-0.49}L^{-0.53}\mu^{0.03}n_0^{~0.00}$ $[0.00]$ & $\sim E^{-0.48}L^{-0.55}\mu^{0.014}n_0^{~0.01}$ $[0.06]$ \\\hline
\end{tabular}
\caption{Scaling of the peak position of the $\td P^{(2)}/\td \theta$ distribution in \eqref{eq:distribution} for $c=-2,-1.5,-1,0$   evaluated through the  single scattering (GLV) approach (see section~\ref{subsec:GLV}). The critical energy is $E_{\rm c} = \mu^{2}n_{0}L^{2}$. In square brackets we show the mass dimension of each fit. When $c=-2$, $\td P^{(2)}/\td \theta$ does not present a single crisp peak. The errors on the power of the parameters are smaller or equal to $\pm 0.02$.}
\label{tbl:GLV}
\end{table}

The results of the power-law fits to the peak position of the $\td P^{(2)}/\td \theta$ distribution computed within the single scattering (GLV) framework described in section~\ref{subsec:GLV} are shown in table~\ref{tbl:GLV}. We have made use of 146 sets of parameters within the following ranges: $E\in[50,1000]$\,GeV, $L\in[2,10]$\,fm, $\mu\in[0.7,1.4]$ GeV, and $n_{0}\in[1,4]$ fm$^{-1}$. The reduced sample size relative to the multiple scattering approaches is due to two main factors: the increase in the computational time needed to obtain $F_{\mathrm{med}}$ with respect to the multiple scattering approaches, and the reduction of the available parameter space due to the unitarity issues inherent to the GLV formalism highlighted in section~\ref{subsec:GLV} (all sets containing points yielding $F_{\rm med}<-1$ were discarded). Even though the angular scale $\theta_{\rm c}$ does not appear naturally in the formula for $F_{\rm med}$ in \eqref{eq:GLVspectrum}, we still expect the coherence transition to occur at the same place as for the Yukawa collision rate model in the multiple scattering approach: $E_{\rm c} = n_{0}\mu^{2}L^{2}$. We observe that the peak position appears slightly sensitive to coherence for $c=-1.5,-1,0$, but can also be reasonably well described by the no-coherence ansatz. To further differentiate the quality of the coherence and no-coherence fits, we performed a reduced-$\chi^2$ analysis for the fits. The standard variance on each peak position was estimated from the convergence of our numerical analysis. The reduced-$\chi^2$ test does favoured the coherence fit by a factor of $5$.

When $c=-2$, $\td P^{(2)}/\td \theta$ does not present a consistent single crisp peak due to the more complex small angle tail of the in-medium enhancement in the EEC distribution. This tail is visible in figure~\ref{fig:combined_bump} where the transition at small angles from agreeing with the vacuum result to having a significant medium enhancement is very slow for the GLV curve. As already mentioned at the end of subsection~\ref{subsec:GLV}, this behaviour is a direct consequence of the phases in \eqref{eq:GLVphases} all being different, as opposed to the multiple scattering cases where all the phases appearing in \eqref{eq:Fmed2} are the same. Similarly, the exponential factors in the $n$-point functions in the multiple scattering case responsible for the coherence transition turn into polynomial for the single scattering case, thus leading to a less pronounced change of regime. Additionally, one can also notice that the peak position is not sensitive to $n_{0}$. This agrees with our expectations since in the single scattering approach $n_{0}$ only appears as a normalisation parameter in $F_{\rm med}$ (see \ref{eq:GLVspectrum}).

\subsection{Visualising Coherence}
\label{subsec:visualising}
\begin{figure}
\centering
\includegraphics[width=0.9\textwidth]{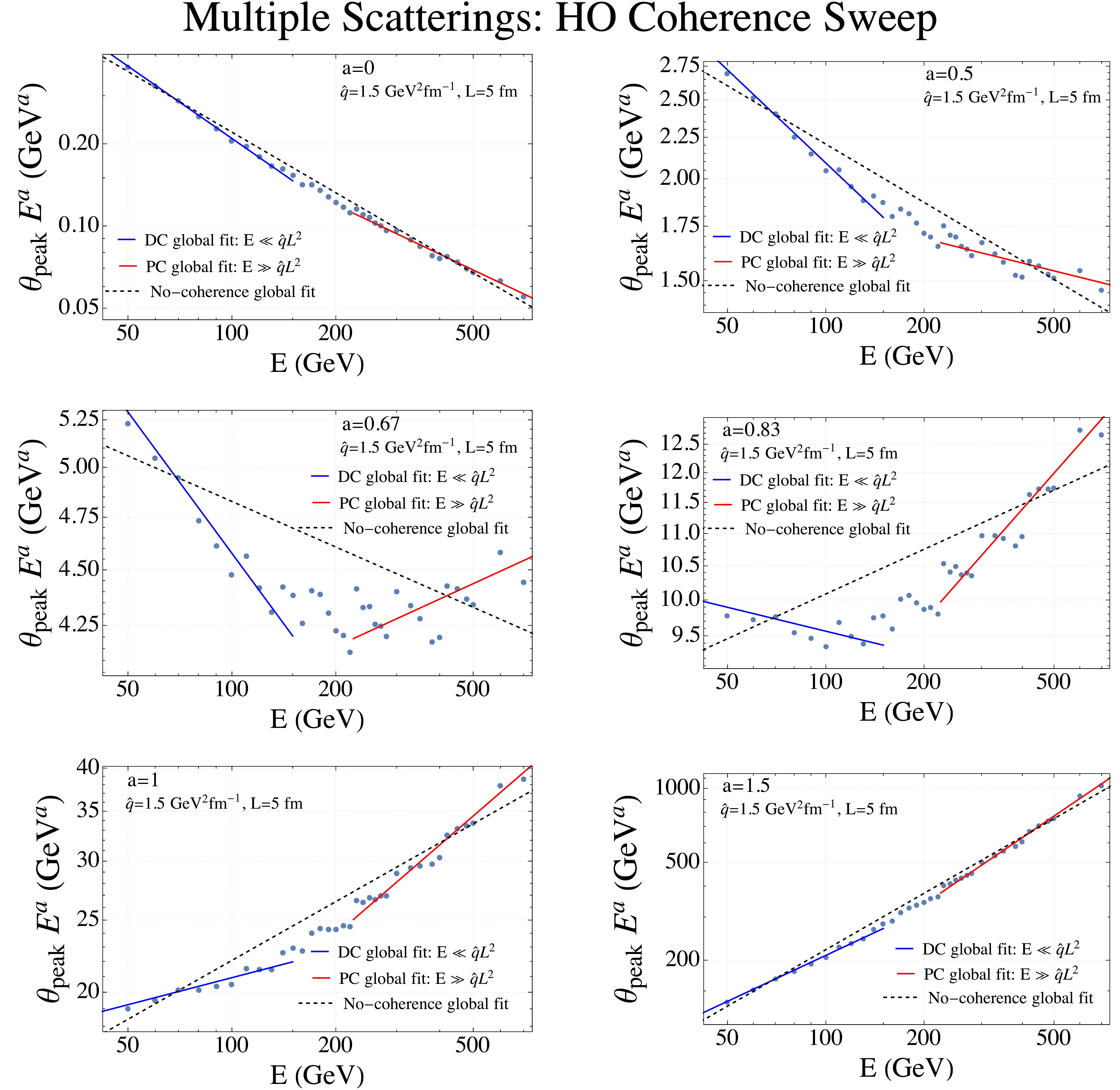}
\caption{Peak angle $\theta_{\rm peak}$ of the ${\rm d}P^{(2)}/{\rm d}\theta$ with $c=-1$ computed within the multiple scattering HO approach with $\hat q=1.5\, {\rm GeV}^2$/fm and $L=5$\,fm multiplied by a variable power of the jet energy $E^a$, where the value of $a$ is indicated in each subplot. The blue (red) solid lines correspond to the coherent fits described at the beginning of section~\ref{sec:procedure} for sets in the $E \ll E_{\rm c}=\hat{q}L^2$ decoherent ($E \gg E_{\rm c}$ partially coherent) region. The black dashed lines correspond to the no-coherence fits.}
  \label{fig:coherenceHO}
\end{figure}
The key results of our numerical coherence analysis are the power-law scalings presented in the previous section and summarised in tables~\ref{tbl:HO},~\ref{tbl:Yukawa},~and~\ref{tbl:GLV}. However, since visualising these scaling laws might be hard, we present in this subsection projections onto particular slices of the parameter space. We choose to isolate the energy dependence of the position of the peak and thus, for each of the jet quenching formalisms, we fix the medium parameters and present the position of the peak angle of the ${\rm d}P^{(2)}/{\rm d}\theta$ distribution scaled by a variable power of the jet energy $E^a$ as function of the jet energy. Changing the power of the jet energy $a$ in the rescaling of the axis allows us to amplify the differences between the two regions, above and below the critical energy, and make the transition visible. 

In figure~\ref{fig:coherenceHO} we show the product $\theta_{\rm peak}E^a$ for several values of $a$ and with $\theta_{\rm peak}$ being the position of the peak angle of the ${\rm d}P^{(2)}/{\rm d}\theta$ distribution for $c=-1$ evaluated through the multiple scattering HO approach. The blue (red) lines are the power laws of the peak angle in the $E \ll E_{\rm c}$ DC region ($E \gg E_{\rm c}$ PC region), where $E_{\rm c}=\hat qL^2$,  previously presented in table~\ref{tbl:HO}, projected for the fixed values of $L$ and $\hat{q}$ indicated in the figure. In the top left panel, where we simply show $\theta_{\rm peak}$ as a function of the jet energy, one can already see that the blue line presents a steeper slope than the red one, indicating that $\theta_{\rm peak}^{\rm DC}$ and $\theta_{\rm peak}^{\rm PC}$ scale differently with the energy, which is a signature coherence transition in  medium-induced radiation. This becomes more evident when increasing the value of the power of the energy, $a$, since the slope of $\theta_{\rm peak}E^a$ for the sets below the critical energy $E_{\rm c}$  clearly changes its sign at a different value of $a$ than that of the sets above $E_{\rm c}$, and thus the no-coherence fit (black dashed) is unable to reproduce the energy dependence of $\theta_{\rm peak}$. The dependence on the critical energy, which coincides with the condition $\theta_{\rm L}=\theta_{\rm c}$, undoubtedly indicates the emergence of a new relevant angular scale. We note that in \cite{Andres:2022ovj} a plot of $\theta_{\rm onset}/\theta_{\rm peak}$ was presented to illustrate the coherence transition. As it was found that $\theta_{\rm onset}\sim E^{-0.5}$ both for the decoherent and partially coherent regions, this figure is equivalent to the top right hand panel in figure~\ref{fig:coherenceHO} when reflected over the $x$-axis.

\begin{figure}
\centering
\includegraphics[width=0.9\textwidth]{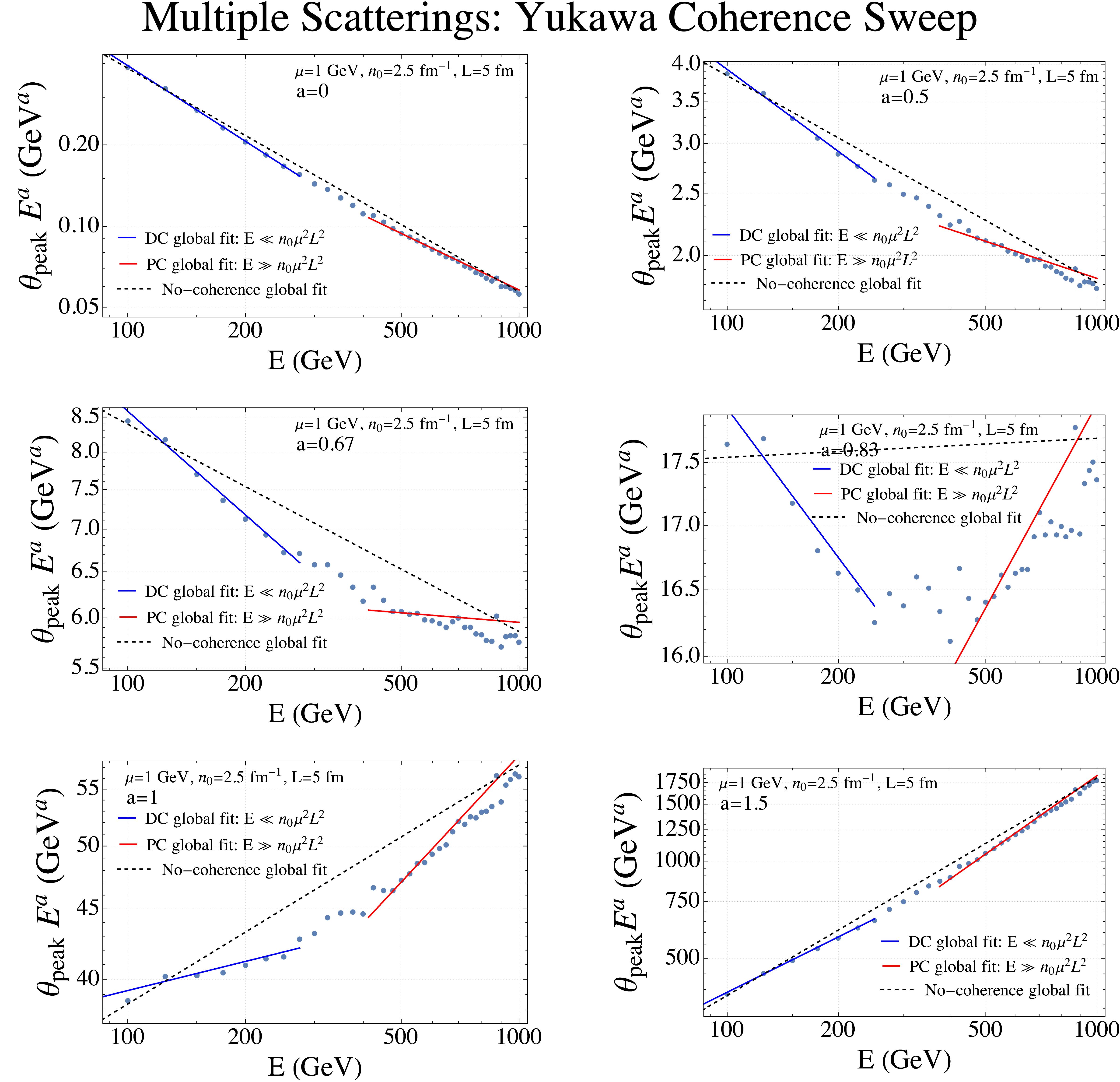}
\caption{Peak angle $\theta_{\rm peak}$ of the ${\rm d}P^{(2)}/{\rm d}\theta$ with $c=-1.5$ computed within the multiple scattering Yukawa approach with $n_0=2.5\,{\rm fm}^{-1}$, $\mu =1$\,GeV, and $L=5$\,fm multiplied by a variable power of the jet energy $E^a$, where the value of $a$ is indicated in each subplot. The blue (red) solid lines correspond to the coherent fits described at the beginning of section~\ref{sec:procedure} for sets in the $E \ll E_{\rm c}=n_0\mu^2L^2$ decoherent ($E \gg E_{\rm c}$ partially coherent) region. The black dashed lines correspond to the no-coherence fits.}
  \label{fig:coherenceYukawa}
\end{figure}
The equivalent results for the multiple scattering formalism with a Yukawa parton-medium interaction model are presented in figure~\ref{fig:coherenceYukawa}. In this case,  $\theta_{\rm peak}$ is the position of the ${\rm d}P^{(2)}/{\rm d}\theta$ distribution with $c=-1.5$, since for this jet quenching approach this was the value of $c$ that yielded the largest sensitivity to the colour coherence dynamics (see section~\ref{subsec:scalings_HOYuk}). By varying the value of $a$, we clearly observe that $\theta_{\rm peak}$ depends on the jet energy in a different way for the data sets below and above the critical energy $E_{\rm c} = n_0\mu^2L^2$. It is worth keeping in mind that, as in the previous figure, the fits shown in  red (blue) were performed including all data sets provided $E \gg E_c$ ($E \ll E_c$), not only the ones shown in the figure, and thus do not always seem to be the best fit to the points on each plot. We can also clearly observe that the no-coherence fits (black dashed lines) do not properly describe the energy dependence of $\theta_{\rm peak}$.

\begin{figure}
\centering
\includegraphics[width=0.9\textwidth]{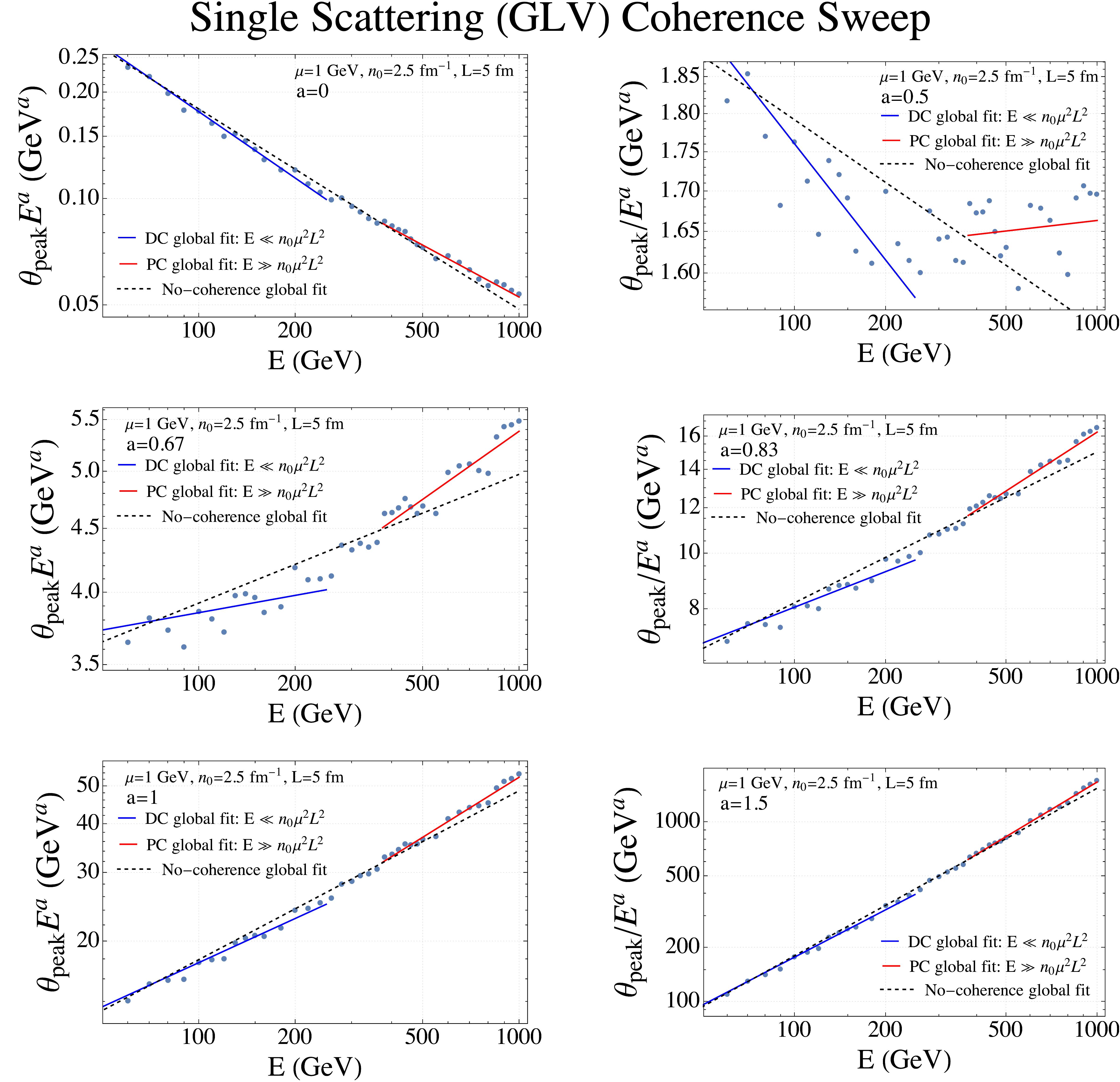}
\caption{Peak angle $\theta_{\rm peak}$ of the ${\rm d}P^{(2)}/{\rm d}\theta$ with $c=-1$ computed within the single scattering (GLV) approach with $n_0=2.5\,{\rm fm}^{-1}$, $\mu =1$\,GeV, and $L=5$\,fm multiplied by a variable power of the jet energy $E^a$, where the value of $a$ is indicated in each subplot. The blue (red) solid lines correspond to the coherent fits described at the beginning of section~\ref{sec:procedure} for sets in the $E \ll E_{\rm c}=n_0\mu^2L^2$ decoherent ($E \gg E_{\rm c}$ partially coherent) region. The black dashed lines correspond to the no-coherence fits.}
\label{fig:coherenceGLV}
\end{figure}

Finally, we present in figure~\ref{fig:coherenceGLV} the equivalent results computed within the single scattering (GLV) approach. We see that the dependence on coherence in this framework is relatively weak when compared to the previous multiple scattering approaches. Indeed, although the top right and center left panels seem to present a clear transition to coherent dynamics, it is still possible within the current precision of our numerical results to reasonably describe all the data sets across all energies with the non-coherent fit (black dashed lines) summarised in table~\ref{tbl:GLV}. It is also notable that the apparent coherent transition is smoother and takes place over a wider range of energies than those in figures~\ref{fig:coherenceHO} and~\ref{fig:coherenceYukawa}, as expected. Varying the value of $E_{\rm c}$ to account for this effect did not improve the quality of the fits for the DC and PC scaling laws presented in section~\ref{subsec:scalingsGLV}.

\section{Discussion}
\label{sec:discuss}

This paper is focused on semi-analytic computations of the EEC distribution measured on a massless quark-initiated jet which propagates through a  static QCD medium of finite size $L$. For the extension to the massive case, see~\cite{Andres:2023ymw}. Our results can be understood as LO+NLL accurate in vacuum physics, and  LO in $\As$ for in-medium splittings, within the jet quenching formalisms considered. From section \ref{sec:theory_mult} onward, the interactions with the medium are always considered in the high-energy approximation, where the momentum transfer in each scattering is purely transverse and thus no energy (longitudinal momentum) is exchanged. Corrections to this approach maximally entering at order $\mathcal{O}(\bar{\mu}^2_{\rm s}/Q^2)$.\footnote{The relevant scale here is the typical momentum transferred per scattering given by the Debye mass, which is of the same order or less than $\bar\mu_{\rm s}.$} This separation between longitudinal and transverse dynamics allows us to have a different power counting for splittings, which we consider to LO only, and interactions with the medium, which are further enhanced by the medium density and are resummed to all orders in the multilpe scattering case. Due to the novelty of the EEC observable and the subtlety of the power counting in its in-medium calculation, we further discuss now both the implications of working at fixed coupling and the parametric size of the subleading corrections. 

The transverse nature of the scatterings with the medium implies that there are no corrections to the energy weighting entering at LO in the medium-induced splittings. We quoted in \eqref{eq:medPart} an error $\mathcal{O}(\bar\mu_{\rm s}/Q)$ due to complete momentum conservation but, unless collisional energy loss is included, such terms would only appear when additional splittings are considered and thus are higher-order in the medium splittings, making them of order $\mathcal{O}(\alpha_s\bar\mu_{\rm s}/Q)$.

By restricting our calculation to LO in the medium splittings we induce an error maximally of the order $\mathcal{O}(\As(\theta Q) \ln \theta ~ \bar{\mu}_{\rm s}/Q)$ from the expansion of $F^{(ij)}_{\rm med}\td \hat{\sigma}^{(ij)}$, as explained in the derivation of eq.~\eqref{eq:master}. This can, in principle, be systematically improved by computing $F^{(ij)}_{\rm med}\td \hat{\sigma}^{(ij)}$ at NLO,  which would entail computing $F^{(q_{1}q_{2})}_{\rm med}\td \hat{\sigma}^{(q_{1}q_{2})}$ and $F^{(g_{1}g_{2})}_{\rm med}\td \hat{\sigma}^{(g_{1}g_{2})}$ at LO, where an additional splitting is needed to create the extra quark or gluon. Additionally, one must also compute $F^{(qg)}_{\rm med}\td \hat{\sigma}^{(qg)}$ at NLO involving diagrams where the initial pair can emit an extra gluon (real contributions) but also diagrams with loops (virtual contributions). Overall, these NLO contributions are usually attributed to causing energy loss in typical heavy-ion jet substructure observables (such as jet shapes or groomed observables), when computed in the soft limit, because the additional wide-angle medium-induced soft parton can be `lost' outside the jet cone. Such observables are exclusive and, in vacuum, the lost gluon is associated with an additional Sudakov double (or sometimes single) soft logarithm potentially providing a leading effect, which could be made worse in presence of the medium. However, this is not the case for energy correlator observables, since the EEC is inclusive and so the parton is not lost, modifying instead  the structure of the correlations at an angle determined by its kinematics. Since the energy weighting in the EEC removes any essential soft divergences and thus logarithms, in order to find these additional correlations the diagrams must be computed beyond the soft limit ($z \rightarrow 0$) for the jet partons. Taken on its own, the soft limit of the diagrams is further suppressed, maximally scaling as 
$\mathcal{O}(\As(\theta Q) \ln \theta ~ \bar{\mu}_{\rm s}/Q \sqrt{\Lambda_{\rm QCD}/Q}).$ In contrast to typical heavy-ion jet substructure observables, the collinear limit of these NLO diagrams (at finite $z$) could be used to find the leading contribution of the NLO correction to small angle correlations, such as those studied in this paper.

Nevertheless, it is well known that a complete calculation of these NLO contributions is out of reach within the formalisms used in this manuscript. Calculations involving multiple medium splittings are available only under the assumption that emissions occur independently without interference \cite{Salgado:2003gb,Blaizot:2013vha,Mehtar-Tani:2017ypq}, which has been argued to be a good approximation only in the case of very soft gluon emissions and thus would not constitute a leading correction in the calculation of the EEC, as explained above. Recent developments in the calculations of multiple splittings with finite energy fraction have shown the increasing difficulty to perform a full calculation even for less differential observables where the angular dependence is not kept \cite{Arnold:2015qya,Arnold:2020uzm}.

The calculation of $F_{\rm med}$ at LO in the medium splittings can still be improved in several ways either by relaxing some of the approximations entering the jet quenching models or calculating correction terms. In the multiple scattering formalism explained in section~\ref{sec:theory_mult} the calculation of $F_{\rm med}$ was performed in a semi-hard approximation at leading colour which can be systematically improved by: implementing finite $N_{\rm c}$ results, which are known within the HO approximation \cite{Isaksen:2020npj} and could be easily extended to include the Yukawa case too with a slight increase in the computational cost, or computing corrections to the straight-line approximation in the in-medium propagators in eq.~\eqref{eq:prop_tilted2} which would allow us to improve the handling of the transverse momentum broadening of the daughter particles.  This correction goes as $\mathcal{O}(\bar{\mu}_{s}/zQ)$ and so has the potential to be large for $z< \bar{\mu}_{s}/Q$. Consequently, this correction will likely be one of the most important to include in the future.

Regarding the single scattering framework, a recursive method to systematically increase the number of in-medium scatterings in the calculation of $F_{\rm med }$ is known \cite{Sievert:2018imd,Sievert:2019cwq}. We note that the resulting expressions are cumbersome, and will thus pose computational challenges in the numerical evaluation of the EEC beyond the single scattering limit implemented in this manuscript. In any case, we expect that gradually increasing the number of scatterings will result into a higher sensitivity to colour coherence.

The same calculation performed here can also be carried out for in-medium gluon initiated jets, albeit with some extra technical difficulties. In this case, two different channels enter at the leading order, $g\to q\bar q$ and $g\to gg$. For the multiple scattering case, $F_{\rm med}$ has been calculated in the semi-hard approach within the HO approximation in \cite{Isaksen:2020npj}, with the $g\to gg$ channel result being slightly more difficult and requiring some additional computational time. For the single scattering case $F_{\rm med}$ can be straightforwardly derived from the formulas for splittings in the first opacity expansion in \cite{Ovanesyan:2011kn}. In all cases we expect the result to present the same features as for the quark-initiated jets since formation times and colour coherence arguments still apply.

Although the static brick model of the QGP employed in this manuscript is too simple to be compared with realistic scenarios, our analysis is of vital importance to understand how particular features of the calculation of the modification of in-medium splittings are imprinted in the EEC. In order to understand how this observable would behave in an actual experimental environment going beyond this simplified static scenario is of the utmost importance. This can be done in two steps: first, the calculation of $F_{\rm med}$ can be performed for an evolving medium, with the local parameters varying according to a simple expansion model or extracted from a hydrodynamic simulation, second, evaluations of $F_{\rm med}$ performed under different dynamic profiles can be combined in the evaluation of the correlator to mimic the different trajectories along the changing underlying event contributing to a possible measurement.

Finally, we note that it would be also extremely interesting to perform a comprehensive analysis on the in-medium EEC within a full parton shower Monte Carlo implementation, which will allow us also to study the resilience of the correlator to medium response. This will be done in a separate publication.

\section{Conclusions}
\label{sec:conclusions}

In this paper we have presented a thorough analysis of how correlators of energy flow operators can be applied to jet substructure studies in heavy-ion collisions. Expanding on the original proposal in \cite{Andres:2022ovj}, we have performed a detailed derivation of the two-point correlator (EEC)  of a heavy-ion jet including medium-induced radiation effects on top of the vacuum structure, which is here known to high (NLL) accuracy. We have considered different jet quenching frameworks, both with and without multiple scattering resummation to test the model dependence of our results.

The advantages of using energy correlators in addition to other jet substructure observables are already well documented in the context of p-p collisions. It is thus natural to extend these studies to the case of heavy-ion collisions, even more so when considering some of their useful properties from the perspective of a heavy-ion environment. In particular, the fact that for energy correlators in the vacuum the presence of any new scales is reflected in different angular regions seems tailor made for addressing multi-scale systems, such as heavy-ion collisions. In the context of jet quenching theory,  colour coherence effects have been argued during the last decade to play an important role in our understanding of jet-medium interactions due to the emergence of a resolution scale which determines if the inner jet structure affects the quenching. This discussion has often been framed in terms of the angular separation between jet constituents where the resolution scale turns into a critical angle below which the medium is not sensitive to jet substructure. It is then natural to formulate the emergence of colour coherence in jet quenching in terms of energy correlators.

Our analysis shows that the EEC can be used to identify the dynamics of colour coherence, with a consistent picture emerging from all the jet quenching formalisms considered. We found that the angular region of the EEC most sensitive to coherence is broadly model independent, appearing at the moderately small angle region of the medium enhancement for all the approaches. We also showed that the signatures of coherence dynamics are weaker in the single scattering GLV framework than in the multiple scattering approaches. This outcome agrees with our expectations, since the appearance of coherence phenomena is due to the loss of colour correlations induced by colour exchanges between the jet and the medium, and thus its effects are enhanced in a multiple scattering setting.

It is remarkable how the EEC provides an observable for heavy-ion collisions which is expected to be largely insensitive to soft degrees of freedom (and can easily be extended to higher powers of the energy weighting to make it even less sensitive), directly reflects the physical phenomena and important scales of the microscopic dynamics, and whose vacuum baseline is well understood and calculable to a high degree of accuracy. As we eagerly anticipate experimental measurements at both RHIC and LHC in the near future, we envisage an exciting opportunity for direct comparisons with theoretical calculations. However, we recognise the importance of implementing first a more realistic model for the medium and jet-medium interactions, which should encompass crucial aspects such as the event geometry, hydrodynamical evolution, and medium response, as outlined at the end of section~\ref{sec:discuss}.

Our work is meant to open the door for a rich program of theoretical and experimental studies of  jet substructure in heavy-ion collisions in terms of energy correlators. We would like to encourage our colleagues to consider building upon this new approach. To this end, we have provided a detailed discussion on how to compute the EEC of an in-medium massless jet, describing the accuracy and limitations of our calculation and the different directions in which it can be systematically improved.

\acknowledgments
We thank Liliana Apolin\'ario and Raghav Kunnawalkam Elayavalli for useful discussions. This work is supported in part by the GLUODYNAMICS project funded by the ``P2IO LabEx (ANR-10-LABX-0038)'' in the framework ``Investissements d’Avenir'' (ANR-11-IDEX-0003-01) managed by the Agence Nationale de la Recherche (ANR), France. This work is also supported by European Research Council project ERC-2018-ADG-835105 YoctoLHC; by Maria de Maetzu excellence program under project CEX2020-001035-M; by Spanish Research State Agency under project PID2020-119632GB-I00; and by Xunta de Galicia (Centro singular de investigación de Galicia accreditation 2019-2022), by European Union ERDF and by OE - Portugal, Fundac\~ao para a Ci\^encia e Tecnologia (FCT) under projects EXPL/FIS-PAR/0905/2021. C.A. has received funding from the European Union’s Horizon 2020 research and innovation program under the Marie Sklodowska-Curie grant agreement No 893021 (JQ4LHC). I.M. is supported by start-up funds from Yale University.

\appendix

\section{Vacuum Resummation and Anomalous Dimensions}
\label{app:app_1}

The following discussion draws on results originally presented in \cite{Hofman:2008ar,Belitsky:2013bja,Chen:2021gdk,Chen:2022jhb}. The primary goal of this appendix is to present the expressions needed to evaluate the vacuum contribution in eq.~\eqref{eq:master}. These are written in a condensed form in appendix~\ref{app:gn}. Concurrently, we aim to present the vacuum resummation via the celestial OPE in a manner accessible to a reader with no prior knowledge of conformal field theories. This style of presentation is largely missing from the literature and we hope will help bridge the gap between the heavy-ion and the energy correlator communities.

\subsection{Recap of the OPE}

We first review the celestial OPE with specific focus on its application to the energy flow operator $\cE(\vec n_1)$ used in this paper. Several of the results we present here hold more generally than the context in which we shall derive them. We will assume that we can compute the $\cE(\vec n_1)$ on a QCD process. Firstly, let us stress that $\cE(\vec n_1)$ is a light-ray operator, which is obtained by integrating a ``usual'' local QFT operator (in this case the energy-momentum tensor) along a light-cone:
\beq
    \cE(\vec n_1) = 
    \lim_{r\rightarrow \infty} \int^{\infty}_{0} \mathrm{d}t \,r^2 n_1^i \,
    T_{0i}(t,r\vec{n}_1) = \frac{1}{4} \lim_{x_{+}\rightarrow \infty} \int^{\infty}_{-\infty} \mathrm{d} x_{-} ~ x^{\, 2}_{+}   ~ T_{\mu\nu}(x_{+} n_{1} + x_{-} \bar{n}_{1}) \bar{n}_{1}^{\mu} \bar{n}_{1}^{\nu}\,.
\eeq
Here $n^{\mu}_{1} = (1, \vec{n}_1)$ and $\bar{n}^{\mu}_{1} = (1, -\vec{n}_1)$. $x^{-} = x \cdot n_{1} /2$, and $x^{+} = x \cdot \bar{n}_{1} /2$ are coordinates along the $n$ light-cone. Consequently, $\cE(\vec n_1)$ does not share the same transformation properties under the Lorentz group as a local QFT operator. 

In the small angle limit ($\vec n_{1}\cdot \vec n_{2}\rightarrow 1$) the composite operator $\cE(\vec n_1) \cE(\vec n_2)$ diverges, since $\td\sigma_{ij}/\td \vec n_{i}\td \vec n_{j}$ diverges when $\vec n_{i}\cdot \vec n_{j}\rightarrow 1$. Thus, we expect to find a Laurent series expansion for the small angle limit, schematically given by
\begin{align}
    \cE(\vec n_1) \cE(\vec n_2) = \sum_{i} \frac{\As}{\pi} C_{i} ~ (1-\vec n_{1}\cdot \vec n_{2})^{\kappa_{i}} \mathbb{O}_{i}(\vec n_2) + (1-\vec n_{1}\cdot \vec n_{2})^{0}, \label{eq:A2}
\end{align}
where $\mathbb{O}_{i}(\vec n_2)$ is a QCD light-ray operator with the same mass dimension as $\cE(\vec n_1) \cE(\vec n_2)$. The coefficients $C_{i}$ are independent of $\vec n_{1}\cdot \vec n_{2}$ and are real process-dependent constants when spin averaged. Necessarily, for terms where $C_{i}\sim 1$, we must have that ${\kappa_{i}} = -1 + \mathcal{O}(\As)$ in order to recreate the universal LO $\td \theta^{2}/\theta^{2}$ collinear divergences of the QCD quark/gluon splitting functions. It is required that the sum is over light-ray operators, so that the left hand side and right hand side have the same causal structure, and therefore transformations, under the Lorentz group. 

A general light-ray operator (with integer spin $J$) can be defined from a local QFT operator, $\mathcal{O}^{\mu_{1}, \dots , \mu_{J}}_{i}$, by
\begin{align}
    \mathbb{O}^{[J]}_{i}(\vec n) &= \lim_{r\rightarrow \infty} \,r^{\Delta_{i} - J} \int^{\infty}_{0} \mathrm{d}t ~~ n_{\mu_{1}} \dots n_{\mu_{J}} \,
    \mathcal{O}^{\mu_{1}, \dots , \mu_{J}}_{i}(t,r\vec{n}) \,, 
    \nonumber \\
    &= \frac{1}{4}\lim_{x_{+}\rightarrow \infty} \int^{\infty}_{-\infty} \mathrm{d} x_{-} ~ x_{+}^{\Delta_{i} - J}   ~ n_{\mu_{1}} \dots n_{\mu_{J}} \,
    \mathcal{O}^{\mu_{1}, \dots , \mu_{J}}_{i} (x_{+} n + x_{-} \bar{n})\,,
\end{align}
where $\Delta_{i}$ is the mass dimension of $\mathcal{O}^{\mu_{1}, \dots , \mu_{J}}_{i}$. From this definition, simple power counting gives that an operator $\mathbb{O}^{[J]}$ has a mass dimension $J-1$. It is therefore the spin-$3$ QCD light-ray operators which will appear in eq.~\eqref{eq:A2}. We will provide the relevant QCD operators, $\mathcal{O}^{\mu_{1}, \mu_{2} , \mu_{3}}$, in the following subsection.

We have pinned down that 
$$\sum_{i} \mathbb{O}_{i}(\vec n_2) \rightarrow \sum_{i} \mathbb{O}^{[3]}_{i}(\vec n_2)\,,$$
in our small angle expansion. In order to compute $\cE(\vec n_1) \cE(\vec n_2)$ we must also fix $C_{i}$ and $\kappa_{i}$. $C_{i}$ must be found by performing a fixed order matching, however, we can find more insight into the form of $\kappa_{i}$ by considering a general boost in the direction $\vec{n}_{2}$. We define the boost so that
\begin{align}
    \Lambda n_{2} = \lambda_{2} n_{2}\,, \qquad \Lambda \bar{n}_{2} = \lambda^{-1}_{2} \bar{n}_{2}\,, \qquad \Lambda n_{1} = \lambda_{1} n'_{1}\,.
\end{align}
Upon applying this boost to eq.~\eqref{eq:A2}, assuming a scaleless theory so the operators $\mathcal{O}^{\mu_{1}, \dots , \mu_{J}}_{i}$ and $T^{\mu\nu}$ are  invariant under reparameterisations, one finds that
\begin{align}
    \Lambda: ~ \cE(\vec n_1) \cE(\vec n_2) \mapsto \cE(\Lambda \vec n_1) \cE(\Lambda \vec n_2) = \lambda_{1}^{-3} \lambda_{2}^{-3} \cE(\vec n_1) \cE(\vec n'_2)\,,
\end{align}
and that
\begin{align}
    \Lambda: ~ (1-\vec n_{1}\cdot \vec n_{2})^{\kappa_{i}} \mathbb{O}_{i}(\vec n_2) \mapsto (\Lambda n_{1}\cdot \Lambda n_{2})^{\kappa_{i}} \mathbb{O}_{i}(\Lambda \vec n_2) = \lambda_{1}^{\kappa_{i}} \lambda_{2}^{\kappa_{i}-\Delta_{i}+1} (n'_{1}\cdot n_{2})^{\kappa_{i}} \mathbb{O}_{i}(\vec n_2)\,.
\end{align}
In the small angle limit of our expansion $\lambda_{1}\rightarrow \lambda_2$ and so for consistency $2 \kappa_{i} = -7 + \Delta_{i}$. The twist of a local operator is given by $\tau_{i} = \Delta_{i} - J$ and so we can rewrite our small angle expansion as a sum over operators of a given twist
\begin{align}
    \cE(\vec n_1) \cE(\vec n_2) = \sum_{i} \frac{\As}{\pi} C_{i} ~ (1-\vec n_{1}\cdot \vec n_{2})^{\frac{\tau_{i}-4}{2}} \mathbb{O}^{[3]}_{i}(\vec n_2) + (1-\vec n_{1}\cdot \vec n_{2})^{0}   \,. \label{eq:OPE}
\end{align}
As massless QCD is scaleless prior to renormalisation, this result will hold with bare operators or when one sets QCD $\beta=0$. Indeed, we can readily see that the leading twist contribution, $\kappa_{i} = (2-4)/2 = -1$, recreates the expected LO QCD divergences and that higher twist contributions will be sub-leading, as also expected. 

The ansatz in eq.~\eqref{eq:OPE} is what is referred to as the light-ray OPE for the energy flow operators. The derivation can be performed more generally by analysing the conformal symmetries of the operators. Upon doing so, one finds that for scaleless (conformal) theories the renormalisation of light-ray operators has the effect of modifying their twist by a scaleless anomalous dimension: $\tau_{i} = 2 + \gamma_{i}(J,\As)$ where $\gamma_{i}$ is the  anomalous dimension for the renomalisation of the operator $\mathcal{O}^{\mu_{1}, \dots , \mu_{J}}_{i}$ \cite{Hofman:2008ar}. This result therefore also holds at fixed QCD coupling. With a running coupling, the renormalisation of $\mathbb{O}^{[3]}_{i}(\vec n_2)$ will introduce a logarithmic dependence on the scale $\mu \sim n_{1}\cdot n_{2} Q$. These terms can be resummed at a given logarithmic accuracy using renormalisation group flow equations and have the effect of complicating the $\mathcal{O}(\As)$ terms in $\kappa_{i}$ if one expands around a fixed point in the running coupling. We will not review the derivation of the conformal result, instead in the following section we will review the more relevant QCD resummation at NLL accuracy.

\subsection{Spin-3 Twist-2 Operators and Anomalous Dimensions}

In QCD, the twist-2 operators for quarks and gluons are given by:
\begin{align}
    \mathcal{O}_q^{\mu_{1}\dots \mu_{J}}&=\frac{1}{2^J}\bar{\psi}\gamma^{\mu_{1}}(iD^{\mu_{2}})\dots (iD^{\mu_{J}}) \psi\,,\\
\mathcal{O}_{g}^{\mu_{1}\dots \mu_{J}}&=-\frac{1}{2^J} F_{c}^{i \mu_{1}}(iD^{\mu_{2}})\dots (iD^{\mu_{J-1}})F_{c}^{i \mu_{J}}\,, \\
\mathcal{O}_{\tilde{g},\lambda}^{\mu_{1}\dots \mu_{J}}&=-\frac{1}{2^J} F_{c}^{i \mu_{1}}(iD^{\mu_{2}})\dots (iD^{\mu_{J-1}})F_{c}^{j \mu_{J}} \varepsilon_{\lambda,i} \varepsilon_{\lambda,j} \,.
\end{align}
In \cite{Chen:2021gdk} it was shown by explicit tree-level Feynman computation that
\begin{align}
&\mathcal{E}(\vec n_1)\mathcal{E}(\vec n_2)= \nonumber \\
-\frac{\As}{8\pi^2}& \frac{1}{2(n_1\cdot n_2)} \left\{ 
\left[ (\gamma_{qq}(2)-\gamma_{qq}(3))+(\gamma_{gq}(2)-\gamma_{gq}(3))\right]\mathbb{O}_q^{[3]}(\vec n_2)\right. \nonumber\\
& \qquad\qquad +\left[ (\gamma_{gg}(2)-\gamma_{gg}(3))+2 n_f(\gamma_{qg}(2)-\gamma_{qg}(3))\right]\mathbb{O}_g^{[3]}(\vec n_2) \nonumber\\
&\qquad\qquad \left. +\frac{1}{2} \left[(\gamma_{g\tilde{g}}(2)-\gamma_{g\tilde{g}}(3))+2 n_f (\gamma_{q\tilde{g}}(2)-\gamma_{q\tilde{g}}(3)) \right]
\left( e^{2i\phi_S} \mathbb{O}_{\tilde{g},-}^{[3]}(\vec n_2) + e^{-2i\phi_S} \mathbb{O}_{\tilde{g},+}^{[3]}(\vec n_2)\right)
\right\} \nonumber \\
&+\mathcal{O}((n_1\cdot n_2)^0)+\mathcal{O}\left(\As^2\right)\,,
\end{align}
from which we can read off $C_{i}$. Here
\beq
\begin{split} \label{eq: gamma_values}
&\gamma_{qq}(J)=C_F\left( 4\left(\psi^{(0)}(J+1)+\gamma_E\right)-\frac{2}{J(J+1)}-3\right)\,,\quad 
\gamma_{qg}(J)=-T_F \frac{2(J^2+J+2)}{J(J+1)(J+2)}\,,\\
&\gamma_{gq}(J)=-C_F \frac{2(J^2+J+2)}{(J-1)J(J+1)}\,, \\
&\gamma_{gg}(J)= 4 C_A \left( \psi^{(0)}(J+1)+\gamma_E -\frac{1}{(J-1)J}-\frac{1}{(J+1)(J+2)} \right)-\beta_0\,,\\
&\gamma_{q\tilde g}(J)= - T_F\frac{8}{(J+1)(J+2)}\,,  \quad
\gamma_{g\tilde g}(J)= C_A \left(\frac{8}{(J+1)(J+2)}+3\right) -\beta_0\,,
\end{split}
\eeq
where $\psi^{(0)}(z) = \Gamma'(z)/\Gamma(z)$ is the digamma function, and $\beta_0 = 11 C_A/3 - 4 n_f T_F/3$ is the one-loop QCD beta function. Note that for readability we have omitted the commonly used $(0)$ superscript on each $\gamma_{ab}$. These anomalous dimensions are computed from the Mellin moments of the regularised collinear splitting functions (i.e. using the plus prescription):
\begin{align}
    \frac{\As}{4\pi} \gamma_{ab}(J) = -2 \int_{0}^{1} \td z ~ z^{J-1} P_{a \leftarrow b}(z)\,.
\end{align}

As previously discussed, at fixed coupling, $\tau_{i}$ is determined by the twist and anomalous dimension of the relevant QCD operators. For the twist-2 operators listed~\footnote{The unpolarised QCD operators mix with each other (see the forthcoming discussion), so the provided anomalous dimensions are for the diagonalised operators $$\mathbb{O}_{q'}^{[J]} = \mathbb{O}_{q}^{[J]} - \frac{2 \gamma_{gq}(J)}{\gamma_{gg}(J)+\gamma_{qq}(J)+A}\mathbb{O}_{g}^{[J]}, \qquad \mathrm{and}  \qquad \mathbb{O}_{g'}^{[J]} = \mathbb{O}_{g}^{[J]} -\frac{\gamma_{gg}(J)+\gamma_{qq}(J)-A}{2 \gamma_{gq}(J)}\mathbb{O}_{q}^{[J]}.$$}
\begin{align}
    \tau_{q'}(J) &= 2 + \frac{\As}{8\pi}  \left(\gamma_{gg}(J)+\gamma_{qq}(J)-A\right) +\mathcal{O}\left(\alpha_s^2\right)\,, \nonumber \\
    \tau_{g'}(J) &= 2 + + \frac{\As}{8\pi}  \left(\gamma_{gg}(J)+\gamma_{qq}(J)+A\right) +\mathcal{O}\left(\alpha_s^2\right)\,, \nonumber \\
    \tau_{\tilde{g},\lambda}(J) &= 2 + \frac{\As}{4\pi} (\gamma_{g\tilde{g}}(J) + 2n_{f}\gamma_{q\tilde{g}}(J)) +\mathcal{O}\left(\alpha_s^2\right)\,, \nonumber \\ 
    \mathrm{where} \quad A &= \sqrt{(\gamma_{gg}(J)-\gamma_{qq}(J))^2+8 n_{f} \gamma_{gq}(J) \gamma_{qg}(J) }\,.
\end{align}
In this paper we focus on a quark jet propagating through a medium. In eq.~\eqref{eq:master} the twist-2 spin-3 quark anomalous dimension, $\gamma(3)$, appears: this is given by $\gamma(3)=\tau_{q}(3)-2$. 

When one includes the running coupling, the $\mathcal{O}\left(\alpha_s\right)$ corrections to $\tau_{i}(3)$ can be computed from the renormalisation of $\mathbb{O}^{[3]}_i$, and solving the renormalisation group equations within a given logarithmic accuracy. This can be done as follows. As per the usual renormalisation group flow, schematically a renormalised operator/parameter $O(\mu)$ defined at a given scale, $\mu$, obeys a group flow equation:
\begin{align}
    \mu^{2}\frac{\td O}{\td \mu^{2}} = - \gamma_{O} O\,,
\end{align}
where $\gamma_{O}$ is the anomalous dimension of $O$. $\gamma_{O}$ is readily computed in dimensional regularisation at one-loop as the coefficient to the $\epsilon^{-1}$ pole in the counter term $\delta_{O}$ for which $O - \delta_{O} O$ is UV finite. When operators $O_{1}$ and $O_{2}$ share the same quantum numbers, such that divergences in Green functions of $O_{1}$ are cancelled by $\delta_{O_{1},O_{1}}O_{1}+\delta_{O_{1},O_{2}}O_{2}$ (and vice versa for $O_{2}$), then evolution equation must be extended as a matrix equation:
\begin{align}
    \mu^{2}\frac{\td \vec{O}}{\td \mu^{2}} = - \hat{\gamma}_{\vec{O}} \vec{O}\,,
\end{align}
where $\vec{O}=(O_{1},O_{2})^{\mathrm{T}}$ and $\hat{\gamma}^{ij}_{\vec{O}}=\delta_{O_{i},O_{j}}\epsilon|_{\epsilon \rightarrow 0}$\,. 

For convenience, we define a vector of twist-2 QCD operators as $$\mathcal{O}^{\mu_{1}\dots \mu_{J}} = (\mathcal{O}^{\mu_{1}\dots \mu_{J}}_{q}, \mathcal{O}^{\mu_{1}\dots \mu_{J}}_{g}, \mathcal{O}^{\mu_{1}\dots \mu_{J}}_{\tilde{g},\lambda})^{\mathrm{T}}\,.$$
By computing at one-loop the Green functions $G^{\mu_{1}\dots \mu_{J}}_{\lambda,\lambda'}(k,\epsilon)= \< k, \lambda \rkl \mathcal{O}^{\mu_{1}\dots \mu_{J}} \lkl k, \lambda' \>$, where $k$ is the momentum of a quark or gluon external line and $\lambda$ its spin/polarisation, the $\mathcal{O}(\As)$ anomalous dimensions can be found. One finds the operators obey the renormalisation group equation
\begin{align}
    \mu^{2}\frac{\td \, \mathcal{O}^{\mu_{1}\dots \mu_{J}}}{\td \mu^{2}} = - \frac{\As(\mu)}{4 \pi} \hat{\gamma}(J) ~ \mathcal{O}^{\mu_{1}\dots \mu_{J}}\,, \qquad 
    \hat{\gamma}(J) =
    \begin{pmatrix}
    \gamma_{qq}(J) ,&  2n_{f} \gamma_{qg}(J),& 0 \\
    \gamma_{qg}(J) , & \gamma_{gg}(J) ,& 0 \\
    0, & 0, & \gamma_{g\tilde{g}}(J)
    \end{pmatrix}\,,
\end{align}
where we have again factorised the $\As/(4 \pi)$ for convenience. As the integrals to form light-ray operators commute with the scale derivative, this also gives the renormalisation of the twist-2 light-ray operators:
\begin{align}
    \mu^{2}\frac{\td \, \mathbb{O}^{[J]}(\vec n_2;\mu)}{\td \mu^{2}} = - \frac{\As(\mu)}{4 \pi} \hat{\gamma}(J) ~ \mathbb{O}^{[J]}(\vec n_2;\mu)\,.
\end{align}
At one loop accuracy in the anomalous dimension and $\beta$ function, i.e. single-logarithmic accuracy in $\As (Q)\ln (Q/\mu)$, this flow equation is solved by
\begin{align}
 \mathbb{O}^{[J]}(\vec n_2;\mu) = \left(\frac{\As(Q)}{\As(\mu)}\right)^{\frac{\hat{\gamma}(J)}{\beta_{0}}}\mathbb{O}^{[J]}(\vec n_2;Q)\,.
\end{align}

We are interested in the 2-point correlator measured on a sample of unpolarised jets initiated by quarks at the hard scale $Q$. The jet content remains unpolarised but otherwise unconstrained at the scale $\mu$. Hence, the resummed OPE must sum over the contributions of $\mathbb{O}^{[3]}_{q}(\vec n_2;\mu)$ and $\mathbb{O}^{[3]}_{g}(\vec n_2;\mu)$. Thus, at single-logarithmic accuracy, the OPE is given by 
\begin{align}
\< \mathcal{E}(\vec n_1)\mathcal{E}(\vec n_2) \>=& -\frac{\As(\mu)}{8\pi^2} \frac{1}{2(n_1\cdot n_2)} \Bigg( (\gamma_{qq}(2)-\gamma_{qq}(3)+\gamma_{gq}(2)-\gamma_{gq}(3))\left[\left(\frac{\As(Q)}{\As(\mu)}\right)^{\frac{\hat{\gamma}(3)}{\beta_{0}}}\right]_{qq}  \nonumber \\
&
+(2 n_{f}(\gamma_{qg}(2)-\gamma_{qg}(3))+\gamma_{gg}(2)-\gamma_{gg}(3))\left[\left(\frac{\As(Q)}{\As(\mu)}\right)^{\frac{\hat{\gamma}(3)}{\beta_{0}}}\right]_{gq}\Bigg) \<\mathbb{O}_q^{[3]}(\vec n_2) \> \nonumber \\
&+\mathcal{O}\left(\frac{\As( Q)^2 \ln (Q/\mu)}{(n_1\cdot n_2)}\right)+\mathcal{O}((n_1\cdot n_2)^0)\,.
\end{align}
A natural scale choice for $\mu$ is the average momentum exchange between the two points in correlator which is $\sim \theta Q$. At the risk of being overly verbose, we can now write $\tau_{i}$ for QCD at next-to-leading (single) log accuracy in the small angle:
\begin{align}
    (n_{1}\cdot n_{2})^{\frac{\tau_{i}-2}{2}} \mathbb{O}^{[3]}_{i}(\vec n_2;\mu) \mapsto \sum_{j}\left[( n_{1}\cdot n_{2})^{-\frac{\hat{\gamma}(J)\ln \left(\As(n_{1}\cdot n_{2} Q^2)/\As(Q^2)\right)}{\beta_{0}\ln n_{1}\cdot n_{2}} }\right]_{ij} \mathbb{O}^{[3]}_{j}(\vec n_2;Q)\,.
\end{align}

\subsection{Summary of $g^{(n)}$ at Fixed and Running Coupling}
\label{app:gn}

We summarise here the results of the discussions in this appendix as expressions for $g^{(1)}$, provided the two-point correlator is measured on a quark jet and $\td\hat{\sigma}^{\mathrm{vac}}$ is given at $\mathcal{O}(\As)$.

At fixed coupling we have that: 
\begin{align}
    &g^{(1)}  =  \\
    &\frac{(\gamma_{gg}(3)+\gamma_{qq}(3)+A)~\theta^{\frac{\As}{8\pi}  \left(\gamma_{gg}(3)+\gamma_{qq}(3)-A\right)+\mathcal{O}\left(\As^2\right)} + 2 \gamma_{gq}(3) ~\theta^{\frac{\As}{8\pi}  \left(\gamma_{gg}(3)+\gamma_{qq}(3)+A\right)+\mathcal{O}\left(\As^2\right)}}{2 \gamma_{gq}(3)+\gamma_{gg}(3)+\gamma_{qq}(3)+A}  +  \mathcal{O}(\theta)\,, \nonumber
\end{align}
where $A$ and $\gamma_{ab}$ where defined in the previous section and are given again below.\footnote{The statement in section~\ref{sec:two-point} that $g^{(1)} = \theta^{\gamma(3)}$ should really be read as a matrix relation due to operator mixing. When expanded, this leads to the slightly more complicated result given above. If one imagines that the QCD quark operator did not mix with gluonic operators (i.e. taking the $\gamma_{gq},\gamma_{qg}\rightarrow 0$ limit), then the given result simplifies to $g^{(1)} = \theta^{\frac{\As}{4\pi}\gamma_{qq}(3)}$. } With a running coupling, at one-loop and at single logarithmic accuracy,
\begin{align}
    g^{(1)}  =& \left(\left[\left(\frac{\As(Q)}{\As( \theta Q)}\right)^{\frac{\hat{\gamma}(3)}{\beta_{0}}}\right]_{qq}+\frac{2 n_{f}(\gamma_{qg}(2)-\gamma_{qg}(3))+\gamma_{gg}(2)-\gamma_{gg}(3)}{\gamma_{qq}(2)-\gamma_{qq}(3)+\gamma_{gq}(2)-\gamma_{gq}(3)}\left[\left(\frac{\As(Q)}{\As( \theta Q)}\right)^{\frac{\hat{\gamma}(3)}{\beta_{0}}}\right]_{gq}\right) \nonumber \\
    & +\mathcal{O}\left(\As( Q)^{n} \ln ( \theta)^{n-1}\big|_{n \geq 1}\right) +  \mathcal{O}(\theta)\,,
\end{align}
where 
\begin{align}
    & \left[\left(\frac{\As(Q)}{\As( \theta Q)}\right)^{\frac{\hat{\gamma}(3)}{\beta_{0}}}\right]_{qq}  =\frac{1}{2 A}\left(\frac{\As(Q)}{\As(\theta Q)}\right)^{\frac{1}{2} (B_{+}-A)} \left(A \left(\left(\frac{\As(Q)}{\As(\theta Q)}\right)^A+1\right)-B_{-} \left(\left(\frac{\As(Q)}{\As(\theta Q)}\right)^A-1\right)\right)\,, \nonumber \\ 
    & \left[\left(\frac{\As(Q)}{\As( \theta Q)}\right)^{\frac{\hat{\gamma}(3)}{\beta_{0}}}\right]_{gq} = \frac{1}{A}\left(\frac{\As(Q)}{\As(\theta Q)}\right)^{\frac{1}{2} (B_{+}-A)} \gamma_{gq}(3) \left(\left(\frac{\As(Q)}{\As(\theta Q)}\right)^A-1\right)\, , \nonumber \\
    &A = \sqrt{(\gamma_{gg}(3) - \gamma_{qq}(3))^2+  8 n_{f}  \gamma_{gq}(3) \gamma_{qg}(3)}\,, \nonumber \\
    &B_{\pm} = \gamma_{gg}(3) \pm \gamma_{qq}(3)\,.
\end{align}
For convenience we here present again:
\beq
\begin{split}
&\gamma_{qq}(J)=C_F\left( 4\left(\psi^{(0)}(J+1)+\gamma_E\right)-\frac{2}{J(J+1)}-3\right)\,,\quad 
\gamma_{qg}(J)=-T_F \frac{2(J^2+J+2)}{J(J+1)(J+2)}\,, \\
&\gamma_{gg}(J)= 4 C_A \left( \psi^{(0)}(J+1)+\gamma_E -\frac{1}{(J-1)J}-\frac{1}{(J+1)(J+2)} \right)-\beta_0\,,\\
&\gamma_{gq}(J)=-C_F \frac{2(J^2+J+2)}{(J-1)J(J+1)}\,.
\end{split}
\eeq
where $\psi^{(0)}(z) = \Gamma'(z)/\Gamma(z)$ is the digamma function, and $\beta_0 = 11C_A/3  - 4n_fT_F/3  $\,.

At leading-order $g^{(2)}=1$. Expressions for $g^{(2)}$beyond fixed order require the incorporation of either track functions or fragmentation functions and will be provided in a subsequent publication.

\section{Additional Figures}
\label{app:figs}

\begin{figure}
\centering
\centering
\includegraphics[width=0.50\textwidth]
{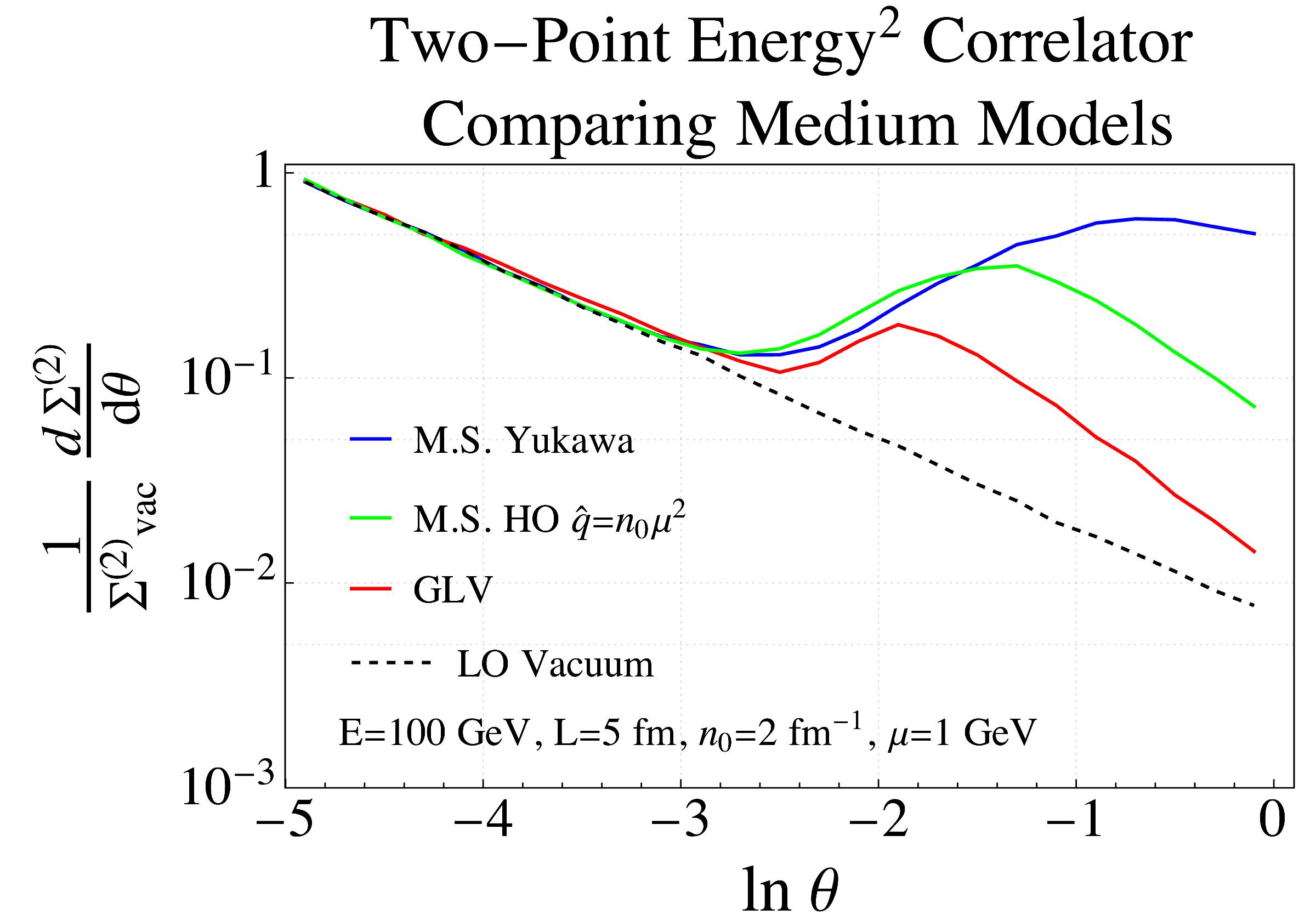}\includegraphics[width=0.50\textwidth]{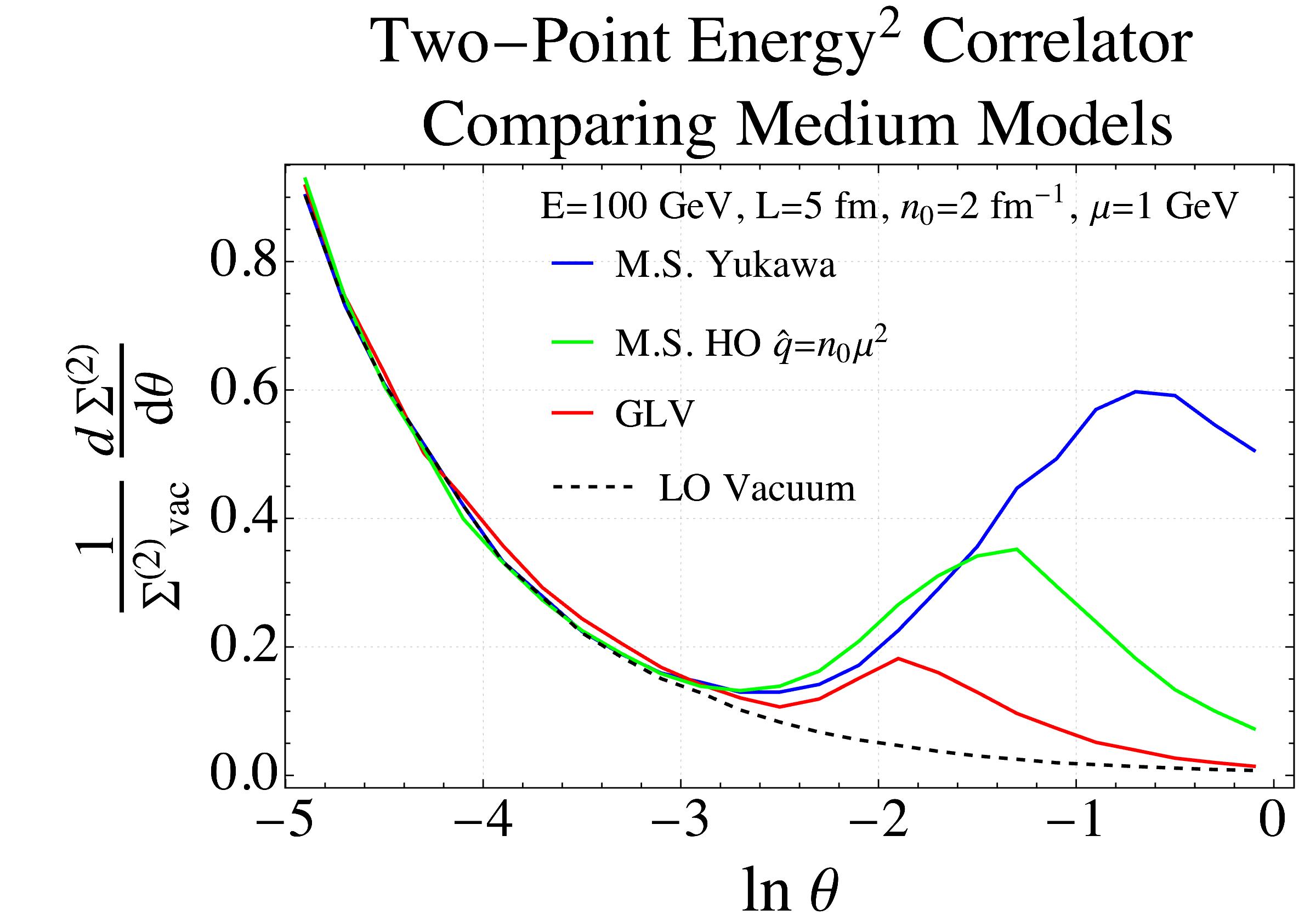}\newline \includegraphics[width=0.50\textwidth]{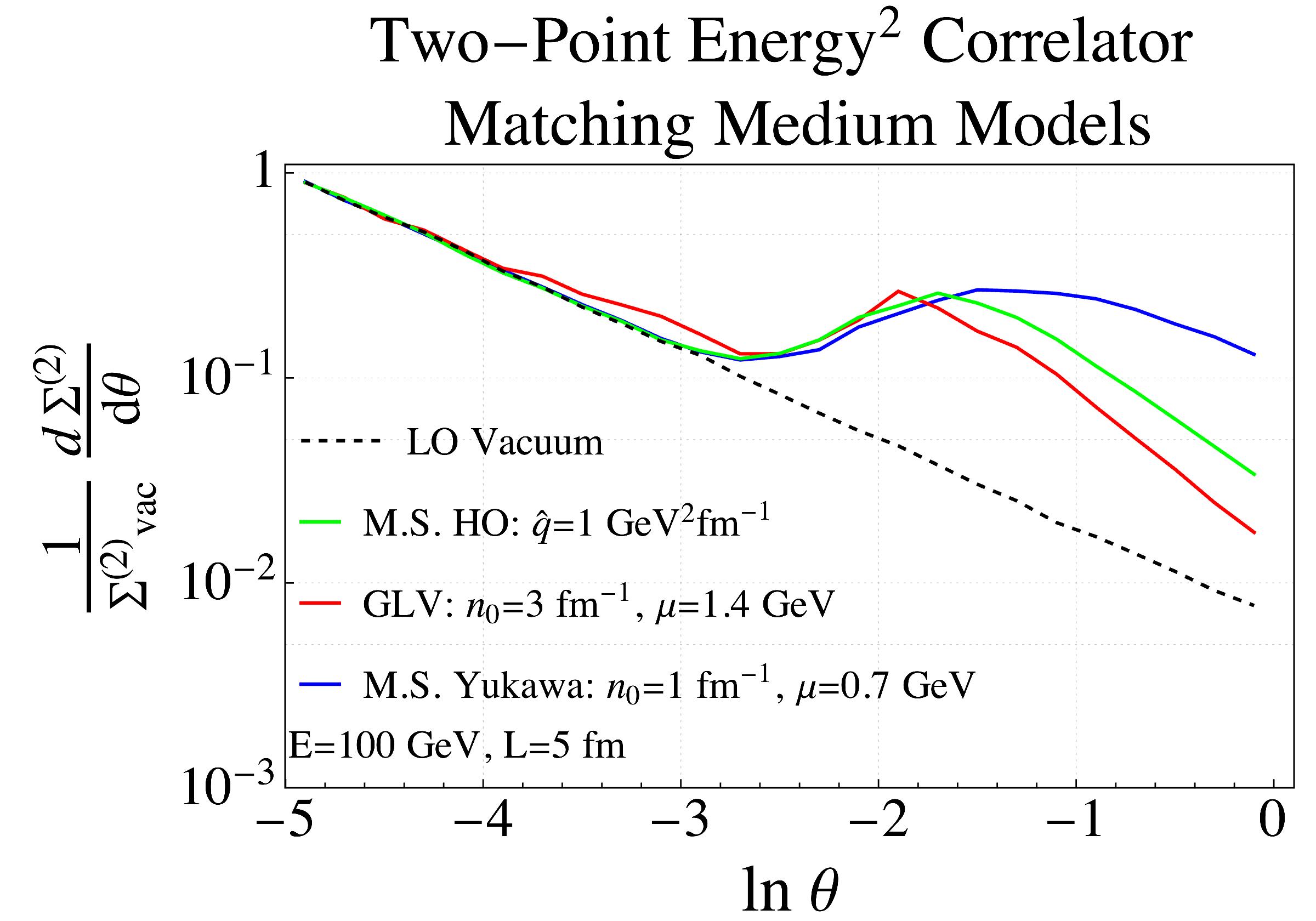}\includegraphics[width=0.50\textwidth]{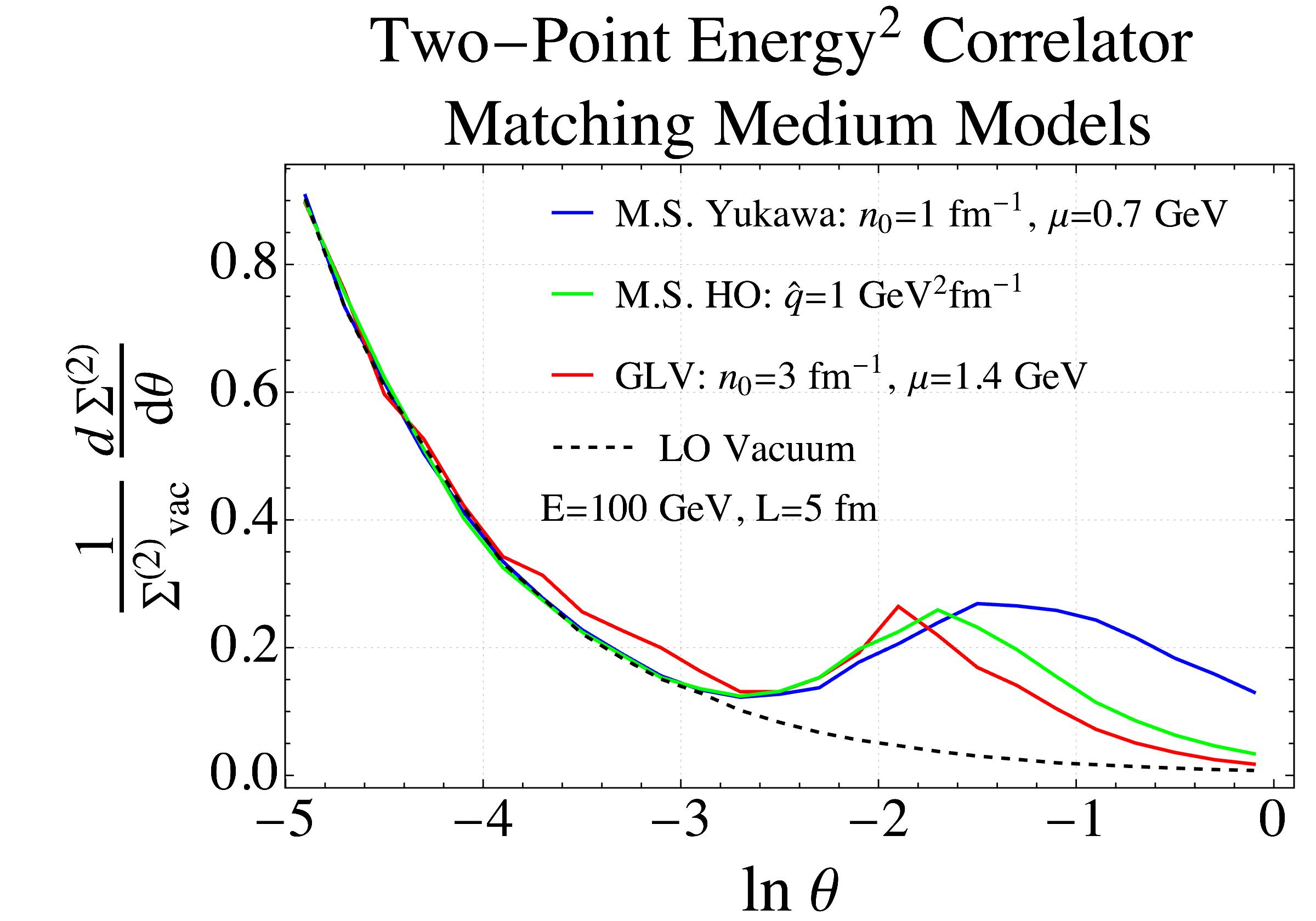}
\caption{Top: The $n=2$ EEC in \eqref{eq:master} of an $E=100$\,GeV quark jet evaluated through the multiple scattering HO (green), multiple scattering Yukawa (blue), and single scattering GLV (red) approaches compared to the vacuum LO result (black dashed). Both plots show the same curves, either with (left) or without (right) log scaling on the y-axis. Bottom: Same as the top panel for different sets of medium parameters chosen so the amplitude and onset of the medium enhancement qualitatively match among the different models. All curves are normalised by the integrated vacuum result $\Sigma^{(2)}_{\rm vac}$. }
\label{fig:combined_bump_n2}
\end{figure}

In this appendix we present additional figures showing the two-point correlator spectra for the different jet quenching formalisms considered in this manuscript, different sets of parameters, and energy weights. 

We first present in figure~\ref{fig:combined_bump_n2} results on the two-point $n=2$ energy correlator, ${\rm d}\Sigma^{(2)}/{\rm d}\theta$, for a quark-initiated jet with initial energy $E=100$\,GeV computed within the three jet quenching formalisms considered in this manuscript. These results for the $n=2$ energy weight qualitatively agree with the corresponding  $n=1$ ones shown in figure~\ref{fig:combined_bump}.
We note that in this figure instead of using the NLL vacuum result as done for the  $n=1$ case, we employ the LO one, which is a good first approximation away
from the $\theta \rightarrow 0$ divergence. As we discuss in appendix~\ref{app:app_1}, including the NLL vacuum for the $n=2$ spectrum requires the inclusion of non-perturbative functions (track or fragmentation functions), which is beyond the scope of this paper.

\begin{figure}
\centering
\subfloat[]{
\includegraphics[width=0.50\textwidth]{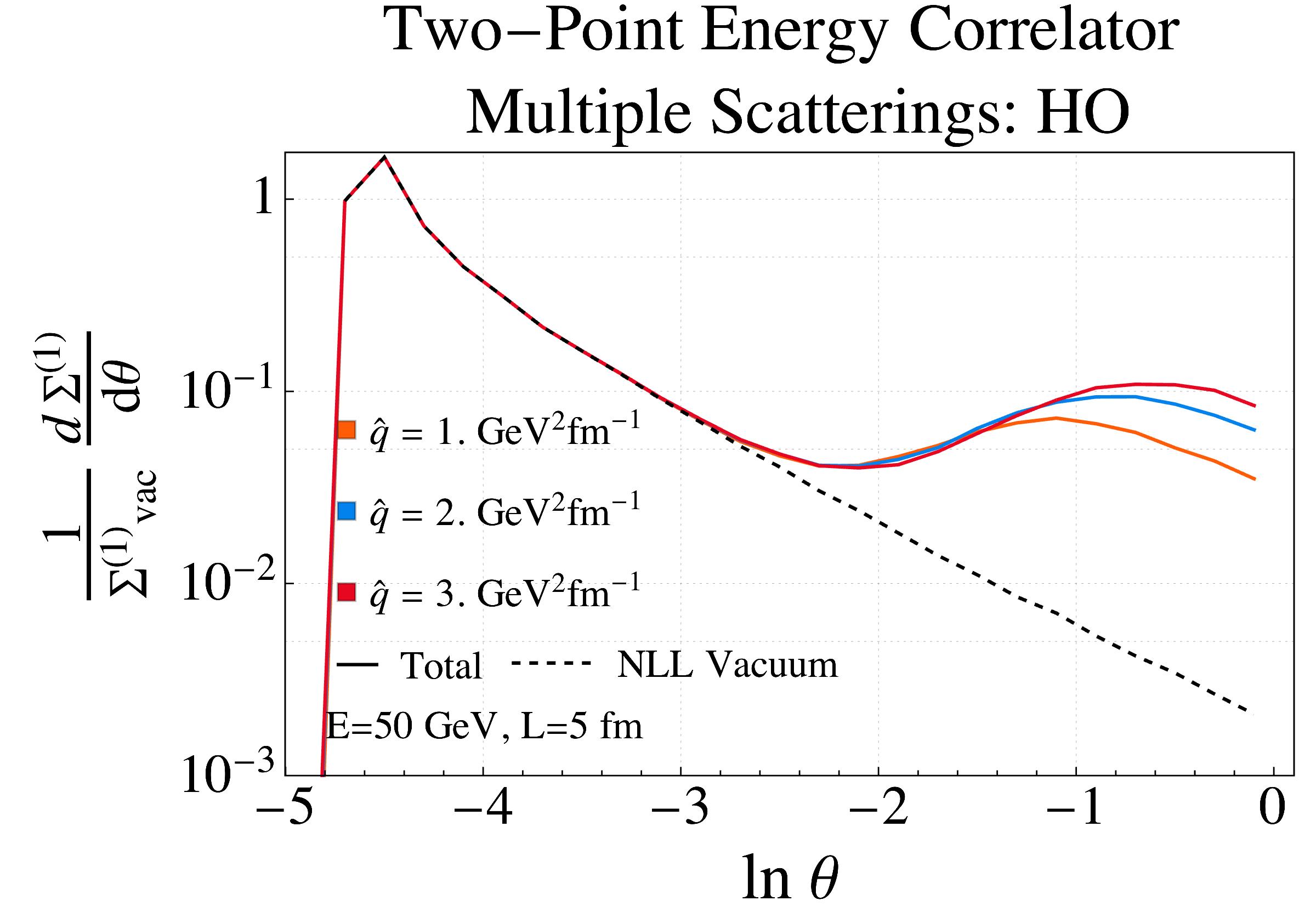}}
\subfloat[]{
\includegraphics[width=0.50\textwidth]{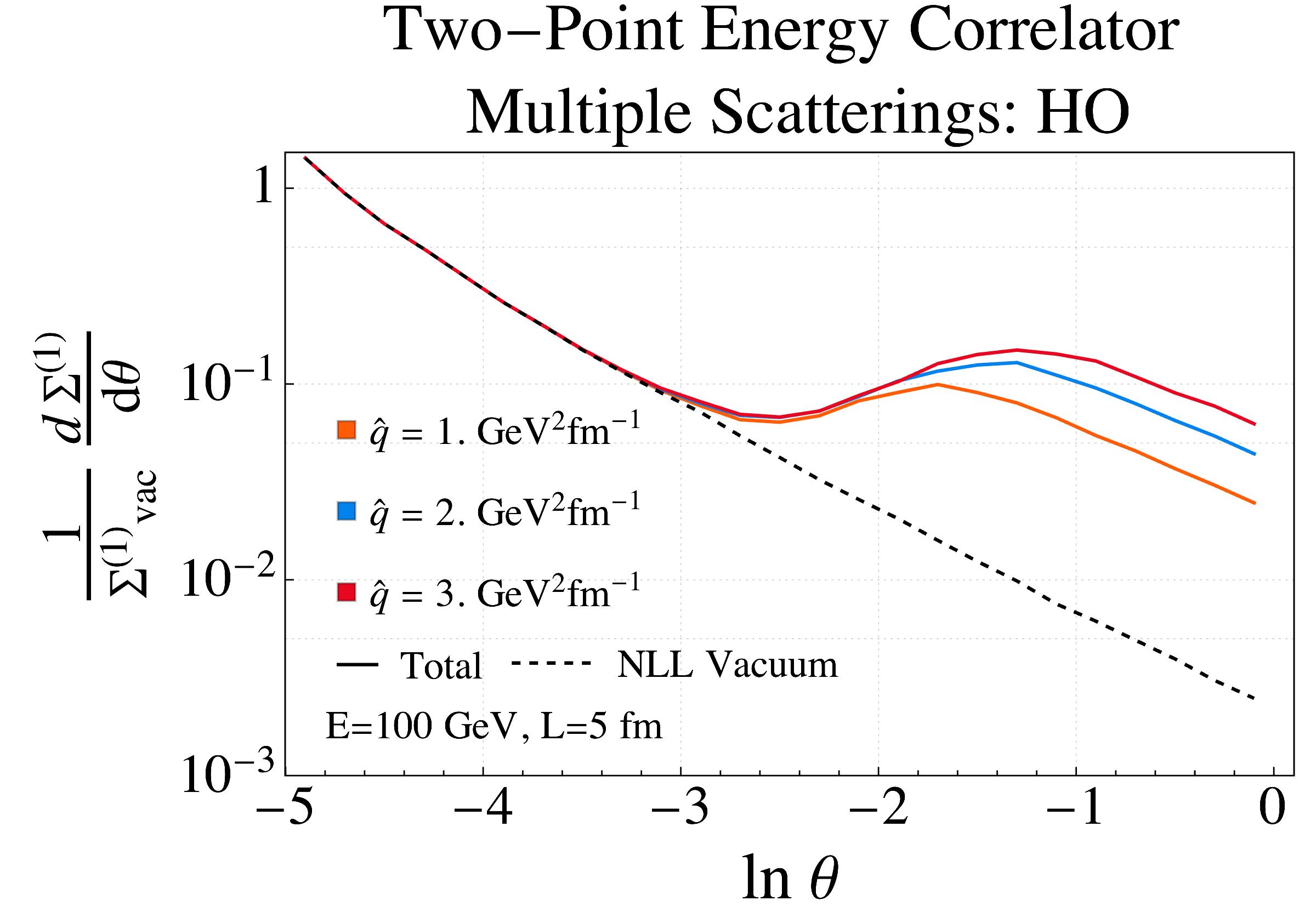}} \newline
\subfloat[]{\includegraphics[width=0.50\textwidth]{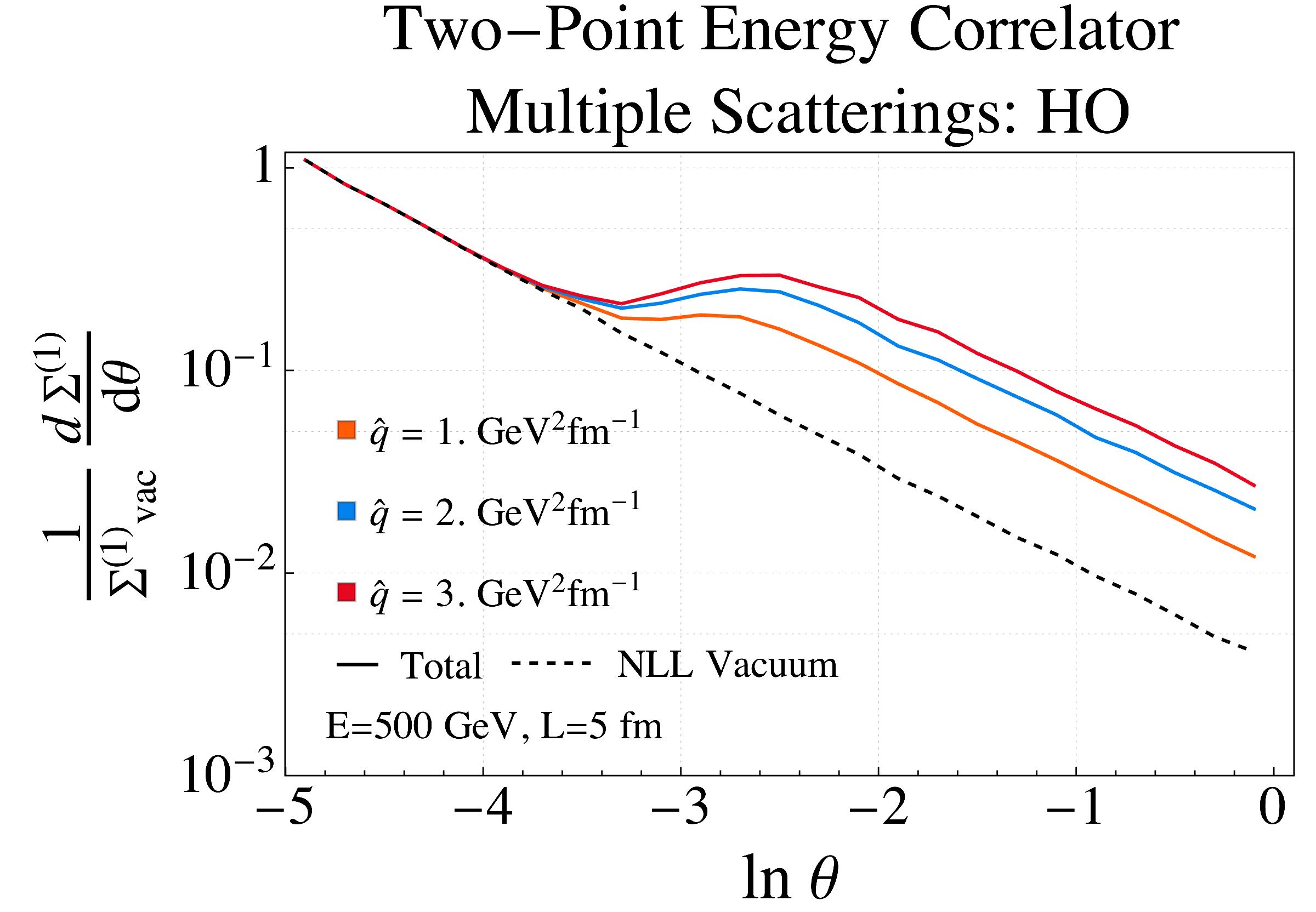}}
\caption{The $n=1$ EEC of an $E=50$ GeV (top left), $E=100$\, GeV (top right), and $E=500$\, GeV (bottom) quark jet computed within the multiple scattering HO approach (see section~\ref{subsec:HO}) for a medium of length $L=5$\,fm  and several values of $\hat{q}$: 1 ${\rm GeV}^2$/fm (orange), 2 ${\rm GeV^2}$/fm (blue), and  3 ${\rm GeV^2}$/fm (red) compared to the vacuum NLL result (black dashed). All curves are normalised by the integrated vacuum result $\Sigma^{(1)}_{\rm vac}$.}
  \label{fig:HO_EEC_app}
\end{figure}

Figure~\ref{fig:HO_EEC_app} shows results on the two-point $n=1$ energy correlator, ${\rm d}\Sigma^{(1)}/{\rm d}\theta$, computed within the multiple scattering approach and the harmonic approximation for quark jets with different initial energies and the same values of the medium parameters ($\hat{q}$ and $L$). As we already discussed in \ref{fig:HOvaryingParameters}, the onset angle of the medium enhancement for a given jet energy $E$ and medium length $L$ is clearly independent of the value of $\hat{q}$. In subfigure (a), we can observe that both the vacuum and in-medium correlators artificially turn over at very small angles, due to the presence of the QCD Landau pole. Indeed, the coupling diverges when $\theta=\Lambda_{\rm QCD}/E$, becoming imaginary for $\theta<\Lambda_{\rm QCD}/E$, and we have removed the resulting imaginary bins from the distribution. The Landau pole signals that for $\theta\sim \Lambda_{\rm QCD}/E$ non-perturbative physics should become a dominant effect, as  already observed in experimental data \cite{Komiske:2022enw}. The Landau pole is not visible in the other subplots of figure~\ref{fig:HO_EEC_app}, since the larger values of the jet energy result in a much smaller value of $\Lambda_{\rm QCD}/E$.

We can also observe in this figure that while the tail of the medium-enhancement to the left of the peak angle agrees for all the $\hat{q}$ values presented in panel (a), it presents a strong sensitivity to $\hat{q}$ in panel (c). This change in the scaling with $\hat q$ is indicative of the emergence of colour coherence when moving from the DC region shown in subfigure (a) to the PC regime shown in subfigure (c).

\begin{figure}
\centering
\subfloat[]{
\includegraphics[width=0.50\textwidth]{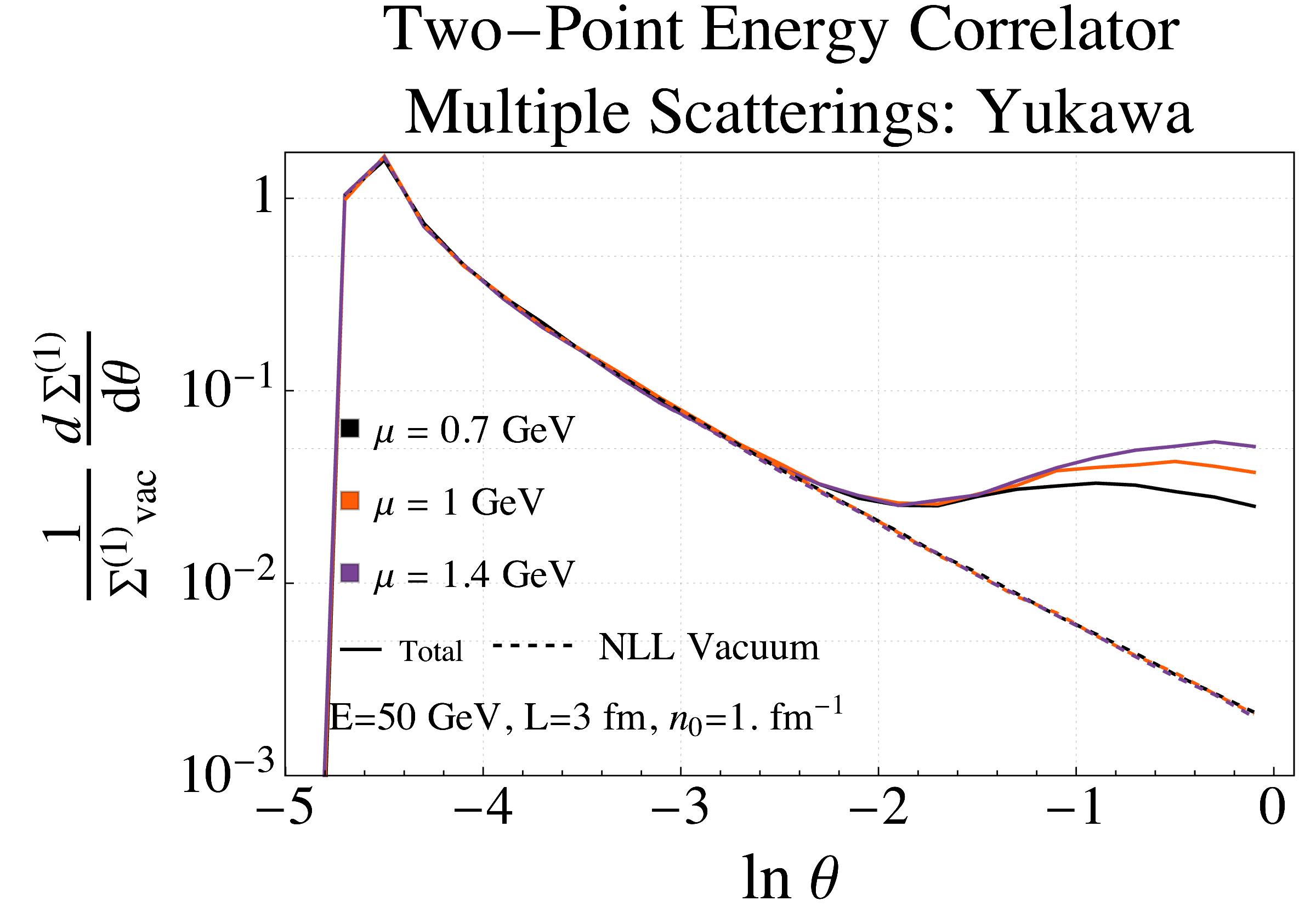}}
\subfloat[]{
\includegraphics[width=0.50\textwidth]{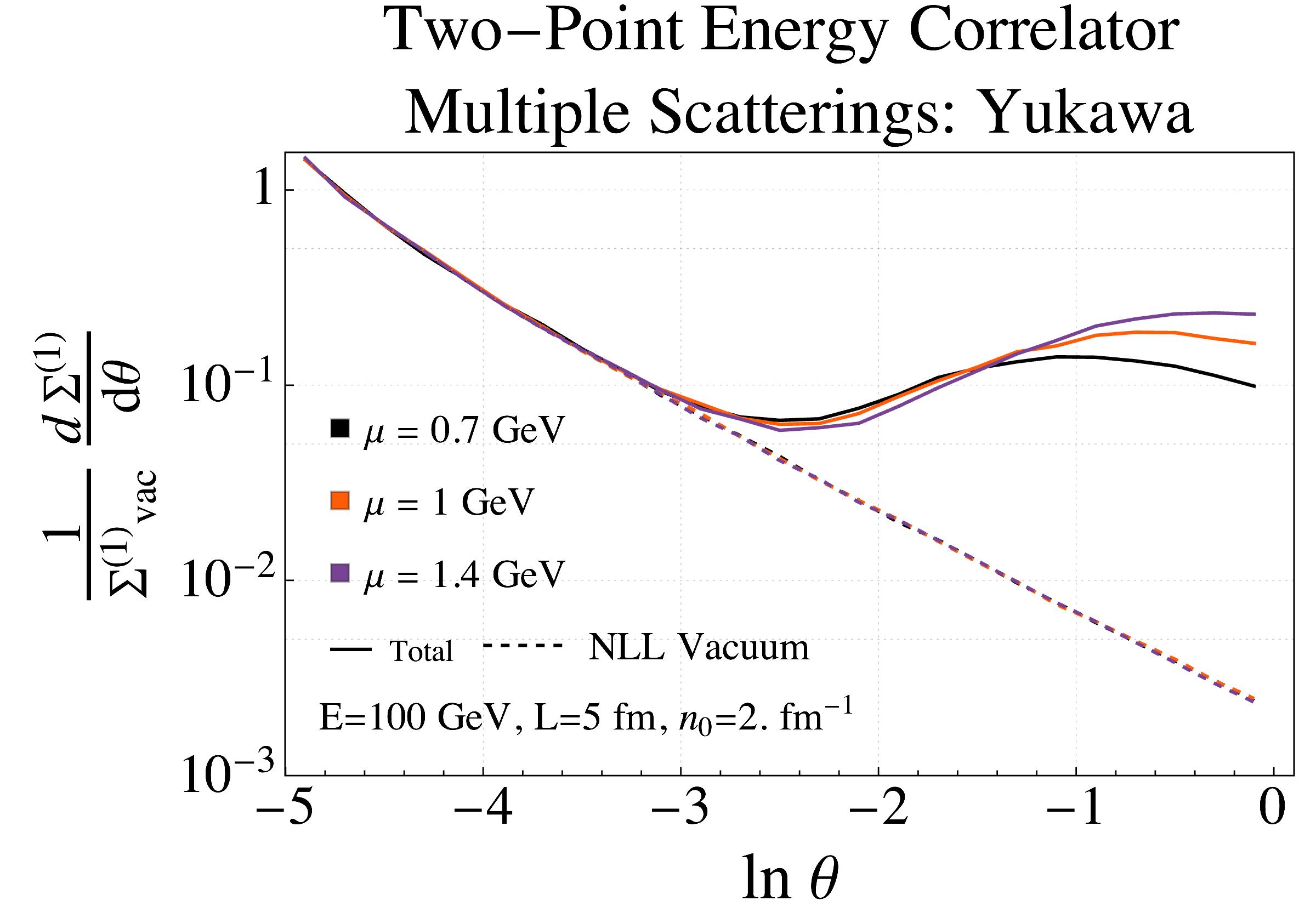}} \newline
\subfloat[]{\includegraphics[width=0.50\textwidth]{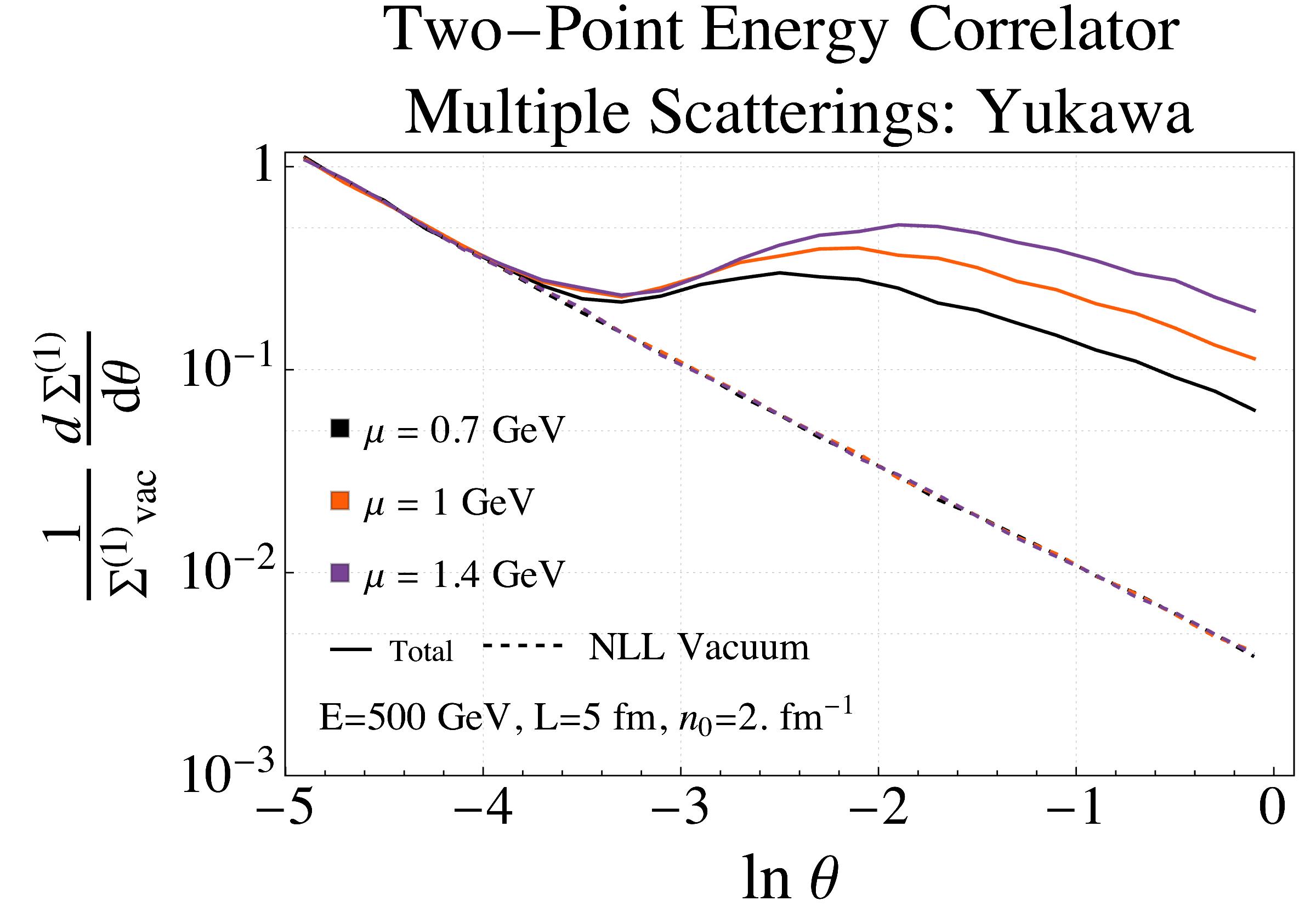}}
\caption{Subfigure (a): The $n=1$ EEC of an $E=50$\,GeV quark jet computed within the multiple scattering Yukawa approach (see section~\ref{subsec:Yuk}) for a medium of $L=3$\,fm, $n_0=1\,{\rm fm}^{-1}$ and several values of $\mu$: $0.7$ GeV (black), $1$\,GeV (orange), and  $1.4$\,GeV (purple) compared to the NLL vacuum  result. Subfigure (b): The $n=1$ EEC of an $E=100$\,GeV quark jet computed within the multiple scattering Yukawa approach for a medium of $L=5$\,fm, $n_0=2\,{\rm fm}^{-1}$ and several values of $\mu$: $0.7$\,GeV (black), $1$\,GeV (orange), and $1.4$\,GeV (purple) compared to the NLL result. Subfigure (c): The same as subfigure (b) for an $E=500$\,GeV jet. All curves are normalised by the integrated vacuum result $\Sigma^{(1)}_{\rm vac}$.}
  \label{fig:Yukawa_EEC_app}
\end{figure}

We present in figure~\ref{fig:Yukawa_EEC_app} the $n=1$ EEC of quark initiated jets with different initial energies computed within the multiple scattering approach with a Yukawa parton-medium interaction model described in section~\ref{subsec:Yuk} for several values of the medium parameters. As for the HO calculation shown  in figure~\ref{fig:HO_EEC_app}, we can see in  panel (a) of this figure how the transition to hadronisation arises at an angular scale $\sim \Lambda_{\rm QCD}/E$. We observe again that the onset angle above which the in-medium EEC curves deviate from the vacuum result is only dependent of the energy of the jet $E$ and the length of the medium $L$. This outcome is in agreement with our expectations, since only splittings with $t_{\rm f} =2/(z(1-z)E\theta)$ smaller than the length of the medium are presumed to have a significant medium modification.

\begin{figure}
\centering
\subfloat[]{
\includegraphics[width=0.50\textwidth]{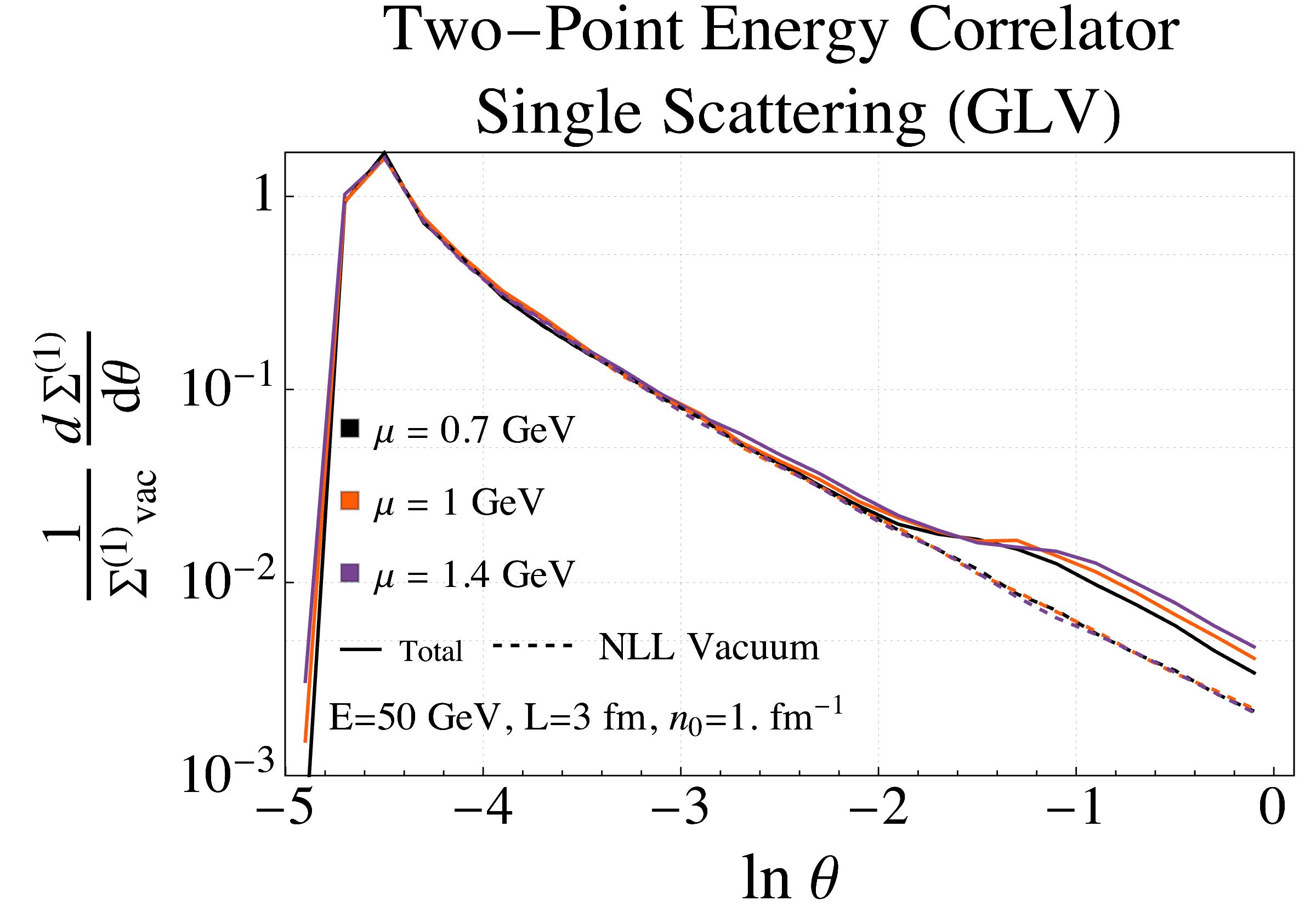}}
\subfloat[]{
\includegraphics[width=0.50\textwidth]{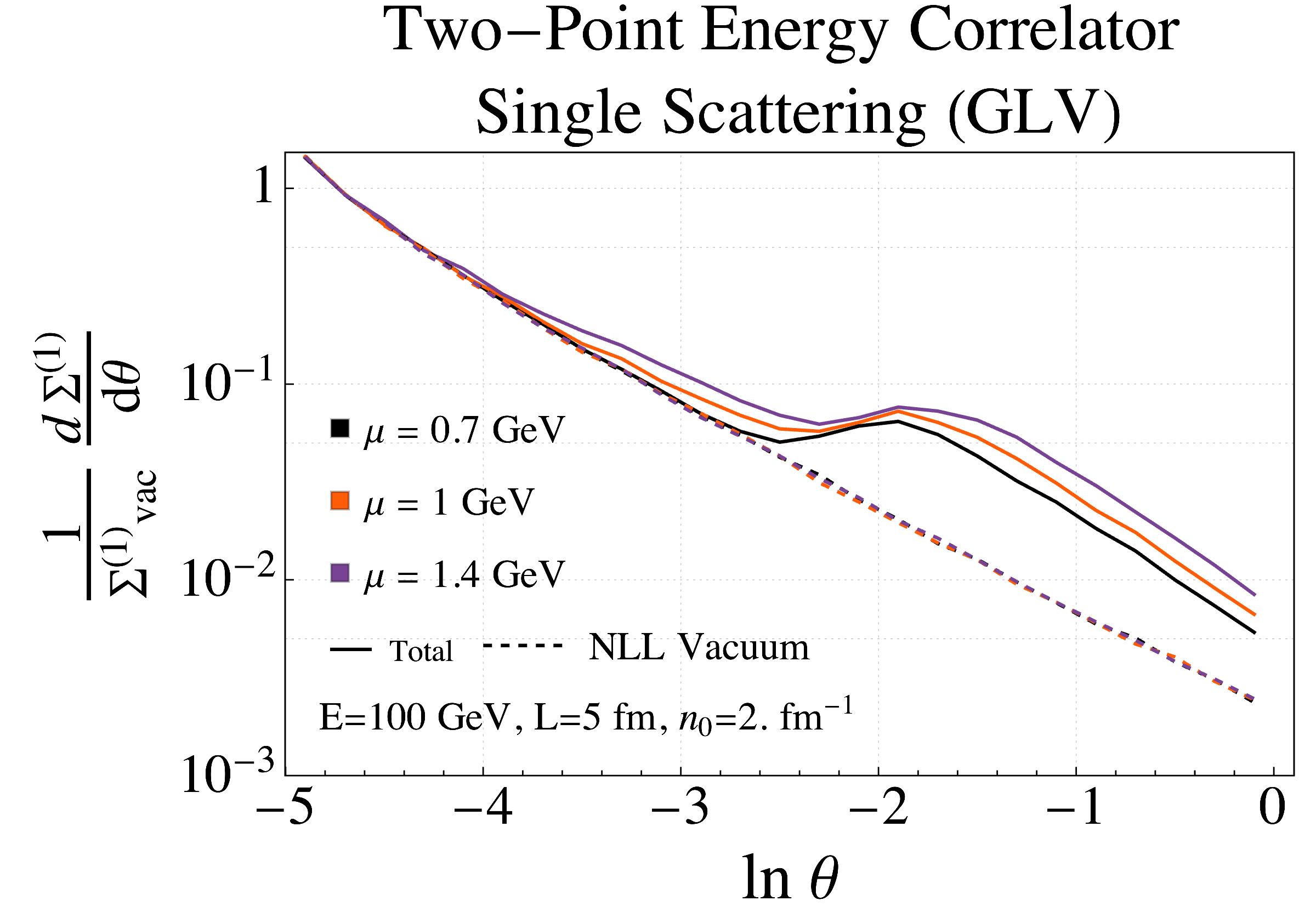}} \newline
\subfloat[]{\includegraphics[width=0.50\textwidth]{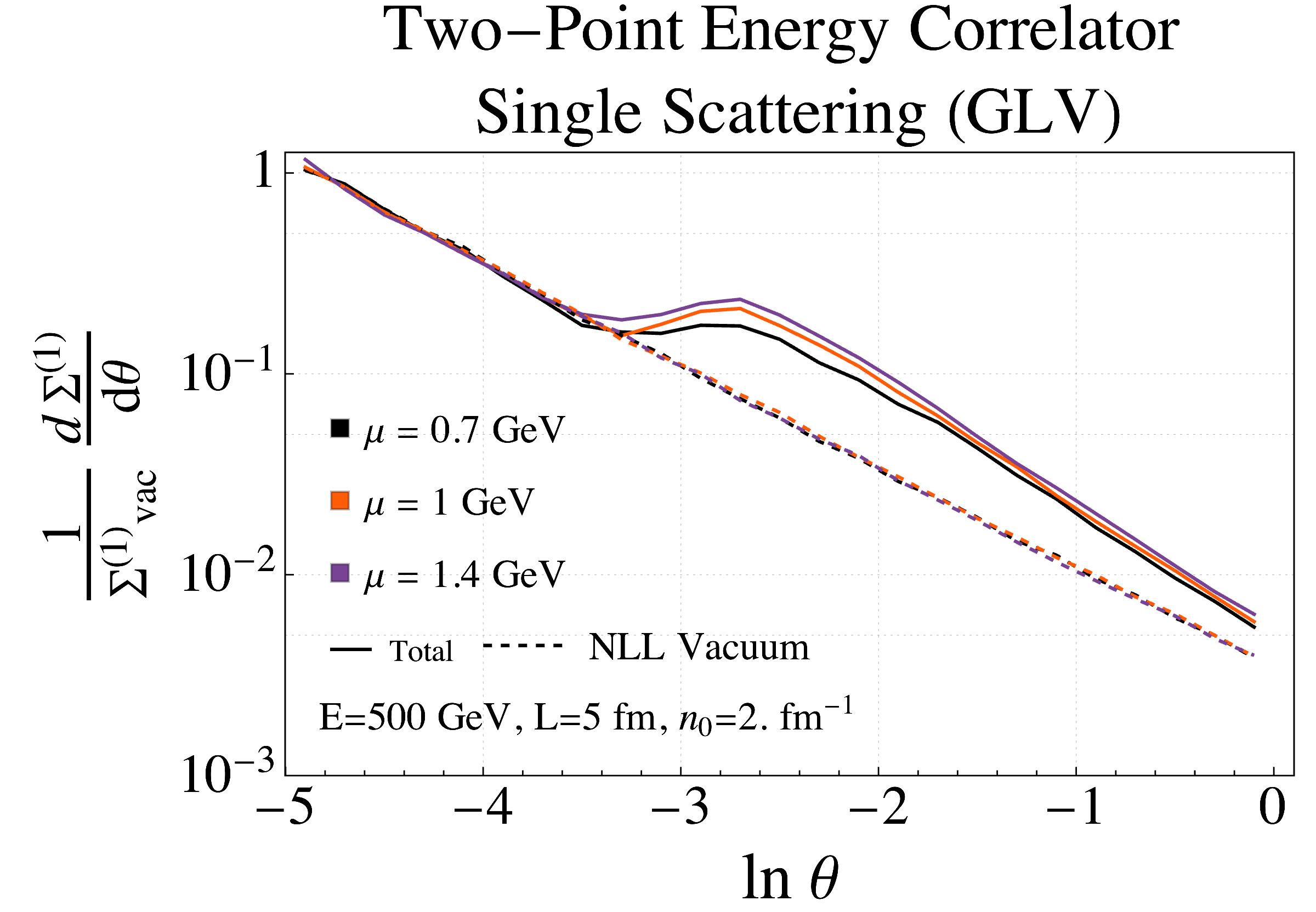}}
\caption{Subfigure (a): The $n=1$ EEC of an $E=50$\,GeV quark jet computed within the single scattering (GLV) approach (see section~\ref{subsec:GLV}) for a medium of $L=3$\,fm, $n_0=1\,{\rm fm}^{-1}$ and several values of $\mu$: $0.7$\,GeV (black), $1$\,GeV (orange), and  $1.4$\,GeV (purple) compared to the NLL vacuum result. Subfigure (b): The $n=1$ EEC of an $E=100$\,GeV quark jet computed within the single scattering (GLV) approach for a medium of $L=5$\,fm,  $n_0=2\,{\rm fm}^{-1}$ and  several values of $\mu$: $0.7$\,GeV (black), $1$\,GeV (orange), and  $1.4$\,GeV (purple) compared to the NLL vacuum result. Subfigure (c): The same as the subfigure (b) for an $E=500$\,GeV quark jet. All curves are normalised by the integrated vacuum result $\Sigma^{(1)}_{\rm vac}$.}
  \label{fig:GLV_EEC_app}
\end{figure}

Finally, we show in figure~\ref{fig:GLV_EEC_app} the $n=1$ EEC of quark initiated jets with different initial energies computed within the single scattering (GLV) approach described in section~\ref{subsec:GLV} for several values of the medium parameters.  We again observe the transition to the non-perturbative region of the EEC at an angular scale $\sim \Lambda_{\rm QCD}/E$. 

\bibliographystyle{JHEP}

\bibliography{references}

\end{document}